\documentclass[11pt]{article}
\usepackage[letterpaper,margin=1in]{geometry}
\usepackage{iftex}
\ifPDFTeX
  \usepackage[T1]{fontenc}
  \usepackage[utf8]{inputenc}
\else
  \usepackage{fontspec}
\fi
\usepackage{amsmath,amssymb,mathtools,mathrsfs}
\usepackage{microtype}
\usepackage{longtable,booktabs,array,calc}
\usepackage{enumitem}
\usepackage[unicode,breaklinks=true,colorlinks=true,linkcolor=blue,citecolor=blue,urlcolor=blue]{hyperref}

\allowdisplaybreaks[2]
\numberwithin{equation}{section}

\title{Mutually unbiased bases:\\[0.6ex]common states, completion, and extension obstructions}
\author{%
	Hao-Yu Sun$^{1,2,3}$\\[0.4ex]
	{\small $^1$ Weinberg Institute, The University of Texas at Austin}\\
	{\small 2515 Speedway, C1600, Austin, Texas 78712, USA}\\
	{\small $^2$ Mathematics Department, Austin Community College}\\ {\small 6101 Highland Campus Drive, Austin, Texas 78752, USA}\\
	{\small $^3$ SpaceXAI, USA}\\
	{\small \href{mailto:hkdavidsun@utexas.edu}{%
			\nolinkurl{hkdavidsun@utexas.edu}}}%
}
\hypersetup{pdfauthor={Hao-Yu Sun}}
\begin{document}
\maketitle
\begin{abstract}\label{abstract}
Which pure states give uniform outcomes in several mutually unbiased measurements, and when can they form another measurement basis? We distinguish common unbiased states, their orthonormal completion, and further extension. For a displayed phase-deformed Fourier pair in dimension \(2m\), a clock symmetry and skew antiunitary complete every prescribed common unbiased state to an explicit orthonormal basis. In dimension six, an amplitude-profile obstruction proves strong unextendibility of a given mutually unbiased triple when its original third-basis rays are invariant under the ternary clock. We give an exact nonproduct example and an application to a companion whose columns are localized at single Fourier frequencies. For every exact Hadamard matrix in a specified closed entrywise neighborhood of a fixed reference matrix, the coordinate and Hadamard bases have exactly 48 common unbiased rays but no third measurement basis, with a quantitative lower bound on the squared Gram defect of every six-column candidate.

We also study auxiliary complex equations for common states, retaining algebraic multiplicities. For a flat unitary of order six with all \(3\times3\) minors nonzero, the projective ray scheme is finite; each singular two-minor contributes two reduced isolated boundary points, and the affine coordinate algebra is finite-dimensional over \(\mathbb C\). In every order, a critical-algebra identity preserves multiplicities; in dimension six it gives finiteness for a fixed phase fibre. Specialization over a discrete valuation ring controls complete reciprocal-orthogonal collections. A nonphysical Tao/Potts collection separates algebraic from physical solutions. These results do not determine the global maximum number of mutually unbiased bases or classify all triples.
\end{abstract}
\clearpage
\begingroup\setlength{\parskip}{0pt}\tableofcontents\endgroup

\section{Introduction}\label{introduction}

An orthonormal basis \(B=\{b_1,\ldots,b_d\}\) of \(\mathbb C^d\) defines a rank-one projective measurement with outcomes \(P_i=|b_i\rangle\langle b_i|\). If the system is prepared in a unit vector \(\psi\), the Born probability of outcome \(i\) is \(\langle\psi,P_i\psi\rangle=|\langle b_i,\psi\rangle|^2\). Two bases \(B,C\) are \emph{mutually unbiased} (MU) when preparing any state of one basis gives a uniform distribution in the other:
\begin{equation}
|\langle b_i,c_j\rangle|^2=\frac1d\qquad(1\le i,j\le d).
\label{eq:measurement-unbiasedness}
\end{equation}
This is an exact finite-dimensional expression of complementarity. Mutually unbiased measurements also organize quantum state determination: their probability data probe orthogonal directions in the space of density matrices \cite{WF,MR}. 

Let \(N(d)\) be the maximum cardinality of a set of pairwise mutually unbiased bases in dimension \(d\). The general bound is \(N(d)\le d+1\), and a \emph{complete set} of \(d+1\) bases exists in prime-power dimensions \cite{WF,MR}. The bound reflects the \(d-1\) independent traceless projector directions contributed by each basis: unbiasedness makes the corresponding subspaces orthogonal inside the \(d^2-1\)-dimensional traceless Hermitian space. Six is the first non-prime-power dimension beyond the trivial one-dimensional case. Triples in dimension six have long been known, while the conjecture \(N(6)=3\) motivated work on their possible extension \cite{BW,J,MW}.

Recent constructions by Cárdenes Wuttig and Tindall improve lower bounds for mutually unbiased families in several composite dimensions \cite{CWTMore}. Our constructive question is to complete every prescribed common-unbiased state for the fixed coordinate and phase-deformed Fourier bases displayed in \hyperref[result:theorem-6.1]{Theorem 6.1}. \hyperref[prescribed-ray-completion-in-dimension-2m]{Section 6} compares this construction with the prescribed-state completion already available for undeformed group-Fourier pairs.

The current context includes two OpenAI preprints dated 24 September 2026. \emph{The maximum number of mutually unbiased bases in dimension six} reports \(N(6)=3\) through a computer-assisted no-four-bases theorem. Its main statement explicitly invokes the arithmetic and compiler conditions of its Appendix A and the documented fresh complete execution of its Section 7 \cite{OAImax}. We cite that preprint result without certifying those execution conditions or claiming community acceptance or complete formal verification. \emph{Exact Fourier certificates for complex Hadamard matrices of order six} has a different conclusion: an individual-matrix Fourier coefficient vanishes outside the Tao cubic equivalence class; a separate exact certificate excludes seven bases and, with Weiner's completion theorem, gives \(N(6)\le5\) \cite{OAIfourier}. It is not another statement of the maximum-three theorem. Neither preprint is a premise of the proofs in this article.

Our question resolves more of the geometry of a given set of measurements than its global cardinality:
\begin{quote}
For given mutually unbiased measurements, which pure states are unbiased to all of them, when can those states form another orthonormal measurement basis, and what prevents further extension?
\end{quote}
A common unbiased pure state is only one possible column. Another basis requires \(d\) such states that are mutually orthogonal. Excluding a fourth full basis does not by itself exclude a single state unbiased to a given triple. Likewise, completing a \emph{prescribed} common state is a stronger request than finding some third basis. Finally, an auxiliary complex solution of the probability equations need not represent a physical pure state. These distinctions organize the results below.

\paragraph{From a prescribed state to a measurement basis.}
For the displayed phase-deformed Fourier pair \((I,K_\eta)\) in dimension \(2m\), consisting of the coordinate basis and the columns of the phase-parameter family displayed below, \hyperref[result:theorem-6.1]{Theorem 6.1} completes \emph{every} prescribed common unbiased state \(g\). A diagonal clock operator \(D\) multiplies the two-coordinate blocks by successive roots of unity and generates \(m\) candidate columns. A phase-dependent skew antiunitary \(J_\eta\), whose square is the negative identity, supplies another \(m\). The construction succeeds because unbiasedness to the fixed coordinate basis gives equal weight to each two-coordinate block, while the skew pairing makes the two local directions orthogonal. These identities check every entry of the matrix of column inner products, or Gram matrix, of
\(E_\eta(g)=[D^rg,D^rJ_\eta g]_{r=0}^{m-1}\).
The precise family, arbitrary unit phases and column order appear in \hyperref[principal-results]{Sections 2.2} and \hyperref[prescribed-ray-completion-in-dimension-2m]{6}. The antiunitary is a mathematical construction on specified vectors. General common-state existence is an attributed input \cite{IW}; the completion retains the chosen state.

A closed neighborhood of a fixed reference matrix illustrates the obstruction to completing candidates. \hyperref[result:theorem-9.3]{Theorem 9.3} treats every exact complex Hadamard matrix \(H\) within entrywise distance \(10^{-6}\) of the reference \(H_\star\) displayed in Appendix~\ref{app:source-neighborhood}. Although the reference belongs to the Karlsson family, the varying matrix need not. The coordinate basis and the columns of \(H/\sqrt6\) have exactly 48 common unbiased rays. Their possible-orthogonality graph joins pairs whose orthogonality has not been excluded by the interval overlap tests; it contains no orthonormal sextuple. For any six-column matrix \(V\) of unit common-unbiased representatives, including repeated ray labels, the squared Gram defect satisfies \(\|V^*V-I\|_F^2>1/1200\). Here \(\|W\|_F^2=\sum_{r,s}|W_{rs}|^2\); for unit columns, zero Gram defect is equivalent to an orthonormal basis. The neighborhood radius in normalized measurement-entry units is \(10^{-6}/\sqrt6\). The computer-assisted proof uses established enumeration and neighborhood methods \cite{BW,J,Crooks}. \hyperref[sec:source-neighborhood]{Section 9.3} gives the graph-to-gap argument, and Appendix~\ref{app:source-neighborhood} supplies the reference matrix and computational proof.

In dimension six, \hyperref[result:theorem-8.2]{Theorem 8.2} excludes any further common unbiased \emph{pure} state for a given pairwise-MU triple \((I,K_\eta,V)\), provided the six original rays of \(V\) are setwise invariant under \(D\). This is \emph{strong unextendibility}. Setwise invariance means that the clock permutes the given basis rays, allowing individual unit phases. The invariance is a condition on the given measurement, not merely a symmetry of the surrounding candidate equations. The proof compares the sine and cosine amplitude factors of the three block angles for two states and of their angle difference. Their probability constraints imply weak triangle inequalities. Heron's identity bounds each profile, while a cosine-difference identity forces incompatible row and column angle sums. Under the assumption of an additional common-unbiased ray, a separate argument connects this contradiction to the original third basis, including an aligned product alternative. An exact example beyond the two classified product-triple models and a localized application are given in \hyperref[applications-and-limits]{Section 9}. The maximally mixed state \(I_6/6\) still has uniform outcomes in every basis; the theorem excludes pure states satisfying those conditions, not all density operators.

\paragraph{Exact algebra for candidate states.}
The complementary part of the article studies the equations before imposing physical conjugation. For a state already unbiased to the coordinate basis, write its entries as \(z_i/\sqrt d\), with \(|z_i|=1\), and fix one global phase. Replacing conjugates by reciprocal variables gives an auxiliary complex algebra. Its solutions include nonphysical points, and its local algebra can retain multiplicities invisible in a list of solutions. This enlarged setting makes exact elimination and the simultaneous constraints on several columns accessible; it does not assign a new observable to every algebraic object.

A positional three-minor is the determinant obtained from any three selected rows and any three selected columns; the selections need not be consecutive or have matching indices. In order six these are the middle minors. For a flat unitary \(C\) of order six with all 400 positional three-minors nonzero, \hyperref[result:theorem-3.1]{Theorem 3.1} proves finiteness of the full common-ray algebra by analyzing its specified projective boundary. Every singular two-minor gives two reduced isolated boundary points. The tangent equations force \(2\delta t=4\delta t\), excluding a component through the boundary; projectivity then rules out a positive-dimensional family of candidates. \hyperref[result:proposition-3.6]{Proposition 3.6} translates the finite physical subset into a finite orthogonality graph: a six-clique gives another measurement basis, and the required cross-overlap edges test two companion bases.

\hyperref[result:proposition-4.1]{Proposition 4.1} identifies the full ray algebra in every order with a logarithmic critical algebra, defined by logarithmic derivative equations on the specified hyperplane complement, preserving local multiplicities. \hyperref[result:proposition-4.2]{Proposition 4.2} uses one row of a fixed diagonal-phase system to deduce finiteness of that entire fibre under the middle-minor hypothesis. \hyperref[result:theorem-5.1]{Theorem 5.1} then controls complete six-column algebraic companions over a discrete valuation ring (DVR). Their ordered reciprocal orthogonality relations, the vanishing bilinear pairings with coordinatewise reciprocals in both column orders, prevent specialization to the boundary and imply module finiteness; the full proof is in \hyperref[appendix-dvr-complete-companions-cannot-escape]{Appendix B}. These are auxiliary algebraic tools for constraining measurement data. The Tao/Potts example shows that even a reciprocal-orthogonal collection of six complex solutions may fail the physical unit-modulus conditions.

The two domains have different hypotheses. In dimension six the displayed \(K_\eta\) has vanishing middle minors, so its physical completion and nonextension are proved by symmetry rather than by \hyperref[result:theorem-3.1]{Theorem 3.1}. Conversely, finiteness for an eligible \(C\) does not guarantee enough mutually orthogonal physical candidates to form another basis. \hyperref[appendix-c13-exact-middle-minor-applicability]{Appendix A} gives an exact middle-minor applicability example.

Our results concern exact measurements and auxiliary complex algebras under the displayed hypotheses. They do not establish unrestricted \(N(6)=3\) or classify all six-dimensional triples, nor claim improved tomography or experimental performance, robustness to measurement noise, or a physical channel implementing the antiunitary construction on an unknown state.

\paragraph{Relation to earlier work and a reading route.}
The recent maximum-number preprint also uses candidate common vectors, orthogonality graphs and all cross-overlap constraints, while its global proof relies on an exhaustive phase cover and computer-assisted exclusions \cite{OAImax}. The Fourier preprint instead uses nonnegative expressions in character sums and rational invariant-moment certificates \cite{OAIfourier}. Under the maximum preprint's stated execution conditions, its no-four-bases theorem also excludes two physical constructed companions for our pair. With our conditional original-ray identification, prescribed-ray completion and full-overlap identities, that external theorem can supply the final nonextension implication in the nonproduct branch; it does not replace those structural steps or prove \hyperref[result:theorem-7.3]{the real scalar theorem below}. We retain the complete local scalar and clock arguments without importing the external computation.

Comparing the external local computation with our neighborhood result requires a matching output record and a checked correspondence covering the entire neighborhood \(\mathcal U\). That correspondence remains unresolved. The neighborhood theorem below has its own computer-assisted proof; this comparison is not a premise of that proof.

The supplementary path-phase estimate in Appendix~\ref{app:prescribed-phase-powers} also uses Newton identities and unit-circle reciprocity, as do parts of the Fourier paper; its six path phases have first sum \(\sqrt6\), rather than being balanced characters of one source or a mixed complete-family invariant. Our statements concern the specified unsaturated boundary scheme, whose defining ideal retains structure at the zero boundary coordinate, a full critical quotient, complete-companion specialization, prescribed-state completion and a pure-state obstruction with an original-ray symmetry hypothesis.

Finiteness antecedents include Haagerup's prime-Fourier support argument and Hopkins--Nicoara's all-square-minor theorem \cite{H,HN}. Logarithmic critical equations and reciprocal spaces are established frameworks \cite{V,BBT}; algebraic unbiasedness through projectors and critical potentials also has close predecessors \cite{BZO,BZH,BZP}. Group-Fourier completion and even-order two-circulant constructions precede our displayed completion \cite{BS,S}. Jaming et al. give Fourier-family symmetries \cite{J}, and earlier work already asserts triplet existence over the full six-dimensional Fourier parameter range \cite{G,GG}. McNulty--Weigert exclude an additional pure state above the two classified product triples \cite{MW}. Our contribution comparisons concern the precise statements and mechanisms just described; their publication history is bounded to the inspected literature, with no worldwide priority assertion.

\hyperref[common-definitions-and-the-domains-of-the-results]{Section 2} fixes the physical and algebraic dictionaries and gives the separate result hypotheses. Readers interested first in measurement construction can then go directly to \hyperref[prescribed-ray-completion-in-dimension-2m]{Section 6}, follow the dimension-six obstruction in \hyperref[six-dimensional-profiles-and-the-scalar-obstruction]{Sections 7}--\hyperref[actual-clock-triples-and-their-conditional-companion-bridge]{8}, and read the examples in \hyperref[applications-and-limits]{Section 9}. \hyperref[boundary-isolation-and-finiteness]{Sections 3}--\hyperref[compatible-algebraic-collections]{5} provide the independent algebraic route from candidate equations to finiteness and specialization. Both routes retain their complete arguments in this article. The appendices supply the exact applicability example, the longer specialization proof, a quantitative prescribed-closing estimate, the source-neighborhood proof and the exact companion-symmetry witness.

\section{Definitions and main results}\label{common-definitions-and-the-domains-of-the-results}

\subsection{Physical rays and reciprocal coordinates}\label{physical-rays-and-reciprocal-coordinates}

A pure state is a ray of unit vectors modulo a global phase. Its rank-one projector, rather than the phase of a chosen vector, specifies the preparation. Throughout we use complete rank-one projective measurements. The Hermitian product is \(\langle x,y\rangle=x^*y\), conjugate-linear in the first entry. Orthonormal bases are represented by the columns of unitary matrices; their column projectors form the measurement outcomes. We call the two fixed given bases the anchors. A physical companion is an orthonormal basis unbiased to both anchors. The entries of the transition matrix \(U^*V\) are probability amplitudes, and their squared moduli are the cross-basis Born probabilities. Two such bases \(U,V\) are mutually unbiased (MU) when every entry of \(U^*V\) has modulus \(n^{-1/2}\). We call a vector or matrix \textbf{flat} when all of its entries have the same modulus. For a unit vector in \(\mathbb C^n\), this modulus is \(n^{-1/2}\); in particular, a unitary \(C\in U(n)\) is a flat unitary exactly when \(|C_{ij}|=n^{-1/2}\) for all \(i,j\). We call \(H=\sqrt n\,C\) the corresponding \textbf{phase Hadamard matrix}: all its entries have unit modulus and \(HH^*=nI_n\). Conversely, dividing any such matrix by the square root of its order gives a flat unitary. In our prose, complex Hadamard matrix uses this same unit-modulus normalization. The word phase refers to the entry normalization, not to a dephasing condition.

A common-unbiased ray of \(I,C\) is a unit vector \(q\) satisfying \(|q_i|=|(C^*q)_j|=n^{-1/2}\), modulo a global unit phase. Its entries in the fixed coordinate basis, called native coordinates, are nonzero, so it has a unique representative \(q=z/\sqrt n\) with \(z_0=1\). This fixes the phase vector, not the unit vector. Such a state has \(|z_i|=1\), because it is unbiased to the coordinate measurement; arbitrary pure states need not have this property. Physical coordinates obey \(w=\bar z=z^{-1}\). To form the auxiliary complex equations, we initially treat \(w\) independently and impose the reciprocal product relations on the affine chart. The resulting nonzero complex \(z_i\) form the algebraic torus; its unit-modulus subset is the physical unit torus. Conjugation and unit modulus must still be imposed to recover physical points.

For a phase Hadamard \(H\), put \[
s_c=\sum_i\bar h_{ic}z_i,\qquad r_c=\sum_i h_{ic}z_i^{-1},\qquad
\mathcal A_H=\mathbb C[z_1^{\pm1},\ldots,z_{n-1}^{\pm1}]/(s_cr_c-n)_c .
\] The algebra associated with \(C\), denoted \(\mathcal A_C\), has equations \((C^*z)_c(C^Tz^{-1})_c-1\), so \(\mathcal A_H=\mathcal A_C\). This equality rescales the generators; it does not identify every complex point with a physical ray. The full ray algebra is the quotient by the stated equations, retaining local multiplicities and nilpotent elements. For a phase Hadamard, matrix dephasing uses row and column unit phases to make its first row and column all ones. A monomial matrix encodes a permutation and nonzero scalings, with exactly one nonzero entry in each row and column. If the scalings are unit phases, it is a unitary monomial. Physical Hadamard equivalence uses unitary monomial row and column operations. The algebraic arguments also allow invertible nonzero scalings; those operations need not preserve physical normalization. Column unit phases preserve the basis rays.

Here \emph{compatibility} refers to the specified orthogonality and cross-overlap equations between candidate columns or bases. It does not mean joint measurability, in which different measurements are marginals of a single parent measurement \cite{IM}. A complete orthonormal basis contains \(d\) vectors; a complete \emph{set of mutually unbiased bases} contains \(d+1\) such bases. Maximum cardinality and inability to extend a particular set are different questions.

Reciprocal orthogonality pairs a column with the coordinatewise reciprocals of another, requiring their bilinear pairing to vanish. Both orders are imposed because no conjugation relation is assumed. An \textbf{algebraic complete companion} in order six is a tuple of six affine complex ray columns satisfying every ordered reciprocal orthogonality equation. Its physical specialization, when present, is an orthonormal common-unbiased basis after division by \(\sqrt6\). Nonphysical tuples remain algebraic objects, not Hilbert-space bases. A set of mutually unbiased bases is \textbf{strongly unextendible} when no unit ray is unbiased to every basis in the set; excluding a further full basis alone is a weaker assertion.

The symbol \(J_\eta\) below denotes an antiunitary used to construct specified basis columns; no physical gate or measured time-reversal operation is asserted. The all-ones matrix in the Potts example is denoted \(\mathbf1\mathbf1^T\) or \(\mathsf J_6\), never \(J_\eta\). The suffix \(C\) on \(Z_C\) denotes its fixed normalized input. A physical orthonormal basis is represented by a unitary column matrix. It is unbiased to the coordinate basis precisely when that unitary is flat. Algebraic flatness over a coefficient ring, used in specialization theory, is a different notion.

\subsection{Principal results}\label{principal-results}

\textbf{Boundary theorem (\hyperref[result:theorem-3.1]{Theorem 3.1}).} Let \(C\in U(6)\) be flat, with all positional three-minors nonzero. Put \(v=C^*z,u=C^Tw,t=z_0w_0\). The ten equations \(z_iw_i=t,v_ju_j=t\), \(i,j=1,\ldots,5\), define \(Z_C\subset\mathbb P^5_z\times\mathbb P^5_w\). Its homogeneous ideal is used without saturation by \(t\), so components supported at the boundary are not discarded. The boundary has exactly \(2Z_2(C)\) points, where \(Z_2(C)\) counts singular positional two-minors. Each boundary point has local ring \(\mathbb C\). The projective scheme is zero-dimensional, and its affine algebra \(\mathcal A_C\) is finite-dimensional over \(\mathbb C\). \hyperref[boundary-isolation-and-finiteness]{Section 3} proves these claims.

\textbf{Critical algebra and fixed-fibre application (\hyperref[result:proposition-4.1]{Propositions 4.1}--\hyperref[result:proposition-4.2]{4.2}).} In every order, the ray algebra of a dephased phase Hadamard \(H\) is the critical quotient of \(\prod_cs_c/\prod_{i>0}z_i\) on the arrangement complement. For the separate dimension-six application, fix flat \(B,C\in U(6)\) with every middle minor of \(C\) nonzero. Let \(z\) vary with \(z_0=1\). The full fixed-phase fibre defined by \(U=B^*\operatorname{diag}(z)C,V=U^{-T},U_{jk}V_{jk}=1/6\) is a quotient of \(\mathcal A_{\bar C}\), hence is finite-dimensional over \(\mathbb C\). Its local lengths need not equal those of the ray algebra. \hyperref[critical-algebra-and-fixed-phase-fibres]{Section 4} proves the ideal maps.

\textbf{Complete-companion specialization (\hyperref[result:theorem-5.1]{Theorem 5.1}).} For a discrete valuation ring (DVR) \(R\), \(6\in R^\times\), a type-II \(C\in GL_6(R)\) with \(C_{ij}(C^{-1})_{ji}=1/6\), and all positional three-minors units, the full six-ray complete-companion algebra is finite as an \(R\)-module. Here a unit is an invertible ring element. Geometric special-fibre emptiness means that residue-field reduction has no points even after extension to an algebraic closure; for a specified finite quotient it forces that quotient to be zero. \hyperref[complete-companions-and-specialization-over-a-dvr]{Section 5.2} defines the ring and complete companions, and \hyperref[appendix-dvr-complete-companions-cannot-escape]{the appendix} supplies the proof. This statement excludes the unconstrained one-ray algebra. Specialization is an algebraic change of coefficient fibre, not time evolution of a quantum state.

\textbf{Prescribed-ray completion (\hyperref[result:theorem-6.1]{Theorem 6.1}).} For \(m\ge1\), \(\omega_m=e^{2\pi i/m}\), arbitrary unit phases \(\eta_j\), and \(a,b\in\{0,1\}\), define \[
(K_\eta)_{(j,a),(\ell,b)}
 =(2m)^{-1/2}\omega_m^{j\ell}(-1)^{ab}\bar\eta_j^b,\qquad
D=\operatorname{diag}_j(\omega_m^j I_2),\qquad
(J_\eta g)_j=\bar\eta_j
\begin{pmatrix}0&1\\-1&0\end{pmatrix}\bar g_j .
\] Every unit \(g\) unbiased to \(I,K_\eta\) belongs to the common-unbiased orthonormal basis \(E_\eta(g)=[D^rg,D^rJ_\eta g]_{r=0}^{m-1}\). Both blocks of \(m\) columns or their interleaving give the same basis with a recorded permutation. \hyperref[prescribed-ray-completion-in-dimension-2m]{Section 6} proves every Gram entry and attributes seed existence. The quantifier covers every prescribed common-unbiased unit state for this explicitly displayed pair, for every integer \(m\ge1\) and every allowed phase choice; it is not a completion theorem for arbitrary MU pairs.

\textbf{Six-dimensional nonextension (\hyperref[result:theorem-8.2]{Theorem 8.2}).} Take \(m=3\) in the displayed pair. If \(I,K_\eta,V\) are pairwise-MU bases and the original six rays of \(V\) are setwise D-invariant, then no unit ray is unbiased to all three. \hyperref[result:theorem-7.3]{The scalar obstruction}, the full overlap calculation in block-phase-and-angle coordinates and the companion-identification argument under an assumed extra common ray are proved in \hyperref[six-dimensional-profiles-and-the-scalar-obstruction]{Sections 7}--\hyperref[actual-clock-triples-and-their-conditional-companion-bridge]{8}. This conclusion is not asserted for general \(m\).

\textbf{Uniform local obstruction (\hyperref[result:theorem-9.3]{Theorem 9.3}).} For every exact phase Hadamard \(H\) in the closed entrywise neighborhood of radius \(10^{-6}\) about the reference \(H_\star\), the pair \((I,H/\sqrt6)\) has exactly 48 common-unbiased rays and no third MU basis. Every six-column matrix of unit common-unbiased representatives has squared Gram defect \(>1/1200\), including matrices with repeated ray labels. \hyperref[sec:source-neighborhood]{Section 9.3} states the domain and graph-to-gap argument; \hyperref[app:source-neighborhood]{Appendix D} gives the reference and computational proof. This local physical result has no middle-minor or clock premise.

\subsection{Scope and applicability of the main results}\label{how-the-domains-differ}

Table~\ref{tab:result-map} summarizes the hypotheses, conclusions, and scope of the principal results, separating the physical completion and obstruction theorems from the auxiliary algebraic results.

\begin{longtable}{@{}
  >{\raggedright\arraybackslash}p{(\linewidth-6\tabcolsep)*\real{0.25}}
  >{\raggedright\arraybackslash}p{(\linewidth-6\tabcolsep)*\real{0.32}}
  >{\raggedright\arraybackslash}p{(\linewidth-6\tabcolsep)*\real{0.29}}
  >{\raggedright\arraybackslash}p{(\linewidth-6\tabcolsep)*\real{0.14}}@{}}
\caption{Principal algebraic and measurement results: hypotheses, conclusions, and proof locations.}\label{tab:result-map}\\
\toprule
Question / result & Dimension and key hypothesis & Conclusion & Where proved \\
\midrule
\endfirsthead
\multicolumn{4}{l}{\tablename~\thetable\ (continued)}\\
\toprule
Question / result & Dimension and key hypothesis & Conclusion & Where proved \\
\midrule
\endhead
\bottomrule
\endlastfoot
\multicolumn{4}{@{}l}{\textit{Constructing or obstructing measurement bases}}\\[2pt]
Complete a chosen common state & \(2m\), \(m\ge1\); displayed Fourier pair \((I,K_\eta)\), arbitrary unit phases & A third MU basis containing the chosen state, for every such choice & \hyperref[result:theorem-6.1]{Thm.~6.1} \\
Exclude any further pure state & Six; actual MU triple, original third-basis rays invariant under \(D\) & No unit ray unbiased to all three measurements & \hyperref[result:theorem-8.2]{Thm.~8.2} \\
Can 48 candidates form a basis? & Six; exact phase Hadamard in the closed entrywise \(10^{-6}\) neighborhood about \(H_\star\) & Exactly 48 rays, no third basis; any six common-unbiased unit columns have squared Gram defect \(>1/1200\) & \hyperref[result:theorem-9.3]{Thm.~9.3}; \hyperref[app:source-neighborhood]{App.~D} \\
Do phase powers preserve composition? & Six; three actual ordered MU transitions & Cubic/quintic closing defects: entrywise and Frobenius lower bounds & \hyperref[app:prescribed-phase-powers]{App.~C} \\
\midrule
\multicolumn{4}{@{}l}{\textit{Constraining the auxiliary algebraic candidates}}\\[2pt]
Are common states a finite set? & Six; flat \(C\), all positional three-minors nonzero & Isolated boundary points; \(\mathcal A_C\) finite-dimensional over \(\mathbb C\) & \hyperref[result:theorem-3.1]{Thm.~3.1} \\
Identify the full critical ideal & Any order; dephased phase Hadamard & Isomorphism with the critical algebra, including multiplicities & \hyperref[result:proposition-4.1]{Prop.~4.1} \\
Constrain a fixed phase fibre & Six; fixed flat \(B,C\), three-minors of \(C\) nonzero, \(z_0=1\) & Finite-dimensional quotient of \(\mathcal A_{\bar C}\); lengths need not be preserved & \hyperref[result:proposition-4.2]{Prop.~4.2} \\
Specialize a complete collection & Six over a DVR; type-II input, \(6\) invertible, three-minors units & Full compatible-collection algebra finite as a module over the DVR & \hyperref[result:theorem-5.1]{Thm.~5.1}; \hyperref[appendix-dvr-complete-companions-cannot-escape]{App.~B} \\
\end{longtable}

Finiteness need not produce an orthogonal basis, and the critical-ideal identity alone does not imply all-order finiteness. Prescribed-state completion concerns the displayed pair, rather than an arbitrary MU pair. The clock and source-neighborhood obstructions are six-dimensional; the latter assumes exact Hadamards and is not an experimental-noise statement. Algebraic specialization does not impose physical conjugation or construct a fourth basis.

For \(m=3\), the displayed \(K_\eta\) has a singular positional three-block: choose rows \((j,0),(j,1),(k,a)\) with \(k\ne j\), and the three columns \((\ell,0)\), \(\ell=0,1,2\). The first two selected rows are identical, for every \(\eta\). Thus the boundary/finiteness row for \(Z_C,\mathcal A_C\) is not invoked for \(K_\eta\). After the harmless column gauge \(\eta_0=1\), \(\sqrt{2m}K_\eta\) is dephased and the all-order critical description still applies. That shared algebraic dictionary does not supply the completion proof or a new combined theorem.

\section{Boundary isolation and finiteness}\label{boundary-isolation-and-finiteness}

For a fixed pair of measurements, can their common-unbiased pure states form a continuous family? We answer this question under the middle-minor hypothesis by studying a larger space of complex solutions. The physical states are selected from that space by conjugation; finiteness of the larger space therefore implies finiteness of the states.

The obstacle is a positive-dimensional family of complex candidates. We place the equations in a projective space, where any such family must meet the boundary of the chosen model. Determining the boundary points is only the first step: a finite boundary could still lie on a curve of solutions. We also show that each boundary point is isolated in the full equations, which rules out every such curve.

\subsection{The specified projective scheme}\label{the-specified-projective-scheme}

The projective model keeps a candidate vector and its reciprocal partner in separate factors. This allows zero coordinates at the boundary, while retaining the nonzero reciprocal coordinates in the affine part. Introduce independent homogeneous coordinates \([z],[w]\in\mathbb P^5\). Set

\[
v=C^*z,\qquad u=C^T w,\qquad t=z_0w_0.
\]

The adjoint in \(v\) and the transpose in \(u\) are essential: \(w\) is an independent complex variable, rather than a conjugate imposed in the defining equations. Define \(Z_C\subset\mathbb P^5\times\mathbb P^5\) by the ten equations

\[
z_iw_i-t=0\ (1\le i\le5),\qquad v_ju_j-t=0\ (1\le j\le5). \tag{3.1}
\label{eq:3.1}\]

A scheme records the polynomial equations together with their local algebra, rather than only their solution points. We use this homogeneous ideal without saturation by \(t\): saturation could remove components supported where that coordinate vanishes. In this section, \textbf{boundary} means the closed subscheme \(Z_C\cap\{t=0\}\) of this specified scheme. It does not mean the closure of the affine solutions, or the set of limits of physical rays. Let \(Z_2(C)\) count singular positional two-by-two submatrices of \(C\), with their row and column index sets retained.

\phantomsection\label{result:theorem-3.1}\textbf{Theorem 3.1 (Boundary isolation and fixed-pair finiteness).} Let \(C\in U(6)\) be flat, and assume that all \(\binom63^2=400\) positional three-by-three minors are nonzero. Then the boundary of \(Z_C\) consists of exactly \(2Z_2(C)\) points. Each point is reduced and isolated in the full scheme \(Z_C\). The whole projective scheme is zero-dimensional. On the open set \(t\ne0\), the coordinate algebra is finite-dimensional over \(\mathbb C\), with no passage to its radical. Its conjugation-fixed complex points correspond exactly to the common-unbiased rays of \(I\) and \(C\).

In particular, there are finitely many such rays. The theorem does not equate their number with the number or the scheme-theoretic length of all complex roots.

The proof first restricts the possible supports at the boundary. A singular two-by-two block then determines each boundary point. Linearizing all ten equations forces a putative transverse displacement to satisfy \(2\delta t=4\delta t\). The remaining displacements are only projective rescalings. Finally, every positive-dimensional projective component would have to meet the boundary, contradicting this isolation.

\subsection{Support restrictions}\label{support-restrictions}

The support of a vector is the set of indices of its nonzero coordinates. The support argument needs all six products in each coordinate system, although only ten equations were used to define the model. The omitted transformed equation follows as a polynomial combination of those ten. Unitarity gives

\[
\sum_{j=0}^5v_ju_j=z^T\overline C C^T w=\sum_{i=0}^5z_iw_i,
\]

and hence

\[
v_0u_0-t=\sum_{i=1}^5(z_iw_i-t)-\sum_{j=1}^5(v_ju_j-t). \tag{3.2}
\label{eq:3.2}\]

Thus all six native and all six transformed products equal \(t\), also scheme-theoretically and in their linearizations.

\phantomsection\label{result:lemma-3.2}\textbf{Lemma 3.2 (Small blocks and supports).} Every three-by-two submatrix of \(C\) has column rank two, and every two-by-three submatrix has row rank two. If \(x=Cy\ne0\), then

\[
|\operatorname{supp}x|+|\operatorname{supp}y|\ge6.
\]

Equality is possible only for support sizes \((4,2)\) or \((2,4)\).

\emph{Proof.} A deficient three-by-two block remains deficient after adjoining any third column, contradicting the assumed three-minor condition. The two-by-three assertion follows by adjoining a row. The same facts hold for the adjoint, transpose and conjugate matrices.

Write \(a=|\operatorname{supp}x|\), \(d=|\operatorname{supp}y|\). If \(d=1\), nonzero entries of \(C\) give \(a=6\); if \(a=1\), apply the same argument to \(y=C^*x\). If \(d=2\) and \(a\le3\), the at least three zero coordinates of \(x\) give a rank-deficient three-by-two block. If \(a=2\) and \(d\le3\), use the adjoint argument. All other pairs have both sizes at least three. The only equality case remaining in that range is \((3,3)\), which is impossible: the three rows outside the support of \(x\) and the three columns supporting \(y\) form an invertible block annihilating \(y\). These observations prove both conclusions. \(\square\)

At the boundary, each native product and each transformed product vanishes. The nonzero coordinates of a vector and of its partner must therefore occupy disjoint supports. At a boundary point put

\[
A=\operatorname{supp}z,\quad B=\operatorname{supp}w,\quad
D=\operatorname{supp}v,\quad E=\operatorname{supp}u.
\]

The product equations imply \(A\cap B=D\cap E=\varnothing\). \hyperref[result:lemma-3.2]{Lemma 3.2} applies to \(z=Cv\) and \(w=\overline C u\), so

\[
|A|+|D|\ge6,\qquad |B|+|E|\ge6.
\]

On the other hand \(|A|+|B|\le6\) and \(|D|+|E|\le6\). All four inequalities must therefore be equalities. In particular \(B=A^c\) and \(E=D^c\), and the only boundary patterns are

\[
(|A|,|B|,|D|,|E|)=(4,2,2,4)\quad\hbox{or}\quad(2,4,4,2). \tag{3.3}
\label{eq:3.3}\]

Thus the boundary problem has only two support patterns, with the large support in one coordinate system paired with the small support in the other. These statements concern nonzero representatives of both projective coordinates. Neither representative can vanish.

\subsection{Boundary parametrization and full local rings}\label{boundary-parametrization-and-full-local-rings}

The small supports now reduce each boundary point to the kernels of a single singular two-by-two block. This identifies the points explicitly before we examine whether a family of solutions can pass through them.

\phantomsection\label{result:lemma-3.3}\textbf{Lemma 3.3 (Boundary parametrization).} Each singular block \(M=C_{B,D}\) with \(|B|=|D|=2\) determines exactly one point of pattern \((4,2,2,4)\). All points of that pattern arise this way.

\emph{Proof.} At such a point, \(z_B=0\) and \(u_D=0\) give

\[
Mv_D=0,\qquad M^Tw_B=0. \tag{3.4}
\label{eq:3.4}\]

The flat block \(M\) has no zero entries; if singular, it has rank one. Both kernels in \hyperref[eq:3.4]{(3.4)} are one-dimensional, and each nonzero kernel vector has both coordinates nonzero. Choose these vectors, extend them by zero, and set \(z=Cv\), \(u=C^Tw\). An extra zero coordinate of \(z\) outside \(B\) would make a three-by-two block deficient. An extra zero coordinate of \(u\) outside \(D\) would make a two-by-three block deficient. Consequently the supports are exactly \(A=B^c\), \(E=D^c\), and every native and transformed product is zero. The choices of the two kernel vectors affect only the two projective scalings. Conversely the supports at a given point recover \(B,D\) and its block, proving injectivity. \(\square\)

The antiholomorphic involution

\[
\tau([z],[w])=([\overline w],[\overline z]) \tag{3.5}
\label{eq:3.5}\]

preserves the equations, since it exchanges \(v\) with \(\overline u\) and \(u\) with \(\overline v\). It exchanges the two patterns in \hyperref[eq:3.3]{(3.3)}, which are disjoint. Thus \hyperref[result:lemma-3.3]{Lemma 3.3} counts the second pattern as well. This is an antiholomorphic involution on complex points, not an assertion that conjugation is a holomorphic morphism of complex schemes.

A useful bound is \(Z_2(C)\le45\). For each of the fifteen row pairs, a singular column pair means equality of the two entry ratios in those columns. No ratio can occur in three columns, by the two-by-three rank statement. The six columns can therefore contribute at most three singular pairs.

The complementary four-by-four block has rank three as well. The following kernel comparison records why its nullity is also one. Write a unitary matrix in the block form

\[
C=\begin{pmatrix}M&P\\Q&N\end{pmatrix}.
\]

If \(y\in\ker N\), then \(C(0,y)=(Py,0)\), so \(M^*Py=0\). The map \(y\mapsto Py\) is injective by unitarity. Conversely, if \(x\in\ker M^*\), then \(C^*(x,0)=(0,P^*x)\) and \(NP^*x=0\). The two maps are inverse on these kernels. Thus \(\dim\ker N=\dim\ker M^*=1\), giving \(\operatorname{rank}N=3\).

To prove isolation, we test every infinitesimal displacement allowed by the full equations. The Zariski tangent space is the space of solutions of these linearized equations, after projective rescalings are removed. A zero tangent space also determines the local ring: there are neither additional nearby directions nor nilpotent structure at that point.

\phantomsection\label{result:lemma-3.4}\textbf{Lemma 3.4 (Full tangent isolation).} The Zariski tangent space of \(Z_C\) at every boundary point is zero. Its local ring there is \(\mathbb C\).

\emph{Proof.} Work first at pattern \((4,2,2,4)\) in projective affine charts. Represent a tangent vector by \(\delta z,\delta w\), set \(\delta v=C^*\delta z\), \(\delta u=C^T\delta w\), and put \(q=\delta t\). By \hyperref[eq:3.2]{(3.2)}, every index is allowed in the linearized equations. Since the supports are complementary, they give

\[
\delta w_i=q/z_i\ (i\in A),\qquad \delta z_i=q/w_i\ (i\in B),
\]

\[
\delta u_j=q/v_j\ (j\in D),\qquad \delta v_j=q/u_j\ (j\in E). \tag{3.6}
\label{eq:3.6}\]

The unchanged vector \(w\) supplies the decisive pairing:

\[
\sum_{i\in B}w_i\delta z_i
=w^T\delta z=(C^Tw)^T\delta v
=\sum_{j\in E}u_j\delta v_j.
\]

Using \hyperref[eq:3.6]{(3.6)}, the left side is \(2q\) and the right side is \(4q\). Hence \(q=0\). This uses the tangent equations of the full scheme; counting boundary points alone would not exclude a component leaving the boundary.

Now \(\delta w_A=\delta z_B=\delta u_D=\delta v_E=0\). The remaining equations reduce to

\[
M\delta v_D=0,\qquad M^T\delta w_B=0.
\]

Their kernels are the same one-dimensional spaces as in \hyperref[eq:3.4]{(3.4)}. Therefore \(\delta z\) is proportional to \(z\) and \(\delta w\) to \(w\). These are precisely the rescalings removed by the projective charts. The tangent vector is zero. At the second pattern the conjugate calculation under \hyperref[eq:3.5]{(3.5)}, or the direct pairing with \(4q=2q\), gives the same conclusion.

Let \(\mathfrak m\) be the maximal ideal of the Noetherian local ring at a boundary point. The zero tangent space means \(\mathfrak m/\mathfrak m^2=0\). Nakayama's lemma gives \(\mathfrak m=0\), so the local ring is its residue field \(\mathbb C\). The point lies on no positive-dimensional component. Deleting the finitely many other irreducible components gives an open neighborhood supported at that point. Each such point is therefore reduced and isolated in the full scheme. \(\square\)

In particular the boundary is a disjoint union of \(2Z_2(C)\) reduced points. Because they are isolated already in \(Z_C\), they need not belong to the scheme-theoretic closure of the affine part.

\subsection{Projectivity, the full affine algebra and physical rays}\label{projectivity-the-full-affine-algebra-and-physical-rays}

The local calculation becomes a global finiteness result because projective families cannot avoid a hyperplane. Under the Segre embedding, which represents a pair of projective vectors by their coordinate products, \(t=z_0w_0\) is a hyperplane coordinate. Every positive-dimensional projective irreducible variety meets a hyperplane. One algebraic proof is as follows. Its homogeneous coordinate domain \(R\) has dimension at least two. If a hyperplane's linear form is identically zero on it, intersection is immediate. Otherwise the form is a nonunit, and the principal ideal dimension theorem gives \(\dim R/(\ell)\ge\dim R-1\ge1\). Empty projective intersection would force the irrelevant ideal to be nilpotent in this graded quotient, making it finite-dimensional and of Krull dimension zero, a contradiction.

If \(Z_C\) had a positive-dimensional component, its reduced irreducible support would therefore meet \(t=0\). At that point \hyperref[result:lemma-3.4]{Lemma 3.4} gives a zero-dimensional local ring of the full scheme, which is impossible on a positive-dimensional component. Hence the projective scheme is zero-dimensional, and has finite support.

On \(t\ne0\), both \(z_0\) and \(w_0\) are nonzero. Rescale separately to \(z_0=w_0=1\). Equations \hyperref[eq:3.1]{(3.1)} then force \(w_i=z_i^{-1}\). Write \(x_0=1\) and

\[
v_j=\sum_{i=0}^5\overline C_{ij}x_i,\qquad
u_j=\sum_{i=0}^5C_{ij}x_i^{-1},\qquad G_j=v_ju_j-1.
\]

The affine algebra is exactly

\[
\mathcal A_C=
\mathbb C[x_1^{\pm1},\ldots,x_5^{\pm1}]/(G_1,\ldots,G_5), \tag{3.7}
\label{eq:3.7}\]

where \(G_0\) is an ideal consequence because \(\sum_jG_j=0\). In this quotient every \(v_j\) is a unit, with inverse \(u_j\).

Finitely many complex points imply that this algebra is finite-dimensional over \(\mathbb C\), even when it contains nilpotents. Present it using the ten ordinary generators \(x_i,x_i^{-1}\). For each generator \(y\), the monic polynomial \(P_y(T)=\prod_p(T-y(p))\), with the product over all complex points, vanishes at every maximal ideal. The Nullstellensatz gives \(P_y(y)^{n_y}=0\) for some positive integer \(n_y\). These monic relations bound the exponent of each generator in a spanning monomial, giving a finite-dimensional span. If there are no points, the algebra is the zero ring. This argument does not assume that the affine roots are reduced; reducedness would mean that their coordinate algebra has no nonzero nilpotent elements.

\phantomsection\label{result:proposition-3.5}\textbf{Proposition 3.5 (Physical points).} The fixed points of \hyperref[eq:3.5]{(3.5)} in \(Z_C\) are in \(t\ne0\) and correspond bijectively to common-unbiased rays.

\emph{Proof.} If a point is fixed, choose representatives with \(w=\lambda\overline z\) for a nonzero scalar \(\lambda\). The product equations give \(\lambda|z_i|^2=t\) for all \(i\). Since \(z\ne0\), their common value cannot be zero; in particular \(z_0\ne0\). In the unique affine representatives, \(w=\overline z\), \(z_0=w_0=1\), and \(|z_i|=1\). The transformed products now read \(|v_j|^2=1\). Thus \(q=z/\sqrt6\) is a unit vector unbiased to \(I\) and \(C\). Conversely every such ray has a unique representative of this form with \(z_0=1\), and supplies this fixed point. Boundary patterns \hyperref[eq:3.3]{(3.3)} also directly show there are no fixed boundary points. \(\square\)

\hyperref[result:proposition-3.5]{Proposition 3.5} recovers the measurement problem by selecting conjugation-fixed roots; other complex roots have no physical-ray interpretation. The algebra carries additional local information. A nonzero nilpotent is an algebra element whose power vanishes; such elements can distinguish a multiple root from a simple root even when the underlying point is the same. An algebra homomorphism to \(\mathbb C\) kills nilpotents, so the physical interpretation selects points, while \hyperref[eq:3.7]{(3.7)} retains the scheme multiplicities.

Finitely many candidate states still need to be assembled into a measurement. A graph makes the remaining test explicit: its vertices are physical candidate rays, and an edge records orthogonality. A clique is a set of vertices joined by every possible edge. Six such candidates give a basis; cross-overlap tests then decide whether two candidate bases are mutually unbiased.

\phantomsection\label{result:proposition-3.6}\textbf{Proposition 3.6 (Companion bases from finite candidate sets).} On the finite physical set \(\mathcal R_C\) of \hyperref[result:proposition-3.5]{Proposition 3.5}, use the unique representatives \(z_0=1\) and define

\[
O(z,z'):\ z^*z'=0,\qquad U(z,z'):\ |z^*z'|^2=6.
\]

Six-vertex \(O\)-cliques correspond bijectively to orthonormal bases unbiased to \(I\) and \(C\), modulo column phases and permutations. Two such cliques correspond to mutually unbiased companion bases exactly when all thirty-six cross pairs satisfy \(U\).

\emph{Proof.} Every vertex has \(z^*z=6\). Six distinct pairwise orthogonal vertices therefore give six unit orthogonal columns \(z/\sqrt6\), hence a complete orthonormal basis of \(\mathbb C^6\). Conversely the columns of any actual companion basis supply six distinct vertices and every \(O\)-edge. Projective rays remove column phases; the unordered clique removes column permutations, and these are exactly the redundancies in this correspondence. For two cliques,

\[
\left|\left\langle z/\sqrt6,z'/\sqrt6\right\rangle\right|^2
=|z^*z'|^2/36,
\]

so mutual unbiasedness is precisely \(U\) for every cross pair. The same actual columns occur in all conditions. A shared vertex has squared unnormalized overlap \(36\) and fails \(U\); repeated vertices cannot belong to an \(O\)-clique. All conditions are invariant under unit column phases. This includes the case of no six-clique. No existence of a companion basis or quartet, and no assertion uniform over varying \(C\), is used. \(\square\)

The conclusion is finiteness for one fixed \(C\), together with an exact test for assembling its physical candidates into bases. It gives no uniform physical count as the matrix varies and no obstruction to a quartet. The nilpotents in \(\mathcal A_C\) and the distinction between all complex points and physical rays remain part of the statement: affine reducedness and a formula for affine length would require further information. Nor does this fixed-matrix argument assert a varying-matrix finite morphism.

\section{Critical algebra and fixed phase fibres}\label{critical-algebra-and-fixed-phase-fibres}

The same candidate equations admit another useful representation: they are the critical equations of a rational function. This function is an auxiliary algebraic tool for describing solutions and their multiplicities; it is not a physical energy or an observable. To retain those multiplicities, we need equality of the ideals, not merely equality of their point sets. We prove that equality first, including the localization needed to define the derivatives. We then use the ray algebra to constrain the diagonal phases that make two fixed measurement bases mutually unbiased.

\subsection{The full ray ideal is a critical ideal}\label{the-full-ray-ideal-is-a-critical-ideal}

The rational function is built from the transformed coordinates of a candidate vector. Its critical equations become meaningful where none of those linear forms vanishes. Let \(H=(h_{pc})_{p,c=0}^{n-1}\) be a dephased phase Hadamard: \[
h_{0c}=h_{p0}=1,\qquad |h_{pc}|=1,\qquad HH^*=nI_n.
\] Let \(K\subseteq\mathbb C\) be a subfield containing the entries of \(H\) and closed under complex conjugation. Set \(z_0=1\), and write \[
\begin{split}
\mathscr R&=K[z_1^{\pm1},\ldots,z_{n-1}^{\pm1}],\\
s_c&=\sum_{p=0}^{n-1}\overline{h_{pc}}z_p,\qquad
r_c=\sum_{p=0}^{n-1}h_{pc}z_p^{-1},\\
G_c&=s_cr_c-n,\qquad
\mathscr A_H=\mathscr R/(G_0,\ldots,G_{n-1}).
\end{split}
\tag{4.1}
\label{eq:4.1}\] The normalization \(C=H/\sqrt n\) makes \(G_c/n\) the normalized common-ray equation. Inverting the linear forms means restricting to the complement of their zero hyperplanes; this restriction is the localization used below. Work on \[
\mathscr U=\operatorname{Spec}\mathscr R[s_0^{-1},\ldots,s_{n-1}^{-1}],
\qquad
\Phi_H(z)=\frac{\prod_{c=0}^{n-1}s_c(z)}{\prod_{i=1}^{n-1}z_i}.
\tag{4.2}
\label{eq:4.2}\]

The logarithmic critical algebra is the coordinate ring of this complement modulo the first derivatives of the logarithm of the displayed rational function with respect to the variable coordinates. These derivatives are rational functions and require no choice of logarithm branch.

\leavevmode\phantomsection\label{result:proposition-4.1}\textbf{Proposition 4.1 (full critical algebra).} The full Laurent algebra \(\mathscr A_H\) is canonically isomorphic to the logarithmic critical algebra of \(\Phi_H\) on \(\mathscr U\). This is an equality of quotient algebras, including local multiplicities and nilpotents. On the unit torus, its points are exactly the dephased vectors \(z/\sqrt n\) common unbiased to \(I_n\) and \(C\). They are precisely the points attaining the bound \[
|\Phi_H(z)|\le n^{n/2}.
\tag{4.3}
\label{eq:4.3}\]

\textbf{Proof.} We compare the two ideals on the arrangement complement, then show that localization has not changed the ray algebra. Put \[
D_i=\frac{\partial\log\Phi_H}{\partial z_i}
=\sum_c\frac{\overline{h_{ic}}}{s_c}-z_i^{-1},
\qquad 1\le i<n.
\] Row orthogonality gives \[
\begin{split}
\sum_c\frac{\overline{h_{ic}}G_c}{s_c}
&=\sum_c\overline{h_{ic}}r_c
-n\sum_c\frac{\overline{h_{ic}}}{s_c}\\
&=nz_i^{-1}-n\sum_c\frac{\overline{h_{ic}}}{s_c}.
\end{split}
\] Hence \[
D_i=-\frac1n\sum_c\frac{\overline{h_{ic}}G_c}{s_c}.
\tag{4.4}
\label{eq:4.4}\] For the reverse inclusion define, for all \(0\le p<n\), \[
k_p=z_p\sum_c\frac{\overline{h_{pc}}}{s_c}-1.
\] Then \(k_i=z_iD_i\) for \(i>0\), and \[
\sum_pk_p=\sum_c\frac{s_c}{s_c}-n=0,\qquad
k_0=-\sum_{i=1}^{n-1}z_iD_i.
\] The preceding orthogonality calculation for every \(p\) gives \[
\sum_c\frac{\overline{h_{pc}}G_c}{s_c}=-n\,\frac{k_p}{z_p}.
\] Multiplying by \(h_{pc}\), summing in \(p\), and using column orthogonality yields \[
G_c=-s_c\sum_p h_{pc}\frac{k_p}{z_p}
=-s_c\sum_{i=1}^{n-1}(h_{ic}-z_i)D_i.
\tag{4.5}
\label{eq:4.5}\] Thus the two ideals agree in the localized ring. To return to the original Laurent quotient, observe that the localization removes no part of the ray algebra: modulo the \(G_c\), each \(s_c\) is already invertible, with inverse \(r_c/n\). This proves the full-algebra assertion.

On \(|z_i|=1\), one has \(r_c=\overline{s_c}\), so \(G_c=0\) means \(|s_c|^2=n\). Since \[
C^*(z/\sqrt n)=H^*z/n,
\] these are exactly the common-unbiased-vector equations. Parseval gives \[
\sum_c|s_c|^2=z^*HH^*z=n^2.
\] The arithmetic--geometric mean inequality now gives \[
\prod_c|s_c|^2\le
\left(\frac{\sum_c|s_c|^2}{n}\right)^n=n^n.
\] The denominator of \(\Phi_H\) has modulus one. Equality holds exactly when all \(|s_c|^2=n\), proving \hyperref[eq:4.3]{(4.3)}. \(\square\)

This description allows hyperplane-arrangement results to be applied when their weight hypotheses hold. An arrangement is the collection of hyperplanes removed to define the complement. Here the projective arrangement has the \(n\) coordinate normals \(e_p\), of weight \(-1\), and the \(n\) conjugate Hadamard-column normals, of weight \(+1\). The omitted affine coordinate hyperplane is the hyperplane at infinity, also of weight \(-1\). The arrangement is essential because the coordinate normals span. A projective edge is a nonempty intersection of projective hyperplanes. It is dense when, in local affine coordinates, the hyperplanes containing it cannot be partitioned into nonempty groups depending on separate coordinate sets. Its weight is the sum of the weights of all containing hyperplanes.

For these fixed weights, the count theorem requires every nonempty dense projective edge to have nonzero total weight. Extend scalars from \(K\) to \(\mathbb C\) to obtain the complex complement \(\mathscr U_{\mathbb C}\). Under the weight condition, \href{https://arxiv.org/html/1001.4553v2}{Varchenko, Lemmas 4.1--4.2} gives isolated critical points whose total Milnor number is \(|\chi(\mathscr U_{\mathbb C})|\). Here \(\chi\) is the Euler characteristic of that complement, and the total Milnor number is the sum of the local critical-point multiplicities.

\hyperref[result:proposition-4.1]{Proposition 4.1} transfers this total algebraic length to the ray algebra. The weight condition must be checked for the input; it is not an automatic consequence of the critical-ideal identity. The length does not by itself establish reducedness, count physical rays, or construct a compatible basis.

\subsection{A fixed phase fibre is a quotient of one ray algebra}\label{a-fixed-phase-fibre-is-a-quotient-of-one-ray-algebra}

Suppose two flat measurement bases are fixed and we change their relative coordinate phases. We ask which phase choices make their transition matrix flat. The phase fibre below consists of those choices in the reciprocal complex equations; its physical points additionally have unit-modulus phases. One row of this system already imposes the common-ray equations, after a diagonal change of coordinates. The other rows add constraints. This is why finiteness follows from a single row, while its local lengths need not survive in the full fibre.

Let \(B,C\in U(6)\) be normalized flat matrices, and fix the scalar gauge \(z_0=1\). Define \[
U=B^*\operatorname{diag}(z)C,\qquad
V=B^{\mathsf T}\operatorname{diag}(z^{-1})\overline C=U^{-\mathsf T}.
\tag{4.6}
\label{eq:4.6}\] The full fixed-phase fibre is the Laurent algebra \[
\mathscr F_{B,C}
=\mathbb C[z_1^{\pm1},\ldots,z_5^{\pm1}]
/(6U_{jk}V_{jk}-1:0\le j,k\le5).
\tag{4.7}
\label{eq:4.7}\] For physical phases the displayed matrix is the transition between the two bases after the diagonal unitary phase change. The reciprocal product equations extend its flatness condition to complex coordinates. No radical is taken. Without \(z_0=1\), scalar rescaling would leave these equations unchanged and prevent the finiteness assertion below.

\phantomsection\label{result:proposition-4.2}\textbf{Proposition 4.2 (one-row comparison).} Fix a row \(j\). Its six equations in \hyperref[eq:4.7]{(4.7)} are exactly the normalized common-ray ideal for \(\overline C\), after the diagonal Laurent automorphism \[
\gamma_i=\frac{\overline{B_{ij}}}{\overline{B_{0j}}},
\qquad x_i=\gamma_i z_i,\qquad w_i=x_i^{-1}.
\tag{4.8}
\label{eq:4.8}\] Consequently, if every positional \(3\times3\) minor of \(C\) is nonzero, \hyperref[result:theorem-3.1]{the middle-minor finiteness theorem} makes \(\mathscr F_{B,C}\) finite-dimensional over \(\mathbb C\).

\textbf{Proof.} All \(\gamma_i\) are nonzero and \(\gamma_0=1\), so \hyperref[eq:4.8]{(4.8)} is a Laurent automorphism preserving \(x_0=w_0=1\). For the common-ray input \(\overline C\), put \[
v=C^{\mathsf T}x,\qquad u=\overline C^{\mathsf T}w.
\] By \hyperref[eq:4.6]{(4.6)}, \[
v_k=\frac{U_{jk}}{\overline{B_{0j}}},\qquad
u_k=\frac{V_{jk}}{B_{0j}}.
\] Here the second identity uses \(w_i=B_{ij}z_i^{-1}/B_{0j}\), since \(B\) is flat. As \(|B_{0j}|^2=1/6\), \[
v_ku_k-1=6U_{jk}V_{jk}-1.
\tag{4.9}
\label{eq:4.9}\] This identifies the ideals themselves. The full fibre is obtained by quotienting this one-row algebra by the other row equations. Complex conjugation preserves nonvanishing of the middle minors, so \hyperref[result:theorem-3.1]{the boundary/finiteness theorem} applies to \(\overline C\). A quotient of a finite-dimensional algebra is finite-dimensional. \(\square\)

The one-row isomorphism preserves the local multiplicities, or algebraic lengths, of the candidate roots. Adding the other row equations can kill nilpotents or reduce those lengths. Thus finiteness survives in the full fibre, while multiplicity preservation does not follow. Its complex points need not be physical phase vectors. Since \(|\gamma_i|=1\), any modulus bounds for the finite one-row set pass unchanged to the fibre. Finding finitely many complex choices, or no choices, supplies no existence theorem for physical phases.

\section{Compatible algebraic collections}\label{compatible-algebraic-collections}

A candidate ray and a set of mutually orthogonal candidates answer different questions. In the complex equations even a full reciprocal-orthogonal collection may fail the unit-modulus condition and therefore fail to be a physical measurement basis. The Tao/Potts example exhibits this distinction explicitly. We then study what happens to complete algebraic collections when their coefficients are specialized. Here compatibility means the stated reciprocal orthogonality and cross-overlap equations between columns; it does not mean joint measurability of quantum measurements.

\subsection{The Tao/Potts illustration}\label{the-taopotts-illustration}

The Tao matrix provides a concrete test of the physical restriction: its reciprocal equations admit vectors whose entries have unequal moduli. Let \(\omega^2+\omega+1=0\), with \(\omega\ne1\), and take the dephased Tao phase matrix \[
T_0=(\omega^{E_{pc}}),\qquad
E=
\begin{pmatrix}
0&0&0&0&0&0\\
0&0&1&1&2&2\\
0&1&0&2&2&1\\
0&1&2&0&1&2\\
0&2&2&1&0&1\\
0&2&1&2&1&0
\end{pmatrix},\qquad C=T_0/\sqrt6.
\tag{5.1}
\label{eq:5.1}\] Each difference of distinct rows of \(E\) contains each residue \(0,1,2\) twice. Their inner products are therefore \(2(1+\omega+\omega^2)=0\), while each row has squared norm six. Thus \(T_0T_0^*=6I_6\). For \(q=-2\pm\sqrt3\), so that \[
q^2+4q+1=0,\qquad q+q^{-1}=-4,
\] two useful algebraic rays are \[
v_q=(1,q,1,1,1,1),\qquad
d_q=(1,q,q,q,q,q).
\tag{5.2}
\label{eq:5.2}\] At the all-ones Tao column the reciprocal overlap products are, respectively, \[
(5+q)(5+q^{-1})=6,\qquad
(1+5q)(1+5q^{-1})=6.
\tag{5.3}
\label{eq:5.3}\] At every other column the coefficient sums vanish. Thus the corresponding products are \[
(q-1)(q^{-1}-1)=2-(q+q^{-1})=6
\] for \(v_q\), and the same value from \((1-q)(1-q^{-1})\) for \(d_q\). Both satisfy the full ideal. Neither is physical because \(|q|\ne1\).

The failure of physicality persists when six candidates are assembled and all ordered reciprocal orthogonality equations are imposed. The same algebraic vectors form such a complete collection. Here raw means before rescaling coordinate zero to one in each column. Set \[
X_{\rm raw}=\mathsf J_6+(q-1)I_6,
\qquad x^{\rm raw}_{ir}=
\begin{cases}q,&i=r,\\1,&i\ne r.\end{cases}
\tag{5.4}
\label{eq:5.4}\] For \(r\ne s\), its raw reciprocal pairings are \[
\sum_i(x^{\rm raw}_{ir})^{-1}x^{\rm raw}_{is}
=4+q+q^{-1}=0;
\qquad
\sum_i(x^{\rm raw}_{ir})^{-1}x^{\rm raw}_{ir}=6.
\tag{5.5}
\label{eq:5.5}\] Every column satisfies the Tao equations: the all-ones-column product is the first expression in \hyperref[eq:5.3]{(5.3)}, while every other column gives \((q-1)(q^{-1}-1)=6\). Normalize coordinate zero to one in each algebraic column by \[
x_{ir}=x^{\rm raw}_{ir}/x^{\rm raw}_{0r}.
\tag{5.6}
\label{eq:5.6}\] The reciprocal overlap products are invariant under rescaling each entire column; the ordered pairings are multiplied by nonzero scalars, so their zeros are preserved. Thus \hyperref[eq:5.6]{(5.6)} gives six dephased common algebraic rays with every ordered off-diagonal pairing zero. Five columns have one \(q\) among coordinates \(1,\ldots,5\); the remaining column is \(d_{q^{-1}}\). The normalized columns are not physical unit vectors.

The matrix \hyperref[eq:5.4]{(5.4)} is the classical Potts type-II matrix; see \href{https://arxiv.org/pdf/0707.1836}{Chan--Godsil, §2}. Direct substitution identifies it here as a point of the compatible-ray algebra. The matrix construction itself is classical. Known physical Tao common-unbiased-vector counts and the lack of a completing orthonormal sextuple belong to the earlier physical catalogue of \href{https://arxiv.org/html/0901.4051}{Brierley--Weigert, §4.1 and Table 1}. The explicit nonphysical collection illustrates why complex-algebra finiteness does not answer physical completion by itself. We do not need a full complex-point count for this example.

For the arithmetic application below, Tao also satisfies the middle-minor hypothesis. Here is the short applicability argument. A singular \(3\times3\) matrix with unimodular entries has proportional rows or columns: after dephasing its first row and column, its determinant is \((a-1)(d-1)-(b-1)(c-1)\). If one of the four factors vanishes, singularity immediately gives a repeated row or column. Otherwise conjugating the determinant equation and using \(\overline{a-1}=-(a-1)/a\), and the analogous identities, gives \(ad=bc\). The original equation then gives \(a+d=b+c\). The two unordered pairs with these equal sums and products coincide, which again gives repeated rows or columns. This is the local lemma underlying the published ordinary-minor criterion \href{https://arxiv.org/html/2608.18053v1}{Cárdenes Wuttig--Tindall, Lemma 8 and Supplement I.1}.

If three entries in two rows of a phase Hadamard are proportional, row orthogonality forces the complementary three row ratios to be the opposite phase: a sum of three unit numbers can attain modulus three only when all are equal. One column from each triple then exposes a \(2\times2\) Hadamard block. The column case is identical. In \(T_0\), all row and column ratios lie in \(\{1,\omega,\omega^2\}\), which contains no two negatives of one another. Thus it has no such \(2\times2\) block, and every \(3\times3\) minor is nonzero.

\subsection{Complete companions and specialization over a discrete valuation ring}\label{complete-companions-and-specialization-over-a-dvr}

Can the equations for complete reciprocal-orthogonal collections define a finite family as the matrix coefficients are specialized? The equations of one candidate alone do not supply enough control. For six candidates, ordered orthogonality gives an inverse identity that prevents specialization to the projective boundary. We formulate that result over a discrete valuation ring, with a generic coefficient field and a residue field obtained by reduction. The coordinates throughout this argument are algebraic; physical coordinates are recovered separately over the complex numbers.

A discrete valuation ring (DVR) is a nonfield local domain whose nonzero ideals are powers of its principal maximal ideal. A generator of that ideal is a uniformizer. Reduction modulo it gives the closed fibre, while passing to the fraction field gives the generic fibre. The valuation measures the order of vanishing along the closed fibre. These are changes of coefficient domain, rather than time evolution of a quantum state. Let \(R\) be any discrete valuation ring with \(6\in R^\times\), fraction field \(K\), and residue field \(k\). Let \[
C\in GL_6(R),\qquad T=C^{-1},\qquad
C_{ij}T_{ji}=1/6\quad(0\le i,j\le5),
\tag{5.7}
\label{eq:5.7}\] and assume every positional \(3\times3\) minor of \(C\) is a unit of \(R\). A unit is invertible in the coefficient ring; nonvanishing in its fraction field alone is not enough. In this subsection \(T\) always means \(C^{-1}\); it is not the phase Tao matrix \(T_0\). The entry-product condition is the type-II identity: it pairs each matrix entry with the transposed entry of the inverse. This is the inverse-orthogonal setting, and it needs no involution on \(R\).

The one-ray algebra uses the same reciprocal overlap products as before, with the inverse matrix replacing the adjoint. Define \[
\begin{split}
F_j(x)&=\left(\sum_iT_{ji}x_i\right)
\left(\sum_iC_{ij}x_i^{-1}\right)-1,\qquad x_0=1,\\
A_R&=R[x_1^{\pm1},\ldots,x_5^{\pm1}]/(F_1,\ldots,F_5).
\end{split}
\tag{5.8}
\label{eq:5.8}\] The identity \(\sum_{j=0}^5F_j=0\) recovers \(F_0\) in the displayed ideal. With six independent ray copies \(x_{:r}\), put \[
q_{rs}=\sum_ix_{ir}^{-1}x_{is},\qquad
P_R=A_R^{\otimes_R6}/(q_{rs}:r\ne s).
\tag{5.9}
\label{eq:5.9}\] All thirty ordered equations are imposed. Without conjugation, the two orders of a pairing are separate constraints. For two complete copies \(x,y\), put \[
\begin{split}
q^{xy}_{rs}&=\sum_ix_{ir}^{-1}y_{is},\qquad
q^{yx}_{sr}=\sum_iy_{is}^{-1}x_{ir},\\
Q_R&=(P_R\otimes_RP_R)/
(q^{xy}_{rs}q^{yx}_{sr}-6:0\le r,s\le5).
\end{split}
\tag{5.10}
\label{eq:5.10}\] All thirty-six mixed equations are imposed. Over physical complex coordinates these are the cross-basis uniform-probability conditions; over the coefficient ring they remain reciprocal product equations.

The geometric special fibre is obtained by residue-field reduction followed by extension to an algebraic closure. Vertical torsion consists of algebra elements annihilated by a power of a uniformizer. Finiteness as a module means that finitely many elements generate it over the coefficient ring.

\phantomsection\label{result:theorem-5.1}\textbf{Theorem 5.1 (finite complete-companion algebras).} Under \hyperref[eq:5.7]{(5.7)} and the three-minor unit hypothesis, \(P_R\) is finite as an \(R\)-module. So are \(Q_R\) and its quotients by further explicitly specified equations. These are conclusions about the full displayed ideals, including zero algebras, nilpotents and vertical torsion. If the complete geometric special fibre of \(Q_R\), or of such a specified quotient, is empty, that algebra is zero over \(R\) and after extension to \(K\).

The auxiliary projective ray scheme over \(R\) is finite. In every geometric fibre its boundary consists of reduced isolated points of the two support types \[
(|\operatorname{supp}z|,|\operatorname{supp}w|,
|\operatorname{supp}Tz|,|\operatorname{supp}C^{\mathsf T}w|)
=(2,4,4,2)\ \text{or}\ (4,2,2,4).
\tag{5.11}
\label{eq:5.11}\] Any set of at most six distinct boundary points \(([z_r],[w_r])\) satisfying \(w_r^{\mathsf T}z_s=w_s^{\mathsf T}z_r=0\) for every \(r\ne s\) has linearly independent tensors \(z_rw_r^{\mathsf T}\). A complete generic companion cannot specialize to that boundary.

\textbf{Proof and mechanism.} The argument has two parts. The middle-minor hypotheses make the projective ray family finite, but a torus chart of that family need not remain finite over the coefficient ring. The ordered compatibility equations supply the missing control: the inverse identity of a complete collection would turn boundary escape into a forbidden dependence of boundary tensors. The full proof, including that independence argument, appears in \hyperref[appendix-dvr-complete-companions-cannot-escape]{Appendix B}. We give the mechanism here. The projective equations are \[
z_iw_i=t,\quad (Tz)_j(C^{\mathsf T}w)_j=t
\quad(i,j=1,\ldots,5),\qquad t=z_0w_0.
\tag{5.12}
\label{eq:5.12}\] The missing crossed zero equation follows from \(\sum_j(Tz)_j(C^{\mathsf T}w)_j=\sum_i z_iw_i=6t\). The ordinary middle-minor support argument and full tangent calculation work in every algebraically closed field of characteristic different from two and three. They give \hyperref[eq:5.11]{(5.11)}, reduced isolated boundary points, and finite projective fibres. The structure morphism of this projective \(R\)-scheme is proper and quasi-finite, hence finite.

The needed independence concerns the rank-one matrices \(z_rw_r^{\mathsf T}\), or outer products, of boundary vectors. It is proved in \hyperref[appendix-dvr6-independence-of-at-most-six-compatible-boundary-matrices]{Appendix B}, using both ordered bilinear compatibility equations, rather than inferred from isolatedness. Type-II row-ratio identities exclude a strip of rank-one \(2\times2\) blocks. The remaining mixed support configuration contradicts \[
\operatorname{tr}\bigl(C[A,E]T[E,A]\bigr)
=\sum_{a\in A,e\in E}C_{ae}T_{ea}=4/6=2/3.
\tag{5.13}
\label{eq:5.13}\] For a complete generic collection, choose primitive projective representatives, with integral coordinates and at least one unit coordinate in each vector, \(z_r,w_r\), and \(t_r=z_{0r}w_{0r}\). The ordered equations give the inverse identity of the complete square column matrix \[
\sum_{r=0}^5\frac{z_rw_r^{\mathsf T}}{t_r}=6I_6.
\tag{5.14}
\label{eq:5.14}\] If some \(t_r\) had positive valuation, multiply by a uniformizer to the largest of those valuations and reduce. The surviving tensors would give a nontrivial dependence among at most six compatible boundary points. The invertible boundary Jacobian prevents two generic members from colliding at one boundary point. Thus \hyperref[eq:5.13]{(5.13)} and the full independence argument rule out escape.

Excluding escape makes the torus locus of the finite projective compatible-tuple scheme closed as well as open. That locus is precisely the scheme of \(P_R\), with its entire scheme structure. Finite tensor products and quotients give \(Q_R\). Complete geometric special-fibre emptiness says its reduction modulo the uniformizer is zero; Nakayama gives the final zero-algebra implication. \hyperref[appendix-dvr-complete-companions-cannot-escape]{The appendix} supplies every support, block, collision, valuation and full-scheme step in this outline. \(\square\)

For \(R=\mathbb C[[\pi]]\), take constant \(C=T_0/\sqrt6\). Its nonzero three-minors are units. The dephased columns \hyperref[eq:5.6]{(5.6)} are constant \(R\)-valued coordinates and satisfy \hyperref[eq:5.8]{(5.8)}--\hyperref[eq:5.9]{(5.9)}, so they give an explicit section of \(P_R\). Thus the complex complete-companion equations have an explicit solution. Its unequal coordinate moduli keep it from being a physical third basis. It satisfies the single-companion equations, with no claim that it is a point of \(Q_R\) or an example of special-fibre emptiness. No modular certificate is exhibited.

The physical interpretation returns when the coefficient matrix is a flat unitary over the complex numbers. A character is an algebra homomorphism to the complex numbers, equivalently a complex solution of the displayed equations. In the physical specialization over \(\mathbb C\), \(T=C^*\). Conjugating coefficients and exchanging each coordinate with its inverse defines an involution of \(A_{\mathbb C}\), \(P_{\mathbb C}\), and \(Q_{\mathbb C}\). Its fixed characters have unit-modulus coordinates. For \(X_{:r}=x_{:r}/\sqrt6\), \[
F_j(x_{:r})=|(C^*x_{:r})_j|^2-1,\qquad
q_{rs}=6(X^*X)_{rs}.
\] Thus fixed characters of \(P_{\mathbb C}\) are precisely ordered companion bases unbiased to \(I_6,C\), dephased by \(X_{0r}=1/\sqrt6\). Fixed characters of \(Q_{\mathbb C}\) add \[
q^{xy}_{rs}q^{yx}_{sr}=|x_{:r}^*y_{:s}|^2=6,
\quad\text{equivalently}\quad |(X^*Y)_{rs}|^2=1/6.
\] This interpretation is compatible with nonreduced algebras; it is a statement about characters, not a reducedness conclusion.

The finiteness assertion expressly excludes the unconstrained one-ray \(A_R\). An open torus chart of a finite projective \(R\)-scheme need not be finite over \(R\). The elementary algebra \(R[x^{\pm1}]/(\pi x-1)=R[1/\pi]\) shows why a special-fibre emptiness argument would fail without module finiteness. Checking only \(k\)-rational points is also insufficient; the hypothesis concerns all geometric points. The argument assumes no flatness over \(R\) and proves neither interior reducedness nor finiteness over a universal parameter base. It supplies no particular good prime, companion count, or physical quartet exclusion.

The preceding sections constrain candidate rays and reciprocal-orthogonal algebraic collections. We now return to physical pure states: given one state with uniform probabilities for two measurements, we construct a whole orthonormal measurement basis containing it. The displayed family admits two symmetries that make this possible. This proof uses those symmetries directly, with separate hypotheses from the middle-minor and DVR results.

\section{\texorpdfstring{Prescribed-ray completion in dimension \(2m\)}{Prescribed-ray completion in dimension 2m}}\label{prescribed-ray-completion-in-dimension-2m}

For the following pair of rank-one projective measurements, every common-unbiased pure state can be made the first column of another orthonormal measurement basis. Each new column has uniform probabilities for both given measurements. The construction uses a cyclic unitary to generate one orbit and a skew antiunitary to generate an orthogonal second orbit. Their actions preserve the rays of the two given bases, and a blockwise calculation proves every Gram entry.

The antiunitary is a map used to construct vectors and establish their inner products. This algebraic construction does not assert a deterministic quantum channel that implements the map on an unknown input state.

Let \(m\ge1\), \(\omega=e^{2\pi i/m}\), \(F_m=(\omega^{j\ell}/\sqrt m)_{j,\ell\in\mathbb Z/m}\), and \(H_2=2^{-1/2}\begin{pmatrix}1&1\\1&-1\end{pmatrix}\). Use interleaved coordinates \((j,a)\), \(j\in\mathbb Z/m\), \(a\in\{0,1\}\). Thus each of the cyclic blocks contains two vector entries. Choose arbitrary unit phases \(\eta_j\), and put \[
(K_\eta)_{(j,a),(\ell,b)}
=\frac{\omega^{j\ell}(-1)^{ab}\bar\eta_j^{\,b}}{\sqrt{2m}},
\qquad D=\operatorname{diag}_j(\omega^jI_2),
\] \[
(Jg)_j=\bar\eta_jL\bar g_j,\qquad
L=\begin{pmatrix}0&1\\-1&0\end{pmatrix}.
\tag{6.1}
\label{eq:6.1}\] Write \(J=J_\eta\) in this section. We place the two cyclic orbits in the \textbf{interleaved column order} \[
E_\eta(g)=[e_{(r,a)}]_{r\in\mathbb Z/m,\ a\in\{0,1\}},
\quad e_{(r,0)}=D^rg,\quad e_{(r,1)}=D^rJg.
\tag{6.2}
\label{eq:6.2}\] Thus its columns are \(g,Jg,Dg,DJg,\ldots,D^{m-1}g,D^{m-1}Jg\).

\phantomsection\label{result:theorem-6.1}\textbf{Theorem 6.1 (prescribed-ray completion).} Every unit \(g\) unbiased to \(I,K_\eta\) belongs to the common-unbiased orthonormal basis \(E_\eta(g)\), for every choice of phases and every \(m\), including even and composite \(m\).

\textbf{Proof.} We first check that the symmetries preserve common unbiasedness, and then check orthogonality. The factorization \[
K_\eta=(I_m\otimes H_2)\operatorname{diag}_j(1,\bar\eta_j)(F_m\otimes I_2)
\] proves unitarity; the displayed entries prove flatness. If \(k_{\ell b}\) is its actual column, direct substitution gives \[
Dk_{\ell b}=k_{\ell+1,b},\qquad
Jk_{\ell0}=k_{-\ell,1},\qquad Jk_{\ell1}=-k_{-\ell,0}.
\tag{6.3}
\label{eq:6.3}\] The unitary \(D\) permutes the anchor rays, while \(J\) exchanges their two binary labels. Both also preserve the coordinate rays of \(I\). Since each \(\eta_j\) is unit and \(L\) is real with \(L^2=-I_2\), \(J\) is antiunitary and \[
J^2=-I,\qquad JD=D^{-1}J.
\tag{6.4}
\label{eq:6.4}\] For any anchor column \(k\), unitarity and antiunitarity give \[
|\langle k,Dg\rangle|=|\langle D^{-1}k,g\rangle|,\qquad
|\langle k,Jg\rangle|=|\langle J^{-1}k,g\rangle|.
\] Thus every iteration of \(D\) or \(J\) sends a common-unbiased vector to another one. Every column of \hyperref[eq:6.2]{(6.2)} is consequently unbiased to both anchors.

Orthogonality has two separate ingredients. Equal entry moduli in the fixed coordinate basis, or native flatness, give each block the same weight, \(\|g_j\|^2=\|Jg_j\|^2=1/m\); skew antiunitarity makes the two local directions orthogonal:

\[
g_j^*Jg_j=\bar\eta_j(\bar g_{j0}\bar g_{j1}-\bar g_{j1}\bar g_{j0})=0.
\] The equal block weights make the within-orbit sums Fourier characters, while the zero local pairing annihilates every mixed sum. For all \(r,s\in\mathbb Z/m\), \[
\langle D^rg,D^sg\rangle
=\langle D^rJg,D^sJg\rangle
=\frac1m\sum_j\omega^{j(s-r)}=\delta_{rs},\qquad
\langle D^rg,D^sJg\rangle=0.
\tag{6.5}
\label{eq:6.5}\] The complete Gram matrix is therefore the identity. The \(2m\) columns form a common-unbiased orthonormal basis whose first column is the prescribed \(g\). \(\square\)

\phantomsection\label{result:corollary-6.2}\textbf{Corollary 6.2 (existence of a third mutually unbiased basis).} Every pair \(I,K_\eta\) belongs to an MU triple.

\textbf{Proof.} \hyperref[result:theorem-6.1]{The completion theorem} needs one seed. \href{https://arxiv.org/abs/1408.5728v3}{Idel--Wolf, Theorem 2} supplies it: for every unitary \(U\), diagonal unitary matrices \(L_0,R_0\) make \(L_0UR_0\) have row and column sums one. Apply it to \(K_\eta^*\). Then \[
L_0K_\eta^*R_0\mathbf1=\mathbf1,\qquad
g=R_0\mathbf1/\sqrt{2m},\qquad
K_\eta^*g=L_0^*\mathbf1/\sqrt{2m}.
\] Both native and anchor coordinates are flat, so \hyperref[result:theorem-6.1]{Theorem 6.1} supplies a third basis. \(\square\)

In Cárdenes Wuttig and Tindall's phased-unitary construction \cite{CWTMore}, selected unit-modulus vectors \(w_a\) give bases \(A_a=\operatorname{diag}(w_a)U_d\), with all pairwise transitions flat. Setting \(\eta_j=1\) in \hyperref[eq:6.1]{(6.1)} gives \(K_1=F_m\otimes F_2\) in our interleaved order; at \(m=6\), this is their dimension-twelve anchor \(U_{12}=F_6\otimes F_2\).

The undeformed group-Fourier construction itself permits completion of every prescribed common state, as in the character-product construction of \cite{BS}. Indeed, if \(U_d\) is a group-Fourier matrix and \(g\) is unbiased to \(I,U_d\), then \(A_g=\operatorname{diag}(\sqrt d\,g)U_d\) is unitary and has first column \(g\) when the constant character is placed first. Its entries are flat. Writing \((U_d)_{xr}=\chi_r(x)/\sqrt d\), one has \((U_d^*A_g)_{rs}=d^{-1}\sum_x\sqrt d\,g_x\overline{\chi_r(x)}\chi_s(x)\). A quotient of two characters is a character, so this is one coordinate of \(U_d^*g\), and common unbiasedness makes its modulus \(1/\sqrt d\). Thus prescribed-state completion already holds at this undeformed anchor; identifying the anchor does not identify this basis with \(E_1(g)\).

\hyperref[result:corollary-6.2]{Corollary 6.2} obtains a first common state from the cited theorem. The comparison target here is \hyperref[result:theorem-6.1]{Theorem 6.1}'s explicit construction at every displayed deformation \(\eta\), retaining any prescribed common state of the fixed pair. Other relevant constructions include even-order two-circulant factorization \cite{S} and full-range dimension-six Fourier triples \cite{G,GG}. Completing a state for one pair need not preserve unbiasedness to additional bases: larger MU families require all cross-basis conditions, so this result supplies no automatic enlargement of those families. Whether a completed triple admits any further common-unbiased ray is the separate question considered next.

\section{Six-dimensional profiles and the scalar obstruction}\label{six-dimensional-profiles-and-the-scalar-obstruction}

\subsection{Seed coordinates and overlaps}\label{raw-coordinates-and-full-overlaps}

Completion raises the next question: can another pure state have uniform probabilities for all columns of a constructed third measurement? We begin by comparing two constructed completions. In dimension six their vectors have three two-entry blocks. Expressing each block through its shared phase and angle separates the overlap conditions into amplitude profiles and phase cancellations. \hyperref[result:theorem-7.3]{The scalar obstruction below} shows that the required profiles cannot coexist; \hyperref[actual-clock-triples-and-their-conditional-companion-bridge]{Section 8} then relates a given clock-invariant third basis to this calculation.

From now through the localized application, \(m=3\), \(\omega=e^{2\pi i/3}\). We may assume \(\eta_0=1\) with all phases retained: for \(\widehat\eta_j=\eta_j/\eta_0\), \[
K_\eta=K_{\widehat\eta}\operatorname{diag}_{(\ell,b)}(1,\bar\eta_0),\quad
J_\eta=\bar\eta_0J_{\widehat\eta},\quad
E_\eta(g)=E_{\widehat\eta}(g)\operatorname{diag}_{(r,a)}(1,\bar\eta_0).
\tag{7.1}
\label{eq:7.1}\] This normalization changes only the second binary anchor columns and the second orbit columns by the displayed phases; it preserves their rays.

For each block, separate the phase shared by its two entries from their signed half-angle difference. Choosing the global phase so that the first shared block phase is one gives the following phase-and-angle representative, called a raw seed: \[
g_{j0}=p_je^{i\theta_j}/\sqrt6,\quad
g_{j1}=p_je^{-i\theta_j}/\sqrt6,\quad
|p_j|=1,\ p_0=1,\ \theta_j\in\mathbb R.
\tag{7.2}
\label{eq:7.2}\] Set \(c_j=\cos\theta_j\), \(s_j=\sin\theta_j\), \(k_j=p_j\bar p_{j+1}\), \(\gamma_j=\eta_j\bar\eta_{j+1}\). The cosine and sine arrays \((c_j)_j,(s_j)_j\) associated with one angle triple form its amplitude profile. All indices here are cyclic modulo three. The phases \(k_j\) and \(\gamma_j\) enter two correlation equations. Each is a zero sum of three neighboring-block products; together with the norm equation below, they enforce flatness of the two Fourier-transformed amplitude sequences. The raw seed is unbiased to both anchors exactly when it satisfies \[
\sum_jc_j^2=\frac32,\qquad
\sum_jc_jc_{j+1}k_j=0,\qquad
\sum_j\gamma_js_js_{j+1}k_j=0.
\tag{7.3}
\label{eq:7.3}\]

\leavevmode\phantomsection\label{result:lemma-7.1}\textbf{Lemma 7.1 (seed coordinates and nonvanishing).} The seed is unbiased to \(I,K_\eta\) exactly when \hyperref[eq:7.3]{(7.3)} holds. Every common-unbiased unit vector has this representation up to its independently recorded global phase. Every seed satisfying \hyperref[eq:7.3]{(7.3)} has \(c_js_j\ne0\) at every \(j\).

\textbf{Proof.} The native entries in \hyperref[eq:7.2]{(7.2)} already have the required modulus and unit norm. To test the other anchor, take its adjoint: \[
(K_\eta^*g)_{m0}=\frac13\sum_j\omega^{-jm}p_jc_j,\qquad
(K_\eta^*g)_{m1}=\frac i3\sum_j\omega^{-jm}\eta_jp_js_j.
\tag{7.4}
\label{eq:7.4}\] For any triple \(z\), define \(N(z)=\sum_j|z_j|^2\) and \(C(z)=\sum_jz_j\bar z_{j+1}\). Expansion yields \[
\left|\frac13\sum_j\omega^{-jm}z_j\right|^2
=\frac{N(z)+\omega^mC(z)+\omega^{-m}\overline{C(z)}}9.
\tag{7.5}
\label{eq:7.5}\] Comparison of the three Fourier coefficients shows that the squared moduli equal \(1/6\) precisely when \(N=3/2\) and both nonconstant coefficients vanish, that is, \(C=0\). Here \(pc=(p_jc_j)_j\) and \(\eta ps=(\eta_jp_js_j)_j\) denote entrywise products. Applying the formula to these two sequences, and using \(\sum s_j^2=3-\sum c_j^2\), gives the squared-norm equation and two correlation equations in \hyperref[eq:7.3]{(7.3)}.

Conversely, the raw coordinates represent every native-flat physical vector, with its own global phase. For such an \(h\), choose real \(u_j,v_j\) with \(\sqrt6h_{j0}=e^{iu_j},\sqrt6h_{j1}=e^{iv_j}\). Set \(\theta_j=(u_j-v_j)/2\), \(q_j=e^{i(u_j+v_j)/2}\), \(\sigma=q_0\), \(p_j=q_j/\sigma\). Then \(h=\sigma g\), \(p_0=1\), and \(\sigma\) is independent of every other seed's phase. Anchor flatness gives \hyperref[eq:7.3]{(7.3)}. The correlation equations also prevent vanishing sine or cosine amplitudes, which will justify the divisions in \hyperref[actual-clock-triples-and-their-conditional-companion-bridge]{Section 8}. If \(|z_j|\le1,N(z)=3/2,C(z)=0\), one zero is impossible. For example, \(z_0=0\) leaves \(z_1\bar z_2=0\), hence at most one nonzero entry and \(N(z)\le1\). Apply this to \(pc,\eta ps\). \(\square\)

The symmetry actions remain visible in these coordinates. The clock \(D^r\) replaces \(p_j\) by \(\omega^{rj}p_j\), multiplying both correlation sums by \(\omega^{-r}\). The antiunitary acts as \[
(Jg)_j=\bar\eta_j\bar p_j(e^{i\theta_j},-e^{-i\theta_j})^T/\sqrt6.
\] For \(\theta'_j=\theta_j+\pi/2\), \(p''_j=\bar\eta_j\bar p_j\), its dephased raw seed \(g''\) satisfies \[
Jg=-ig'',\qquad c'_j=-s_j,\quad s'_j=c_j,\quad
k'_j=\bar\gamma_j\bar k_j.
\tag{7.6}
\label{eq:7.6}\] Thus \(J\) interchanges the cosine and sine correlation equations, with conjugation, and interchanges their squared norms. The factor \(-i\) in \hyperref[eq:7.6]{(7.6)} is the global phase relating the actual antiunitary image to its dephased raw representative.

Collect the two local directions \(g_j\) and \((Jg)_j\) into the two-column unitary matrix \[
W_j(\theta,p;\eta)=\frac1{\sqrt2}
\begin{pmatrix}
p_je^{i\theta_j}&\overline{\eta_jp_j}e^{i\theta_j}\\
p_je^{-i\theta_j}&-\overline{\eta_jp_j}e^{-i\theta_j}
\end{pmatrix}.
\tag{7.7}
\label{eq:7.7}\] It has \([g_j\mid(Jg)_j]=W_j/\sqrt3\), \(\det W_j=-\bar\eta_j\). The transition matrix between two completed measurements is assembled from the overlaps of these local two-dimensional frames. Its cyclic labels make those overlaps three Fourier sums. In \(\mathcal F_m\), \(m\in\mathbb Z/3\) is a Fourier label; the dimension remains fixed at six. For two seeds satisfying \hyperref[eq:7.3]{(7.3)}, with independent phase data \((\alpha,p),(\beta,q)\), set \[
\mathcal F_m=\frac13\sum_j\omega^{jm}
W_j(\alpha,p;\eta)^*W_j(\beta,q;\eta).
\tag{7.8}
\label{eq:7.8}\]

\phantomsection\label{result:lemma-7.2}\textbf{Lemma 7.2 (36-to-12 reduction).} In the exact interleaved order \hyperref[eq:6.2]{(6.2)}, \[
(E_\eta(g)^*E_\eta(h))_{(r,a),(s,b)}
=(\mathcal F_{s-r})_{a,b}.
\tag{7.9}
\label{eq:7.9}\] Each of the twelve entries on the right occurs three times among the 36 on the left. Thus the constructed bases are MU exactly when all twelve squared moduli are \(1/6\).

\textbf{Proof.} The \(r\)-th column pair has native block \(\omega^{jr}W_j/\sqrt3\). Taking inner products gives the exponent \(j(s-r)\) and factor \(1/3\) in \hyperref[eq:7.9]{(7.9)}. Each difference \(s-r\) occurs for three ordered pairs, so each Fourier entry occurs three times.

The determinant-one form of each local frame product also couples the two columns of the Fourier sums. The products \(W_j(\alpha,p)^*W_j(\beta,q)\) have determinant one, hence form \(\begin{pmatrix}u_j&v_j\\-\bar v_j&\bar u_j\end{pmatrix}\). Consequently \[
(\mathcal F_m)_{11}=\overline{(\mathcal F_{-m})_{00}},\qquad
(\mathcal F_m)_{01}=-\overline{(\mathcal F_{-m})_{10}}.
\tag{7.10}
\label{eq:7.10}\] Flatness of the first column at all three Fourier labels therefore gives flatness of the second, with label reversal \(m\mapsto-m\). For a selected physical ray \(h=\sigma h_0\), \[
(E_\eta(g_0)^*h)_{(r,a)}=\sigma(\mathcal F_{-r})_{a,0}.
\tag{7.11}
\label{eq:7.11}\] Thus unbiasedness of one selected ray to \(E_\eta(g_0)\) supplies the first-column conditions, which imply all twelve conditions through \hyperref[eq:7.10]{(7.10)}. No original fourth basis is assumed in this calculation.

The two seeds retain independent global phases: if \(g=\rho g_0\), then \(E_\eta(g)=E_\eta(g_0)B_\rho\), where the diagonal of \(B_\rho\) is \(\rho\) on \(a=0\) and \(\bar\rho\) on \(a=1\). Thus actual overlaps are \(B_\rho^*[E_\eta(g_0)^*E_\eta(h_0)]B_\sigma\). Further unitary column monomials act by \(M^*[E_\eta(g_0)^*E_\eta(h_0)]N\), retaining their phases and labels. \(\square\)

\subsection{The scalar obstruction}\label{the-full-scalar-obstruction}

The phases can make a sum of three complex numbers vanish only if their magnitudes satisfy the triangle inequalities. We can therefore discard the phase choices and ask whether the necessary amplitude conditions are consistent. \hyperref[result:theorem-7.3]{The following theorem} shows that three profiles---two seed profiles and their difference---cannot all satisfy them. Its proof first obtains strict bounds from weak triangle inequalities, then compares the same nine inverse cosines by rows and by columns. The triangle lengths and comparison angles are auxiliary quantities derived from the state coordinates.

\phantomsection\label{result:theorem-7.3}\textbf{Theorem 7.3 (three amplitude profiles).} There are no real \(\alpha,\beta\in\mathbb R^3\) such that each \(\theta\in\{\alpha,\beta,\beta-\alpha\}\) satisfies \[
\sum_j\cos^2\theta_j=\frac32
\tag{7.12}
\label{eq:7.12}\] and both length triples \((|\cos\theta_j\cos\theta_{j+1}|)_j\), \((|\sin\theta_j\sin\theta_{j+1}|)_j\) satisfy all weak triangle inequalities, including zeros and equality cases.

\textbf{Proof.} We begin with the restrictions on a single permitted profile. Put \[
v_j=\cos2\theta_j,\quad
P_j=\cos^2\theta_j=(1+v_j)/2,\quad
Q_j=\sin^2\theta_j=(1-v_j)/2,\quad
r=\sum_jv_j^2,\quad t=v_0v_1v_2.
\] Here \(P,Q\) are squared amplitudes; the seed phases are \(p_j\). The squared-norm equation gives \(\sum v_j=0\), \(\sum P_j=\sum Q_j=3/2\). Here \(e_1(P)=P_0+P_1+P_2\), \(e_2(P)=P_0P_1+P_1P_2+P_2P_0\), and \(e_3(P)=P_0P_1P_2\) are the elementary symmetric polynomials, with the same convention for \(Q\). The squared sides of the cosine-product triangle are \(P_0P_1,P_1P_2,P_2P_0\); its four Heron factors multiply to \[
2\sum_{\rm cyc}(P_0P_1)(P_1P_2)
-\sum_{\rm cyc}(P_0P_1)^2=4e_1(P)e_3(P)-e_2(P)^2.
\] Every weak triangle inequality makes the corresponding Heron factor nonnegative; equality and zero sides are allowed. Hence \(6e_3(P)-e_2(P)^2\ge0\), and likewise for \(Q\). Using \(\sum v_j=0\) gives \[
e_2(P)=e_2(Q)=\frac34-\frac r8,\quad
e_3(P)=\frac{1-r/2+t}{8},\quad
e_3(Q)=\frac{1-r/2-t}{8}.
\] Multiplication by 64 yields \(12-12r-r^2\pm48t\ge0\), so \[
r+4|t|+\frac{r^2}{12}\le1.
\tag{7.13}
\label{eq:7.13}\] In particular, \[
r+2|t|<1.
\tag{7.14}
\label{eq:7.14}\] For \(r>0\), the upper bound is \(1-2|t|-r^2/12<1\); for \(r=0\), \(t=0\). Also, \[
|v_j|<1/\sqrt2\quad\text{for each }j.
\tag{7.15}
\label{eq:7.15}\] Otherwise a sign reversal and index choice allow \(v_0=x\ge1/\sqrt2\). Write \(u=v_1v_2\); then \(v_1+v_2=-x\), \(r=2(x^2-u)\), \(|t|=x|u|\). If \(u\le0\), \(r+4|t|\ge2x^2\ge1\). If \(u\ge0\), \(r+4|t|=2x^2+(4x-2)u\ge1\), since \(x>1/2\). Both contradict \hyperref[eq:7.13]{(7.13)}, as \(r>0\) supplies the extra positive \(r^2/12\).

These strict estimates apply independently to each of the three profiles. Now form the real array whose row \(j\) is \[
X_j=\cos2\alpha_j,\quad Y_j=\cos2\beta_j,\quad
Z_j=\cos2(\beta_j-\alpha_j).
\] Each column satisfies \hyperref[eq:7.14]{(7.14)}--\hyperref[eq:7.15]{(7.15)}, while each row links the three angles at one native block. The cosine difference identity gives in each row \[
X_j^2+Y_j^2+Z_j^2-2X_jY_jZ_j=1.
\tag{7.16}
\label{eq:7.16}\] Indeed, square \(Z-XY=\sin2\alpha\,\sin2\beta\).

The row identity forces a negative product. To prove this, put \(a=|X_j|,b=|Y_j|,c=|Z_j|\), all below \(k=1/\sqrt2\). If the product were nonnegative, \hyperref[eq:7.16]{(7.16)} would give \(f(a,b,c)=1-a^2-b^2-c^2+2abc=0\). But \(f>0\) on \([0,k)^3\). Separate concavity lets us bound \(f\) from below by applying a chord bound in each coordinate. The resulting lower bound is the trilinear interpolation of its eight vertex values on \([0,k]^3\). Those values, by number of coordinates equal to \(k\), are \[
1,\qquad 1/2,\qquad0,\qquad(\sqrt2-1)/2.
\] The vertex values are nonnegative, and strict membership in the cube gives the positive origin value weight \((1-a/k)(1-b/k)(1-c/k)>0\). This contradicts \(f=0\), also excluding a zero row product. Thus every array entry is nonzero, and each row obeys \[
a^2+b^2+c^2+2abc=1,\qquad0<a,b,c<1/\sqrt2.
\tag{7.17}
\label{eq:7.17}\] Let \(A=\arccos a,B=\arccos b,C=\arccos c\), in \((0,\pi/2)\). The positive root of \hyperref[eq:7.17]{(7.17)} is \[
c=-ab+\sqrt{(1-a^2)(1-b^2)}.
\] Its positivity gives \(a^2+b^2<1\), hence \(\pi/2<A+B<\pi\). Therefore \(c=-\cos(A+B)=\cos(\pi-A-B)\), and \[
\arccos a+\arccos b+\arccos c=\pi.
\tag{7.18}
\label{eq:7.18}\] The same inverse cosines behave differently down a column. There, \hyperref[eq:7.14]{(7.14)} says \(a^2+b^2+c^2+2abc<1\). The increasing quadratic \(z^2+2abz+a^2+b^2-1\), \(z\ge0\), has the unique positive root \[
z_0=-ab+\sqrt{(1-a^2)(1-b^2)}
=\cos(\pi-\arccos a-\arccos b).
\] Here \(a^2+b^2<1\), and the strict column inequality gives \(c<z_0\). Thus \[
\arccos a+\arccos b+\arccos c>\pi.
\tag{7.19}
\label{eq:7.19}\] All entries used in the inverse cosines are positive by the row argument, so no endpoint has been excluded without justification. Summing the same nine arccosines by rows gives \(3\pi\), whereas summing them by columns gives more than \(3\pi\). This contradiction excludes the three profiles. The original signed angles have not been normalized; the arccosines are auxiliary comparison angles. \(\square\)

We now return from the scalar profiles to the measurement overlaps. The two candidate states supply the first two profiles. If their completions were mutually unbiased, the transition matrix would supply the angle-difference profile as well, contradicting \hyperref[result:theorem-7.3]{Theorem 7.3}.

\phantomsection\label{result:corollary-7.4}\textbf{Corollary 7.4 (no mutually unbiased pair of constructed completions).} Two seeds satisfying \hyperref[eq:7.3]{(7.3)} at the same unit phases \(\eta\) cannot have \[
|(\mathcal F_m)_{ab}|^2=1/6
\quad(m\in\mathbb Z/3,\ a,b\in\{0,1\}).
\tag{7.20}
\label{eq:7.20}\] In particular two constructed completions \(E_\eta(g),E_\eta(h)\) cannot be MU.

\textbf{Proof.} The correlation equations for each seed already provide the amplitude conditions for \(\alpha\) and \(\beta\): each is a zero sum of three complex numbers, so its three magnitudes obey the weak triangle inequalities. It remains to extract the same conditions for their difference from the overlap equations. Put \(\delta=\beta-\alpha\), \(a_j=\bar p_jq_j\), \(b_j=\eta_jp_jq_j\). Direct multiplication gives \[
W_j(\alpha,p)^*W_j(\beta,q)
=\begin{pmatrix}
a_j\cos\delta_j&i\bar b_j\sin\delta_j\\
ib_j\sin\delta_j&\bar a_j\cos\delta_j
\end{pmatrix}.
\tag{7.21}
\label{eq:7.21}\] The overlap sums use the positive Fourier exponent in \hyperref[eq:7.8]{(7.8)}, giving \[
\left|\sum_j\omega^{jm}z_j\right|^2
=N(z)+\omega^{-m}C(z)+\omega^m\overline{C(z)}.
\] The three values are \(3/2\) exactly when \(N=3/2,C=0\). Apply this to the 00 and 10 entries in \hyperref[eq:7.21]{(7.21)}. Condition \hyperref[eq:7.20]{(7.20)}, with the \(1/3\) factor, yields \[
\sum_j\cos^2\delta_j=3/2,\quad
\sum_j\cos\delta_j\cos\delta_{j+1}a_j\bar a_{j+1}=0,\quad
\sum_j\sin\delta_j\sin\delta_{j+1}b_j\bar b_{j+1}=0.
\] The unit phases turn the two zero sums into weak triangle inequalities for \(\delta\). Thus all three profiles satisfy the forbidden conditions of \hyperref[result:theorem-7.3]{Theorem 7.3}. Only a necessary subset of the twelve overlap equations was needed for this contradiction; the hypothesis still comprises all twelve. By \hyperref[result:lemma-7.2]{Lemma 7.2}, the contradiction excludes two MU constructed completions, including independently phased columns. \(\square\)

\section{Clock-invariant triples and the original companion basis}\label{actual-clock-triples-and-their-conditional-companion-bridge}

\hyperref[result:theorem-7.3]{The scalar obstruction} compares bases produced by the construction, whereas the extension question starts with a third basis already given. We must therefore use its original columns. When its rays are invariant under the clock, they split into two three-column orbits. The anchor equations determine the second orbit through a two-row correlation-coefficient matrix. Under the assumption of an extra common unbiased ray, its rank reduces the given basis to the constructed form or an aligned product form. The product branch has its own obstruction to an extra pure state, which we prove first. Together these alternatives yield strong unextendibility of the given triple.

\subsection{The exceptional aligned product triple}\label{the-exceptional-aligned-product-triple}

The exceptional branch consists of three product measurements with pairwise unbiased local qubit (dimension two) and qutrit (dimension three) bases. A hypothetical extra pure state would have uniform joint probabilities for all three. \hyperref[result:lemma-8.1]{The lemma below} derives the contradiction through its reduced states and conditional qubit states. Product-triple nonextension is prior work; see \href{https://arxiv.org/abs/1203.6887v2}{McNulty--Weigert, Theorem 1}.

\phantomsection\label{result:lemma-8.1}\textbf{Lemma 8.1 (aligned product obstruction).} Let \(A_0,A_1,A_2\) be three pairwise MU qubit bases, and \(B_0,B_1,B_2,B_3\) four pairwise MU qutrit bases. No unit vector in \(\mathbb C^2\otimes\mathbb C^3\) is unbiased to all vectors of the three product bases \(A_j\otimes B_j\), \(j=0,1,2\).

\textbf{Proof.} Uniform joint probabilities determine the reduced density matrices strongly enough to constrain a pure \(h\). Its conditional qubit states will then contradict the three qubit bases. Suppose \(h\) exists and let \(\rho_2,\rho_3\) be its reduced density matrices. Summing the joint probabilities \(1/6\) over the other factor gives \[
\langle a_{jr},\rho_2a_{jr}\rangle=1/2,\qquad
\langle b_{jt},\rho_3b_{jt}\rangle=1/3
\quad(j=0,1,2).
\tag{8.1}
\label{eq:8.1}\] Put \(\sigma_j=|a_{j0}\rangle\langle a_{j0}|-|a_{j1}\rangle\langle a_{j1}|\). They are traceless Hermitian, square to \(I_2\), and have \(\operatorname{Tr}(\sigma_j\sigma_\ell)=0\) for \(j\ne\ell\), by local unbiasedness. Here the Hilbert--Schmidt product of matrices \(A,B\) is \(\operatorname{Tr}(A^*B)\). Their squared norms are two, so they form an orthogonal real basis of the three-dimensional traceless Hermitian qubit space. The uniform qubit marginals in \hyperref[eq:8.1]{(8.1)} annihilate every traceless component of \(\rho_2\), forcing \[
\rho_2=I_2/2.
\tag{8.2}
\label{eq:8.2}\]

For each qutrit basis let \(\mathcal D_j\) be its two-dimensional real space of traceless Hermitian diagonal matrices. If \(P_{jt}=|b_{jt}\rangle\langle b_{jt}|\), then \[
\operatorname{Tr}[(P_{jt}-I_3/3)(P_{\ell s}-I_3/3)]
=|\langle b_{jt},b_{\ell s}\rangle|^2-1/3=0
\quad(j\ne\ell).
\] The four diagonal spaces are pairwise orthogonal and their dimensions sum to eight, exhausting the traceless Hermitian qutrit space. The first three uniform marginals in \hyperref[eq:8.1]{(8.1)} therefore leave \(\rho_3-I_3/3\) in \(\mathcal D_3\). Consequently \(\rho_3\) is diagonal in the \textbf{given} fourth basis \(B_3\), rather than merely in an unspecified eigenbasis.

We now use purity to recover the state from these marginal restrictions. Writing \(h=\sum M_{rt}e_r\otimes f_t\) gives \(\rho_2=MM^*\), \(\rho_3=(M^*M)^T\), with the same nonzero eigenvalues. The Schmidt rank is the rank of this coefficient matrix \(M\). By \hyperref[eq:8.2]{(8.2)}, \(\rho_3\) has spectrum \(\{1/2,1/2,0\}\). Exactly two original \(B_3\) columns \(v_a,v_b\) support it. Expansion in the given basis \(B_3\) yields orthonormal qubit vectors \(u,v\) with \[
h=(u\otimes v_a+v\otimes v_b)/\sqrt2.
\tag{8.3}
\label{eq:8.3}\] Because \(\rho_3\) is already diagonal in \(B_3\), the repeated eigenvalue \(1/2\) does not require replacing the given qutrit columns.

For each of the first three product bases, conditioning on a qutrit outcome now gives three qubit states with equilateral relative phases. Fix \(j=0,1,2\) and set \(A_t=\langle b_{jt},v_a\rangle\), \(B_t=\langle b_{jt},v_b\rangle\), \(\zeta_t=B_t/A_t\). All denominators have modulus \(1/\sqrt3\). The normalized conditional qubit vector is \[
\psi_t=(\sqrt3A_t)(u+\zeta_tv)/\sqrt2,\qquad|\zeta_t|=1.
\] Orthogonality gives \(\sum_t\bar A_tB_t=\frac13\sum_t\zeta_t=0\). Three unit numbers with zero sum are a rotated equilateral triple: \[
\{\zeta_0,\zeta_1,\zeta_2\}=\zeta_*\{1,\omega,\omega^2\}.
\tag{8.4}
\label{eq:8.4}\] To see this, divide by one entry. The others sum to \(-1\); their real parts are \(-1/2\), making them \(\omega,\omega^2\) in one order.

The conditional probability is \((1/6)/(1/3)=1/2\), so every \(\psi_t\) is unbiased to \(A_j\). Hence \(\langle\psi_t,\sigma_j\psi_t\rangle=0\). The two diagonal entries of \(\sigma_j\) in the orthonormal basis \(u,v\) sum to zero. With \(m_j=\langle u,\sigma_jv\rangle\), expansion gives \[
0=\operatorname{Re}(\zeta_tm_j).
\] Equation \hyperref[eq:8.4]{(8.4)} forces both \(\operatorname{Re}(\zeta_*m_j)=0\) and \(\operatorname{Re}(\omega\zeta_*m_j)=0\). If \(\zeta_*m_j=iy\), \(y\in\mathbb R\), the latter is \(-\sqrt3y/2=0\); hence \(m_j=0\). For each \(j\), the three phase directions force the off-diagonal entry \(m_j\) to vanish. Thus all three \(\sigma_j\) are diagonal in the same qubit basis \(u,v\). Each is a real multiple of \(|u\rangle\langle u|-|v\rangle\langle v|\), and squaring to identity makes the multiple \(\pm1\). Two then have Hilbert--Schmidt product \(\pm2\), contradicting their orthogonality. The conclusion concerns pure rays; no mixed-state obstruction is claimed. \(\square\)

\subsection{The clock-invariant triple argument}\label{the-actual-clock-argument}

\phantomsection\label{result:theorem-8.2}\textbf{Theorem 8.2 (clock-invariant triple has no extra pure state).} Let \(I,K_\eta,V\) be given pairwise MU bases in dimension six. If the six \textbf{original rays} of \(V\) are setwise \(D\)-invariant, no unit vector is unbiased to all three.

The argument does not identify every clock-invariant companion unconditionally. Under the assumption of an extra common ray, it establishes, for any chosen original column \(g\) of \(V\), \[
V=E_\eta(g)M_g
\tag{8.5}
\label{eq:8.5}\] with a unitary column monomial recording column permutations and phases. The alternative is the aligned product branch, already excluded by \hyperref[result:lemma-8.1]{Lemma 8.1} under the extra-ray assumption.

\textbf{Proof.} We first locate the two original clock orbits and describe their complement. The rank of the correlation-coefficient matrix then determines the companion, apart from the product alternative. Normalize \(\eta_0=1\) by \hyperref[eq:7.1]{(7.1)}, retaining all phases. Each native-flat vector has three distinct \(D\)-rays: if \(D^rg\) were proportional to \(g\), its three nonzero blocks would force \(\omega^{rj}\) to be constant, hence \(r=0\). Thus the six invariant rays of \(V\) are exactly two selected three-orbits. Choose any original seed \(g\) from one. Its orbit is orthonormal by \hyperref[result:theorem-6.1]{Theorem 6.1}.

\subsubsection*{Complement geometry and the correlation matrix}\label{complement-geometry-and-closure-rank}

Fourier inversion extracts the native block projections from one orbit. It therefore describes both its span and its orthogonal complement: with \(\mathcal L_g=\operatorname{span}\{D^rg\}\), \[
\mathcal L_g=\bigoplus_j\mathbb Cg_j,\qquad
\mathcal L_g^\perp=\bigoplus_j\mathbb C(Jg)_j.
\tag{8.6}
\label{eq:8.6}\] Local skew orthogonality identifies the second line in each nonzero block. Hence any complementary flat vector has \[
w_j=t_j(Jg)_j,\qquad |t_j|=1,
\tag{8.7}
\label{eq:8.7}\] because both vectors have block norm squared \(1/3\). The equal block weights also force three orthogonal flat vectors in this complement to form a full \(D\)-orbit, even without separately assuming invariance. For two, write \(w'_j=u_jw_j\), divide by the independent phase \(u_0\), and obtain \[
0=\langle w,w'\rangle=(1+u_1+u_2)/3.
\] The remaining units are \(\omega,\omega^2\) in some order; hence \(w'\) is proportional to \(Dw\) or \(D^2w\). Three columns exhaust that orbit. Their phases and order remain recorded in a unitary column monomial.

The complement geometry fixes the local direction of every remaining column, but leaves one phase per block. To determine which complementary orbit can satisfy the anchor equations, use the representative \hyperref[eq:7.2]{(7.2)} of \(g\), with its first shared block phase set to one. Its independent physical phase is restored through \hyperref[result:lemma-7.2]{Lemma 7.2}. The coefficients of the two correlation equations form the matrix \[
A_j=c_jc_{j+1}k_j,\quad
B_j=\gamma_js_js_{j+1}k_j,\quad
R(g)=\begin{pmatrix}A_0&A_1&A_2\\B_0&B_1&B_2\end{pmatrix}.
\tag{8.8}
\label{eq:8.8}\] The anchor equations say that both rows sum to zero. \hyperref[result:lemma-7.1]{Lemma 7.1} makes every entry nonzero, so \(R(g)\) has rank one or two. The seed \(b=iJg\) of \hyperref[eq:7.6]{(7.6)}, with the first shared block phase again set to one, satisfies \[
A_j(b)=\overline{B_j(g)},\qquad
B_j(b)=\overline{A_j(g)}.
\tag{8.9}
\label{eq:8.9}\] Consequently \(R(g)\) and \(R(b)\) have the same rank. Keeping the independent global phase separate, write a complementary seed as \(w_j=t_jb_j\), \(t_0=1\). Its norm equation is automatic; its two correlation equations are exactly \[
R(b)\ell=0,\qquad
\ell_j=t_j\bar t_{j+1},\quad|\ell_j|=1,\quad
\ell_0\ell_1\ell_2=1.
\tag{8.10}
\label{eq:8.10}\] At rank two, the kernel of \(R(b)\) is uniquely determined: it is the line through \((1,1,1)\). Thus \(\ell_j=z\) for all \(j\), \(z^3=1\), and \(t_j=t_0z^{-j}\). The complementary orbit is exactly \(O(Jg)\), proving \hyperref[eq:8.5]{(8.5)} in this case.

\subsubsection*{Rank one: a common coordinate map}\label{rank-one-an-exact-common-coordinate-map}

At rank one, the two correlation equations no longer give independent restrictions on the complementary phases. Their proportionality instead forces a special product structure. We first show that the sine and cosine magnitudes are equal in every block. Proportional nonzero rows give \[
\gamma_j\frac{s_j}{c_j}\frac{s_{j+1}}{c_{j+1}}=\kappa\ne0.
\tag{8.11}
\label{eq:8.11}\] All divisions follow from \hyperref[result:lemma-7.1]{Lemma 7.1}. The three equal absolute tangent products force \(|s_j/c_j|=u>0\) for every \(j\). Then \(c_j^2=1/(1+u^2)\), and the squared norm \(3/2\) forces \(u=1\). Write \(\epsilon_j=s_j/c_j\in\{1,-1\}\). Equation \hyperref[eq:8.11]{(8.11)} becomes \(\gamma_j=\kappa\epsilon_j\epsilon_{j+1}\); their product gives \(\kappa^3=1\). Choose \(t\in\mathbb Z/3\) with \(\kappa=\omega^{-t}\), set \(\sigma_j=\epsilon_0\epsilon_j\), and use \(\eta_0=1\) to solve the recurrence: \[
\eta_j=\sigma_j\omega^{tj},\qquad \sigma_0=1.
\tag{8.12}
\label{eq:8.12}\]

Let \(X=\begin{pmatrix}0&1\\1&0\end{pmatrix}\), and let \(P_\sigma\) be block \(I_2\) if \(\sigma_j=1\), block \(X\) if \(\sigma_j=-1\). This single real coordinate permutation commutes with \(D\). Applying it to all bases preserves their mutual unbiasedness. Since \(XH_2=H_2\operatorname{diag}(1,-1)\), \[
P_\sigma K_\eta=K_1R_t,\qquad
R_te_{(m,0)}=e_{(m,0)},\quad
R_te_{(m,1)}=e_{(m-t,1)}.
\tag{8.13}
\label{eq:8.13}\] The identity anchor is restored by its matching column permutation. Also \(XLX=-L\), giving \[
P_\sigma J_\eta P_\sigma^{-1}=D^{-t}J_1.
\tag{8.14}
\label{eq:8.14}\] Thus \(P_\sigma E_\eta(g)\) is \(E_1(P_\sigma g)\) with only the second orbit relabeled \(r\mapsto r-t\). The same \(P_\sigma\) acts on all three original bases and any hypothetical \(h\). Its inverse is \(P_\sigma\) itself; the displayed column permutations are reversed when returning to the original labels.

The transformed raw angle is \(\theta'_j=\sigma_j\theta_j\), with the same phase witness. Its tangent is the constant sign \(\epsilon_0\). The cosine correlation equation is unchanged; its sine counterpart is the old one multiplied by \(\omega^t\). Hence \(\theta'_j=\epsilon\pi/4+n_j\pi\) with a single \(\epsilon\in\{1,-1\}\). Absorb \((-1)^{n_j}\) into the phase witness, then extract its first entry as a separate global phase to restore \(p_0=1\). These operations retain the physical vector and all phase choices. For \[
u_\epsilon=(e^{i\epsilon\pi/4},e^{-i\epsilon\pi/4})^T/\sqrt2,
\] the actual transformed seed is, up to that phase, \[
g=(p/\sqrt3)\otimes u_\epsilon,
\tag{8.15}
\label{eq:8.15}\] in the native qutrit--qubit order, and the remaining seed condition is \[
|p_j|=1,\quad p_0=1,\quad\sum_jp_j\bar p_{j+1}=0.
\tag{8.16}
\label{eq:8.16}\]

\subsubsection*{The six quadratic-phase sequences and the product alternative}\label{exhaustive-chirp-alternatives-and-the-actual-product-input}

After this common coordinate change, only the qutrit phases remain to be classified. There are six dephased possibilities, in two clock orbits, so the product branch can be checked without replacing any given basis. The remaining qutrit witnesses can be listed exhaustively from \hyperref[eq:8.16]{(8.16)}. Put \(k_j=p_j\bar p_{j+1}\). These units sum to zero and have product one. The equilateral argument in \hyperref[result:lemma-8.1]{Lemma 8.1} makes them a phase times \(1,\omega,\omega^2\); their product makes that phase a cube root. Thus \(k\) is exactly a permutation of \(1,\omega,\omega^2\). Each of the six permutations determines a unique dephased \(p\): \[
p_0=1,\qquad p_1=\bar k_0,\qquad p_2=\bar k_0\bar k_1.
\] The last edge follows from \(\prod k=1\). Clock multiplication rotates every \(k_j\) by \(\omega^{-r}\), so there are precisely two clock orbits: \[
\begin{aligned}
O(p_+)&=\{(1,1,\omega),(1,\omega,1),(1,\omega^2,\omega^2)\},\\
O(p_-)&=\{(1,1,\omega^2),(1,\omega,\omega),(1,\omega^2,1)\},
\end{aligned}
\tag{8.17}
\label{eq:8.17}\] with \(p_+=(1,1,\omega)\) and \(p_-=\bar p_+\). These six vectors are the dephased qutrit quadratic-phase, or \textit{chirp}, sequences: equivalently \(p_j=\omega^{aj^2+bj}\) with \(a\in\{1,2\}\) and \(b\in\mathbb Z_3\). Each dephased phase sequence has \(p_0=1\), \(\|p\|=\sqrt3\) and \(p_j^3=1\); its physical qutrit vector is \(p/\sqrt3\). Arbitrary physical global phases remain separate.

The complementary qubit line is \(u_{-\epsilon}\). Every complementary seed has, up to its own phase, \[
w=(q/\sqrt3)\otimes u_{-\epsilon},
\] with \(q\) satisfying \hyperref[eq:8.16]{(8.16)}, hence in one of the two lists. The antiunitary factor is \[
L\bar u_\epsilon=i\epsilon u_{-\epsilon},\qquad
J_1g=i\epsilon(\bar p/\sqrt3)\otimes u_{-\epsilon}.
\tag{8.18}
\label{eq:8.18}\] If \(q\in O(\bar p)\), its orbit is canonical; \hyperref[eq:8.14]{(8.14)} gives \hyperref[eq:8.5]{(8.5)} at the original \(\eta\). Otherwise \(q\in O(p)\). Relabeling the original complementary columns and recording their phases gives the transformed product triple \[
I_3\otimes I_2,\qquad Q\otimes H_2,\qquad
B_p\otimes Y_\epsilon,
\tag{8.19}
\label{eq:8.19}\] where \(Q=F_3\), \(B_p=\operatorname{diag}(p)Q\), \(Y_\epsilon=[u_\epsilon,u_{-\epsilon}]\). The two chirp lists leave exactly these alternatives. Only the first gives the canonical companion; the second gives the aligned product triple.

The product branch needs a fourth qutrit basis in order to apply \hyperref[result:lemma-8.1]{Lemma 8.1}. Consider the four bases \[
I_3,\quad Q,\quad B_p,\quad B_{\bar p}
\tag{8.20}
\label{eq:8.20}\] These are pairwise MU. Flatness handles \(I_3\). The entries \[
(Q^*\operatorname{diag}(p)Q)_{rs}
=\frac13\sum_jp_j\omega^{j(s-r)}
\] have squared modulus \(1/3\) by \hyperref[eq:8.16]{(8.16)} and the Fourier identity; likewise for \(\bar p\). Finally \[
B_p^*B_{\bar p}=Q^*\operatorname{diag}(\bar p^{\,2})Q
=Q^*\operatorname{diag}(p)Q
\] because \(p_j^3=1\) for the dephased phase sequences above. The three qubit bases \(I_2,H_2,Y_\epsilon\) are pairwise MU by their displayed entries and \(\cos^2(\pi/4)=\sin^2(\pi/4)=1/2\). A fixed tensor shuffle puts \hyperref[eq:8.19]{(8.19)} into the order of \hyperref[result:lemma-8.1]{Lemma 8.1}. \hyperref[result:lemma-8.1]{Lemma 8.1} therefore excludes every common unit ray in this aligned product branch.

\subsubsection*{Completion of the contradiction}\label{completion-of-the-contradiction}

Suppose a common unit ray \(h\) exists for the original triple \(I,K_\eta,V\). The product alternative is excluded by the common unitary coordinate change and \hyperref[result:lemma-8.1]{Lemma 8.1}. Thus rank two or canonical rank one gives \hyperref[eq:8.5]{(8.5)} for the chosen original \(g\), at the original \(\eta\). Identification \hyperref[eq:8.5]{(8.5)} now lets us apply the completion symmetries to the original companion as well as to the anchors. The canonical companion rays have the actions \[
De_{(r,a)}=e_{(r+1,a)},\quad
Je_{(r,0)}=e_{(-r,1)},\quad Je_{(r,1)}=-e_{(-r,0)}.
\] For each companion ray \(e\), \[
|\langle e,Dh\rangle|=|\langle D^{-1}e,h\rangle|,\qquad
|\langle e,Jh\rangle|=|\langle J^{-1}e,h\rangle|=1/\sqrt6.
\] The first value is also \(1/\sqrt6\); iteration proves that every column of \(E_\eta(h)\) is unbiased to \(V\). \hyperref[result:theorem-6.1]{Theorem 6.1} makes this an orthonormal basis unbiased to the two anchors. Since \(V=E_\eta(g)M_g\), the two constructed completions form the mutually unbiased pair forbidden by \hyperref[result:corollary-7.4]{Corollary 7.4}. Thus the original triple has no extra common unit ray. \(\square\)

The product alternative shows why the hypothetical extra ray is essential to the canonical identification within the proof. At \(\eta=(1,1,1)\), take \(p=(1,1,\omega)\), all first-seed angles \(\pi/4\), and all complementary-seed angles \(3\pi/4\) with the \textbf{same} \(p\). Both seeds satisfy \hyperref[eq:7.3]{(7.3)}, with correlation summands \((1,\omega^2,\omega)/2\). Their two clock orbits form an MU companion, while the canonical complement, with its first shared block phase set to one, has phase sequence \(\bar p=(1,1,\omega^2)\) in the other orbit. This is the product branch of \hyperref[eq:8.19]{(8.19)}, not \(E_\eta(g)\). \hyperref[result:lemma-8.1]{Lemma 8.1} excludes an extra ray, consistently with both the conditional identification and the conclusion of \hyperref[result:theorem-8.2]{Theorem 8.2}.

\section{Applications and limits}\label{applications-and-limits}

\subsection{An exact nonproduct application}\label{an-exact-nonproduct-application}

The original-ray clock hypothesis of \hyperref[result:theorem-8.2]{Theorem~8.2} also admits triples beyond the product models. We give an exact common state whose completion yields three mutually unbiased measurements with no additional common-unbiased pure state. First we verify its two correlation equations. We then distinguish two claims: the constructed state vectors are entangled in the fixed tensor structure, and the whole triple is inequivalent to the classified product triples even under the allowed common global transformations.

Put \(\zeta=e^{\pi i/3}=-\omega^2\), and \[
t=(1+3i)/\sqrt{10},\quad \eta=(1,1,t),\quad
p=(1,\omega,\zeta),
\] \[
c=(\sqrt{2/3},\sqrt{2/3},1/\sqrt6),\quad
s=(1/\sqrt3,1/\sqrt3,\sqrt{5/6}),\quad
g_j=p_j(c_j+is_j,c_j-is_j)^T/\sqrt6.
\tag{9.1}
\label{eq:9.1}\] These are positive-angle witnesses for \hyperref[eq:7.2]{(7.2)}, since \(c_j^2+s_j^2=1\). Write \(a=\sqrt{2/3}\), \(b=1/\sqrt6=a/2\), \(u=1/\sqrt3\), \(v=\sqrt{5/6}\). The sequences \(pc=(a,a\omega,b\zeta)\) and \(\eta ps=(u,u\omega,v\zeta t)\) each have squared norm \(3/2\), and their exact correlations are \[
C(pc)=a^2\omega^2+ab\omega\bar\zeta+ab\zeta
=\zeta(-a^2+2ab)=0,
\] \[
C(\eta ps)=u^2\omega^2+uv\omega\overline{\zeta t}+uv\zeta t
=\zeta[-u^2+uv(t+\bar t)]=0.
\] The latter uses \(t+\bar t=2/\sqrt{10}\) and \(2uv/\sqrt{10}=1/3=u^2\). Equations \hyperref[eq:7.4]{(7.4)}--\hyperref[eq:7.5]{(7.5)}, with the negative anchor exponent, thus prove that \(g\) is unbiased to \(I,K_\eta\). \hyperref[result:theorem-6.1]{Theorem 6.1} therefore constructs the physical triple \(I,K_\eta,E_\eta(g)\). Its companion consists of the two \(D\)-orbits, so \hyperref[result:theorem-8.2]{Theorem 8.2} excludes every additional common unit ray.

All six constructed columns have Schmidt rank two in the fixed native \(\mathbb C^3\otimes\mathbb C^2\) order. The first-orbit qubit row directions at \(j=0,2\) differ, since \(s_0/c_0=1/\sqrt2\), \(s_2/c_2=\sqrt5\). Clock phases do not change rank. The second-orbit row directions \((c_j+is_j,-c_j+is_j)\) also differ at these indices.

Schmidt rank depends on the chosen tensor structure. To exclude global equivalence as well, compare transition-matrix phase ratios with the two product-triple models in \href{https://arxiv.org/abs/1203.6887v2}{McNulty--Weigert, Lemma 1}. In those two displayed normal forms, every pair-transition entry phase is a root of unity of order dividing 24. This follows from their local factors. The nonzero entries of the coordinate, Fourier and quadratic-phase qutrit bases have cube-root phases. Overlaps between the Fourier and quadratic-phase bases are three-term sums \(\omega^r(2+\omega)\) or \(\omega^r(2+\omega^2)\), divided by \(3\); their unit phases are the same sums divided by \(\sqrt3\), hence twelfth roots of unity. Qubit Pauli overlaps have eighth-root phases. Their products, including the indirect qutrit choices in the second model, therefore have twenty-fourth-root phases. For a two-by-two submatrix of a flat transition, its phase cross-ratio is the product of the diagonal entries divided by the product of the off-diagonal entries. Every \(2\times2\) phase cross-ratio of every transition in those models is consequently a root of unity.

Our anchor transition, using rows \((0,0),(2,0)\) and columns \((0,0),(0,1)\), has \[
\frac{K_{(0,0),(0,0)}K_{(2,0),(0,1)}}
{K_{(0,0),(0,1)}K_{(2,0),(0,0)}}=\bar t.
\tag{9.2}
\label{eq:9.2}\] This is not a root of unity: otherwise \(t+\bar t\) would be an algebraic integer, whose square \(2/5\) would be a rational algebraic integer and thus an integer, a contradiction. Common unitary transformations preserve transition matrices; row/column phases cancel from these ratios and permutations reorder them. Common antiunitaries conjugate the ratios; exchanging two anchors adjoints the transition and conjugates or inverts the corresponding ratios. Thus this triple is inequivalent to both product models even allowing common unitary/antiunitary transformations and basis reordering. The example therefore lies beyond both prior product models under the stated global equivalences. It establishes an application of the clock obstruction, without asserting a new triple family or a classification of all clock triples.

\subsection{Application to a Fourier-localized companion}\label{coherent-application-to-an-actual-localized-companion}

\hyperref[result:theorem-8.2]{The clock theorem} can also be applied to a triple written in different coordinates. In this application, each original column of the given companion has support at one frequency of a finite Fourier transform. A single common unitary carries that companion to the coordinate basis and carries the other two given bases to the displayed pair and a clock-invariant companion. Tracking this common change of coordinates preserves the original measurement rays and supplies the hypothesis of \hyperref[result:theorem-8.2]{Theorem 8.2}.

Use \(Q=F_3\), \(T=Q\otimes I_2\) with positive spectral-to-native formula \[
(Tz)_{n,a}=\frac1{\sqrt3}\sum_j\omega^{jn}z_{j,a}.
\] Choose \[
U_j=\begin{pmatrix}\alpha_j&\beta_j\\
\bar\beta_j&-\bar\alpha_j\end{pmatrix},\quad
|\alpha_j|^2+|\beta_j|^2=1,\qquad
S=T\operatorname{diag}_j(U_j)T^*.
\tag{9.3}
\label{eq:9.3}\] For arbitrary unit-modulus \(\nu_j\), define the given companion with column labels \((i,a)\) by \[
(T^*C_\nu)_{j;(i,a)}
=\delta_{j,-i}(1,(-1)^a\nu_j)^T/\sqrt2.
\tag{9.4}
\label{eq:9.4}\]

\phantomsection\label{result:corollary-9.1}\textbf{Corollary 9.1 (Fourier-localized companion).} If \(I,S,C_\nu\) are given pairwise MU bases, no unit vector is unbiased to all three.

\textbf{Proof.} We first identify the localized companion, then apply one common change of coordinates. Each \(U_j\) is unitary by its displayed norm condition. The two columns at each frequency in \hyperref[eq:9.4]{(9.4)} are orthonormal, so \(C_\nu\) is unitary; its native entries have modulus \(1/\sqrt6\). Put \[
B_j=\frac1{\sqrt2}\begin{pmatrix}1&1\\\nu_j&-\nu_j\end{pmatrix},
\quad B=\operatorname{diag}_jB_j,\quad
P e_{(i,a)}=e_{(-i,a)}.
\] Then \hyperref[eq:9.4]{(9.4)} is precisely \(C_\nu=TBP\). The factorization identifies the common unitary \(A=B^*T^*=P C_\nu^*\) that sends the original companion to coordinate rays. Apply it to all three bases and any proposed ray. Direct multiplication gives \[
AC_\nu=P,\qquad AI=B^*T^*,\qquad
AS=B^*\operatorname{diag}_j(U_j)T^*.
\tag{9.5}
\label{eq:9.5}\] Right multiplication of the first anchor by \(P\) restores \(I\). Since \(B_j^*=H_2\operatorname{diag}(1,\bar\nu_j)\) and \(Q^*=QP_{\rm rev}\), where \(P_{\rm rev}e_n=e_{-n}\) and \(P=P_{\rm rev}\otimes I_2\), \[
B^*T^*=K_\nu P.
\] The right column permutation restores the second anchor to \(K_\nu\), with label \(m=-n\); it changes the order of its original rays. For \(e=\nu_0\ne1\), normalize \(\eta_j=\nu_j/e\) by the second-binary-column phases in \hyperref[eq:7.1]{(7.1)}. When \(\nu_0=1\), this normalization is the identity and \(\eta=\nu\).

The transported third basis is \[
V=B^*\operatorname{diag}_j(U_j)(Q^*\otimes I_2),\qquad
V_{j;(n,a)}=\omega^{-jn}B_j^*U_je_a/\sqrt3.
\] Multiplication by \(D\) changes \(\omega^{-jn}\) to \(\omega^{-j(n-1)}\), sending the original column \((n,a)\) exactly to \((n-1,a)\). Thus the transported third basis has \(D\)-invariant original rays. All three MU conditions and any common-ray condition survive the same \(A\), so \hyperref[result:theorem-8.2]{Theorem 8.2} applies to the transported original triple \(I,K_\eta,V\). The inverse common operation is \(A^*=TB\), with the displayed right permutations and phases reversed. Hence an extra ray for the original triple would give one for the transported triple, a contradiction. Every transition is between the given bases. \(\square\)

This application requires the displayed \(C_\nu\) and all three pairwise-MU hypotheses. The common-unitary calculation preserves its original column labels and gives an explicit inverse. It asserts neither uniqueness of this companion nor automatic Fourier localization of arbitrary companions.

\setcounter{equation}{5}
\subsection{Forty-eight candidates with no third measurement basis}\label{sec:source-neighborhood}

Prescribed-state completion in \hyperref[prescribed-ray-completion-in-dimension-2m]{Section 6} depends on its displayed Fourier pair. A contrasting pair has many common-unbiased pure states but no way to choose six orthogonal ones. We prove this distinction near the reference Hadamard matrix \(H_{\star}\) displayed in \eqref{eq:source-neighborhood-center}. Although that reference matrix comes from Karlsson's construction, the varying matrices need not belong to the same family. With entry labels fixed, define the closed entrywise neighborhood
\begin{equation}
\mathcal U=\{H\in\mathbb C^{6\times6}: |H_{ij}|=1,\ HH^*=6I,
\ \max_{i,j}|H_{ij}-(H_{\star})_{ij}|\le10^{-6}\}.
\label{eq:source-neighborhood}
\end{equation}
Every \(H\in\mathcal U\) is an exact unit-entry Hadamard matrix and need not be dephased. Its measurement basis is \(H/\sqrt6\), so the entrywise radius in normalized-basis units is \(10^{-6}/\sqrt6\). The neighborhood is not a model of noisy measurements.

\leavevmode\phantomsection\label{result:theorem-9.3}\textbf{Theorem 9.3 (Uniform local nonextension).}
For every \(H\in\mathcal U\), the pair \((I,H/\sqrt6)\) has exactly 48 common-unbiased projective unit rays. Any three choices of those rays, including repetitions, have a pair of unit representatives with overlap modulus strictly greater than \(1/120\). For every matrix \(K\) satisfying
\begin{equation}
|K_{ij}|=1,\qquad |(H^*K)_{ij}|^2=6\quad(0\le i,j\le5),
\label{eq:source-neighborhood-companion}
\end{equation}
one has
\begin{equation}
\|K^*K-6I\|_F^2>\frac3{100},\qquad
\|(K/\sqrt6)^*(K/\sqrt6)-I\|_F^2>\frac1{1200}.
\label{eq:source-neighborhood-gap}
\end{equation}
Here \(K\) is not assumed orthogonal. Consequently the pair cannot extend to a third mutually unbiased orthonormal basis and cannot occur as an edge of a quartet. Two supplementary character sums, defined and evaluated in \hyperref[sec:source-marks]{Appendix D.1}, also stay nonzero on \(\mathcal U\).

\textbf{Proof.} The uniform root enclosures and complete interval exhaustion in \hyperref[app:source-neighborhood]{Appendix D} establish exactly 48 rays. On their labels, the interval overlap tests retain a graph \(K_{4,4}\sqcup20K_2\): one complete bipartite graph with four vertices on each side, together with twenty disjoint pairs. It has 48 vertices and 36 possible orthogonality edges. Every actual orthogonal pair must be an edge; a retained edge need not be orthogonal. The graph is triangle-free, and every nonedge has overlap modulus \(>1/120\). Three distinct labels therefore contain such a nonedge, while equal labels have overlap modulus one.

Each column of \(K/\sqrt6\) is one of these rays, up to its unrestricted global phase. Form a graph on the six column positions, joining two positions only when their labels are distinct and adjacent in the retained graph. Repeated labels introduce no triangle. A triangle-free graph on six vertices has at most nine edges: for every edge \(uv\), \(\deg(u)+\deg(v)\le6\). Summing over edges and applying Cauchy gives
\begin{equation}
4e^2\le6\sum_v\deg(v)^2\le36e,
\end{equation}
so \(e\le9\) when \(e>0\); the zero-edge case is immediate. At least six of the fifteen column pairs thus have overlap modulus \(>1/120\). Writing the unit columns as \(v_j\), including repetitions,
\begin{equation}
\|K^*K-6I\|_F^2
=36\cdot2\sum_{j<k}|\langle v_j,v_k\rangle|^2
>36\cdot2\cdot\frac6{120^2}=\frac3{100}.
\end{equation}
Division by 36 gives the normalized gap. A third orthonormal basis would have zero Gram defect, contradicting this strict bound. The supplementary character sums are treated separately in \hyperref[sec:source-marks]{Appendix D.1}. \(\square\)

The reference matrix uses Karlsson's construction \cite{K}. Fixed-matrix enumeration and certified nonorthogonality are established methods \cite{BW}; Jaming et al. prove locally stable Fourier-family ray counts and uniform quartet exclusions \cite{J}. Crooks reports uniform Karlsson-family no-third certificates using closely related interval, coverage and graph machinery \cite{Crooks}. The contribution here is the quantitative result on the closed neighborhood \(\mathcal U\), with its exact ray count and a bound for arbitrary six-column candidates.

The maximum-number preprint also treats varying matrices, but its stated graph target excludes two cross-compatible six-cliques \cite{OAImax}. Comparing its local computation with this theorem requires a matching output record and a checked correspondence covering the entire neighborhood \(\mathcal U\); that comparison remains unresolved. Neither OpenAI preprint is a premise of the independent proof in Appendix D. The theorem also has no middle-minor or original-ray clock hypothesis.

\subsection{Common states, completion, and extension}\label{what-candidate-constraints-and-compatibility-each-supply}

The results distinguish three stages of the measurement problem: finding common-unbiased states, assembling them into an orthonormal basis, and extending a set of mutually unbiased bases. For a fixed eligible \(C\), the finite physical ray set in \hyperref[boundary-isolation-and-finiteness]{Section 3} yields a finite graph recording orthogonality and unbiasedness. A compatible six-clique is a basis, and two compatible cliques with every required cross edge are mutually unbiased bases. Finiteness of the candidate set alone supplies neither clique. The full complex algebra additionally contains multiplicities and nonphysical points that this physical graph does not record.

The Tao/Potts illustration makes this distinction concrete: a complete reciprocal-orthogonal complex collection exists even though its columns are not physical phase vectors. \hyperref[result:theorem-5.1]{The DVR theorem} controls complete algebraic collections under specialization, using compatibility to rule out boundary escape. It supplies no general physical completion operation.

The displayed \(K_\eta\) family answers the prescribed-ray completion question through its two symmetries, with no middle-minor premise. In dimension six it has a vanishing positional three-minor, as shown in \hyperref[how-the-domains-differ]{Section 2.3}. \hyperref[result:theorem-3.1]{The boundary theorem} is therefore not invoked for \(K_\eta\); the all-order critical dictionary remains applicable after dephasing. \hyperref[result:theorem-8.2]{The six-dimensional clock theorem} then answers the extension question only for actual triples with the stated original-ray invariance. Its conditional canonical identification and product branch remain essential.

Symmetry of an anchor pair need not preserve a chosen companion's rays.
Appendix~\ref{app:nonsetwise-witness} gives an actual MU triple
\((I,U,C)\) with \(R=\operatorname{diag}(P_3,P_3)\), using
\(P_3e_0=e_1\), \(P_3e_1=e_2\), \(P_3e_2=e_0\), and \(RU=UR\).
Thus \(R\) permutes the rays of both anchors. Its representation
\(A=C^*RC\) has thirty nonzero entries and is not monomial, so \(R\)
does not preserve the original \(C\) ray set. This illustrates why the
given-companion ray-invariance hypothesis in
\hyperref[result:theorem-8.2]{Theorem~8.2} must be supplied for that
theorem. The example supplies no simultaneous anchored correspondence
to the specific clock data of that result.

A further compatibility estimate concerns prescribed entrywise phase
powers of the three ordered transitions of an actual MU triangle.
The cubic and quintic transforms cannot both retain the prescribed closing
edge. Appendix~\ref{app:prescribed-phase-powers} gives the complete
quantitative proof of an entrywise defect and a squared Frobenius bound.
This estimate concerns transformed transitions and leaves the original MU
triangle intact.

These hypotheses support different conclusions. The C13 example illustrates the middle-minor hypothesis; the nonproduct and localized examples illustrate the clock hypothesis. \hyperref[result:theorem-9.3]{Theorem 9.3} supplies a local exact count and Gram gap on the closed entrywise neighborhood \(\mathcal U\), under separate physical hypotheses rather than the middle-minor or \hyperref[result:theorem-5.1]{DVR theorem}. \hyperref[appendix-c13-exact-middle-minor-applicability]{Appendix A} gives the C13 construction and rational minor certificate, and \hyperref[appendix-dvr-complete-companions-cannot-escape]{Appendix B} gives the DVR proof. The remaining appendices give the phase-power estimate, neighborhood computation and symmetry example.

Together these results do not classify all order-six Hadamard matrices or mutually unbiased triples, or establish a global maximum for the number of bases. General affine reducedness, a uniform common-ray count for all matrices, arbitrary-companion localization and unconditional canonical companion classification remain outside the general reductions.

Strong unextendibility here excludes an additional pure state with uniform outcome probabilities for all three measurements. It does not exclude mixed states with that property: the maximally mixed density matrix gives uniform probabilities for every orthonormal measurement.

\section{Computational material and reproducibility}\label{sec:computational-material}

The source archive accompanying this article contains the complete \texttt{main.tex} and the public supplementary directory \texttt{anc/}. After extracting the archive, keep that directory structure. The single guide \path{anc/README.md} maps each result to its exact inputs, checker, expected output and complete guarded command. The checks below distinguish exact finite calculations, stored-data consistency and full interval exhaustion.

\paragraph{Middle-minor applicability.}
\path{anc/verify_c13.py} reconstructs the fixed phase data, propagates directed rational enclosures, and checks all 400 positional three-minors and the normalized comparator distance of \hyperref[appendix-c133-the-inequalities-and-normalization]{Appendix A.3}. Its input rectangles are in \path{anc/inputs/complete_set_advisory/neighborhood_checks.json}; \path{anc/certificate.json} contains the exact output enclosures. A replay records fresh rational enclosures and strict bounds. Agreement with the advisory rectangles is a consistency check, whereas the directed enclosure calculation and complete index coverage establish the inequalities. \path{anc/CONTRACT.md} is the fixed-input and reproduction specification.

\paragraph{The uniform Hadamard neighborhood.}
For \hyperref[result:theorem-9.3]{Theorem~9.3}, the public guide gives the inputs and commands under \path{anc/source_neighborhood/}. The input checker reconstructs \(H_{\star}\) from its radical formula, validates every consumed coefficient rectangle, and verifies the regenerated fixed-input header and arithmetic controls. Stored-record inspection checks all 96 small/large root records, all 1128 unordered pair records, the possible-edge graph and agreement between the input verification and full-run records; it does not replay discarded coverage boxes. The interval program's roots-only calculation establishes local existence, uniqueness and pair bounds, but does not exclude other roots. Complete coverage requires its successful full interval exhaustion, started from the entire closed five-angle cube without a checkpoint in an empty output directory, together with the sound predicates and proof in \hyperref[app:source-neighborhood]{Appendix D}. Hashes, counters and stored-record consistency do not replace that exhaustion.

\paragraph{The exact symmetry example.}
\path{anc/nonsetwise/checker.py}, invoked with a fresh \texttt{--output} path, reconstructs the radical matrices and checks every Gram/covariance identity, support position, pair profile and cross ratio in \hyperref[app:nonsetwise-witness]{Appendix E}. Its exact expected data are \path{anc/nonsetwise/witness.json}. The output distinguishes failure of the original companion's ray invariance from its rebased representation; it establishes no simultaneous correspondence to the clock hypotheses of \hyperref[result:theorem-8.2]{Theorem~8.2}.

\paragraph{Displayed bounds and analytic support.}
\path{anc/check_displayed_bounds.py}, invoked with a fresh \texttt{--output} path, reads the saved exact certificate/root/pair fields and writes the endpoint map \path{anc/numerical_exposition.json}. Integer and rational comparisons check the displayed powers of two and strict thresholds, including division of the phase-vector overlap by six to obtain the unit-ray overlap. This short check neither recomputes the source enclosures nor exhausts the phase cube. The phase-power estimate in \hyperref[app:prescribed-phase-powers]{Appendix C} is supported by its complete analytic proof.

The Python checkers use the standard library. The interval exhaustion also requires a C++17 compiler and Boost headers, with the conservative flags specified in the guide. The unchanged launcher \path{anc/run_cas_guarded.py} supports macOS and Linux; Windows is unsupported. It requires inspection of the full process tree, a shared singleton lock and permission to set the documented process-priority level. The guide gives the precise platform requirements, and the tested recipes impose aggregate memory and wall-time ceilings. Those execution safeguards are not extra hypotheses about quantum measurements. A failed, interrupted or incomplete replay does not establish complete coverage.

\section*{Acknowledgment}\label{ai-assistance}
\addcontentsline{toc}{section}{Acknowledgment}
We acknowledge the use of GPT-5.6, GPT-6, GPT-6.1 by OpenAI and Opus 5.5 by Anthropic to assist with analytical derivations and to suggest edits to the manuscript. All AI-assisted derivations were
independently checked by the author, and all results and arguments were written up and verified by the author.

\appendix
\setcounter{equation}{0}
\section{An example with no vanishing 3 by 3 minors}\label{appendix-c13-exact-middle-minor-applicability}

This appendix gives a pair of mutually unbiased measurements to which \hyperref[result:theorem-3.1]{Theorem 3.1} applies. We construct a fixed matrix \(H\) from the blocks of \href{https://arxiv.org/pdf/1003.4177v1}{Karlsson, Theorem 11}; its example identifier is C13. The algebra below proves that \(H\) is well-defined, has unit-modulus entries and satisfies \(HH^*=6I_6\). Thus \(C_{13}=H/\sqrt6\) is flat and unitary. A rational enclosure certificate separately proves that every positional three-minor is nonzero.

The example also separates proximity from equivalence: \(C_{13}\) lies within \(7/100000\) in Frobenius norm of the normalized Fourier-family comparator \(H_0/\sqrt6\) constructed in \hyperref[appendix-c133-the-inequalities-and-normalization]{Appendix A.3}, but is not equivalent under unitary monomial transformations to any member of that displayed family or its transpose.

Set

\begin{equation}
\rho=\frac1{100000},\quad c=\frac{1-\rho^2}{1+\rho^2},\quad
s=\frac{2\rho}{1+\rho^2},\quad e=\frac{3+4i}{5},\quad z_1=i,
\end{equation}

\begin{equation}
a=-\frac12+\frac{i\sqrt3}{2}(c+\overline e s),\qquad
b=-\frac12+\frac{i\sqrt3}{2}(-c+es),
\end{equation}

\begin{equation}
F_2=\begin{pmatrix}1&1\\1&-1\end{pmatrix},\qquad
A=\begin{pmatrix}a&b\\\overline b&-\overline a\end{pmatrix},\qquad B=-F_2-A.
\end{equation}

The Fourier blocks in this appendix use unit-modulus normalization.

For a first row \((p,q)\) define

\begin{equation}
M_{p,q}(\zeta)=\frac{q^2\zeta-p^2}{(\overline p)^2\zeta-(\overline q)^2},\qquad
M_A=M_{a,b},\quad M_B=M_{-1-a,-1-b}.
\end{equation}

Define \(\zeta_3=M_A(-1)\), \(\zeta_4=M_B(-1)\), \(\zeta_2=M_A^{-1}(\zeta_4)\). For \(j=2,3,4\), choose \(z_j=\sqrt{\zeta_j}\) with positive real part. Put

\begin{equation}
Z_j=\begin{pmatrix}1&1\\z_j&-z_j\end{pmatrix}\ (j=1,2),\qquad
Z_j=\begin{pmatrix}1&z_j\\1&-z_j\end{pmatrix}\ (j=3,4),
\end{equation}

\begin{equation}
H=\begin{pmatrix}
F_2&Z_1&Z_2\\
Z_3&Z_3AZ_1/2&Z_3BZ_2/2\\
Z_4&Z_4BZ_1/2&Z_4AZ_2/2
\end{pmatrix},\qquad C_{13}=H/\sqrt6. 
\label{eq:C13.1}\end{equation}

\subsection{The phase identities and their denominators}\label{appendix-c131-the-phase-identities-and-their-denominators}

The Möbius formulas and square-root choices must be defined at the chosen parameters before they can be used in the blocks. We check their denominators, the compatibility relation between the four blocks, and the square-root convention in that order.

Write

\begin{equation}
r=\frac{2\sqrt3s}{5},\quad
\alpha=\frac{\sqrt3}{2}(c+3s/5),\quad
\beta=\frac{\sqrt3}{2}(-c+3s/5).
\end{equation}

Then \(a=-1/2+r+i\alpha\), \(b=-1/2-r+i\beta\). The identities \(c^2+s^2=1\) and \(|e|=1\) give

\begin{equation}
2r^2+\alpha^2+\beta^2=3/2,
\qquad |a|^2+|b|^2=|-1-a|^2+|-1-b|^2=2. 
\label{eq:C13.2}\end{equation}

Moreover

\begin{equation}
|a|^2-|b|^2=\frac{s}{5}(9c-4\sqrt3),\qquad
|-1-a|^2-|-1-b|^2=\frac{s}{5}(9c+4\sqrt3).
\end{equation}

They are strictly positive: \(s>0\), \(c>9/10\), and \(4\sqrt3<7\). Thus both Möbius maps are invertible. On the unit circle their denominators cannot vanish, since vanishing would equate the two coefficient moduli. Their numerators cannot vanish there either. If \(N=q^2\zeta-p^2\) and \(|\zeta|=1\), the denominator is \(-\zeta\overline N\); hence the maps take the unit circle to itself. The inverse of \(M_A\) is also defined on that circle. In particular all three \(\zeta_j\) are well-defined unit numbers.

The fourth compatibility relation follows algebraically, without an enclosure calculation. Let \(p=-1-a\), \(q=-1-b\), and represent the two Möbius maps by

\begin{equation}
T_A=\begin{pmatrix}b^2&-a^2\\(\overline a)^2&-(\overline b)^2\end{pmatrix},\qquad
T_B=\begin{pmatrix}q^2&-p^2\\(\overline p)^2&-(\overline q)^2\end{pmatrix}.
\end{equation}

Here \(\operatorname{adj}\) denotes the classical adjugate. The trace of \(\operatorname{adj}(T_A)T_B\) is

\begin{equation}
2\operatorname{Re}(a^2(\overline p)^2-b^2(\overline q)^2)=0. 
\label{eq:C13.3}\end{equation}

Indeed, put \(n_a=|a|^2\), \(n_b=|b|^2\), \(x=\operatorname{Re}a\), \(y=\operatorname{Re}b\). By \eqref{eq:C13.2}, \(n_a+n_b=2\), while \(x+y=-1\). Thus \((n_a+x)+(n_b+y)=1\) and

\begin{equation}
(n_a+x)^2-(n_b+y)^2=2(n_a+x)-1
=(\operatorname{Im}a)^2-(\operatorname{Im}b)^2.
\end{equation}

Expanding \(a\overline p=-n_a-a\) and \(b\overline q=-n_b-b\) proves \eqref{eq:C13.3}. By the two-by-two Cayley--Hamilton identity, the invertible trace-zero matrix \(T_A^{-1}T_B\) has scalar square. Its projective transformation is an involution. Consequently \(M_B\circ M_A^{-1}\circ M_B=M_A\), giving

\begin{equation}
M_A(-1)=\zeta_3,\quad M_B(-1)=\zeta_4,\quad
M_A(\zeta_2)=\zeta_4,\quad M_B(\zeta_2)=\zeta_3. 
\label{eq:C13.4}\end{equation}

The square-root convention is also legal algebraically. The numbers \(a^2+b^2\) and \(p^2+q^2\) have common real part \(-1+4r^2\ne0\), so their ratios to their conjugates, \(\zeta_3\) and \(\zeta_4\), are not \(-1\). Their imaginary parts differ by \(-2(\alpha+\beta)\ne0\); with the same nonzero real part, their conjugate ratios are distinct. Hence \(\zeta_2\ne-1\), since otherwise \(M_A(\zeta_2)=\zeta_3=\zeta_4\). For any unit \(\zeta\ne-1\), the chosen root is

\begin{equation}
\sqrt\zeta=R+i\frac{\operatorname{Im}\zeta}{2R},\qquad
R=\sqrt{(1+\operatorname{Re}\zeta)/2}>0.
\end{equation}

It has square \(\zeta\) and modulus one. Every phase in \eqref{eq:C13.1} is therefore exact and well-defined.

\subsection{Flatness and unitarity}\label{appendix-c132-flatness-and-unitarity}

We first prove the scaled orthogonality identity \(HH^*=6I_6\), which makes \(C_{13}\) unitary. We then check the entry moduli separately, using the Möbius compatibility relations.

The forms of \(A,B\) and \eqref{eq:C13.2} give \(AA^*=BB^*=2I_2\). Since \(A+B=-F_2\), expansion gives \(AB^*+BA^*=-2I_2\). Every \(Z_j\) satisfies \(Z_jZ_j^*=Z_j^*Z_j=2I_2\).

Block multiplication now gives \(HH^*=6I_6\). The upper/middle cross block is \((F_2+A^*+B^*)Z_3^*=0\), and the upper/lower block is the same with \(Z_4^*\). The middle/lower cross block is \(Z_3[I_2+(AB^*+BA^*)/2]Z_4^*=0\). A lower diagonal block is \(Z_jZ_j^*+Z_j(AA^*+BB^*)Z_j^*/2=6I_2\), and the upper diagonal block is \(6I_2\).

The scaled orthogonality identity alone does not prove flatness. For unit-modulus \(z,w\), define the two block orientations by
\[
Z_{\rm row}(z)=\begin{pmatrix}1&1\\z&-z\end{pmatrix},\qquad
Z_{\rm col}(w)=\begin{pmatrix}1&w\\1&-w\end{pmatrix}.
\]
Consider \(U=Z_{\rm col}(w)AZ_{\rm row}(z)/2\). Its first entry is \((P+wQ)/2\), where

\begin{equation}
P=a+bz,\qquad Q=\overline b-\overline a z,
\qquad |P|^2+|Q|^2=4.
\end{equation}

Direct multiplication gives

\begin{equation}
P\overline Q=\overline z(b^2z^2-a^2),\qquad
\overline P Q=\overline z((\overline b)^2-(\overline a)^2z^2).
\end{equation}

If \(w^2=M_A(z^2)\), these imply \(P\overline Q+w^2\overline P Q=0\), and hence \(\operatorname{Re}(P\overline Q\overline w)=0\). The first entry has modulus one. The block satisfies \(UU^*=2I_2\), so \(U/\sqrt2\) is unitary and each row and column of \(U\) has squared norm two. These norms give modulus one for its other three entries. The same argument applies to \(B\). Relations \eqref{eq:C13.4}, with \(z_1^2=-1\), give precisely the required relation for each of the four inner blocks. The outer blocks are visibly flat. Thus \(H\) is an exact complex Hadamard matrix and \(C_{13}\) is flat and unitary.

\subsection{The inequalities and normalization}\label{appendix-c133-the-inequalities-and-normalization}

The remaining hypothesis consists of four hundred strict determinant inequalities. The verifier propagates rational enclosures containing the entries and determinants. It first evaluates the phase data and Möbius expressions in \(\mathbb Q(\sqrt3,i)\), representing each value by four rational coefficients on \(1,\sqrt3,i,i\sqrt3\). It then encloses the values in rational center-radius balls at the specified 128-bit precision and constructs the three complex square roots with positive real part used for \(z_2,z_3,z_4\).

For real balls \(m+[-r,r]\) and \(n+[-s,s]\), multiplication uses center \(mn\) and radius \(|m|s+|n|r+rs\). Outward rounding enlarges the enclosure: any error from rounding the center is compensated in the radius. Reciprocals use the ordered rational endpoint reciprocals after excluding zero. Square roots use integer square roots and exact squared-endpoint comparisons. Applying these rules to the real and imaginary parts encloses the exact complex operations. No binary floating-point number is used for a mathematical assertion.

There are twenty row triples and twenty column triples. Each of the resulting four hundred submatrices is evaluated by validated Gaussian elimination with certified nonzero pivots and recorded row swaps. Exact elimination allows the eliminated coordinate to be set to zero; this is not a tolerance truncation. If a nonzero pivot cannot be certified, the checker fails rather than declaring rank deficiency. The determinant is the signed product of the pivots. Complete index coverage is asserted and exported.

Controls before the example check exact field conjugates and inverses, signed interval products, directed square roots, a determinant requiring a row swap and an exactly singular determinant. All four hundred original advisory enclosure rectangles are compared only after the verifier's strict bounds pass; their overlap is a consistency check, not a premise of those bounds.

Let \(\beta_{\rm det}\) be the minimum of the 400 certified squared-minor lower endpoints in the exact supplementary certificate. It is a rational number with denominator \(2^{128}\), reproduced from the recorded enclosures by the displayed-bound check in \hyperref[sec:computational-material]{Section~10}. The certificate covers all twenty row triples times all twenty column triples, and proves for every such \(I,J\) that

\begin{equation}
|\det H_{I,J}|^2\ge
\beta_{\rm det}>10^{-8}. 
\label{eq:C13.5}\end{equation}

Thus \(|\det H_{I,J}|>10^{-4}\). For the normalized matrix the corresponding bound is \(|\det (C_{13})_{I,J}|^2>10^{-8}/216\), because a three-minor is multiplied by \(6^{-3/2}\). Equation \eqref{eq:C13.5} concerns \(H\), not \(C_{13}\).

An explicit Fourier comparator is obtained from \eqref{eq:C13.1} by using \(\omega=e^{2\pi i/3}\),

\begin{equation}
A_0=\begin{pmatrix}\omega&\omega^2\\\omega&-\omega^2\end{pmatrix},\quad
B_0=\begin{pmatrix}\omega^2&\omega\\\omega^2&-\omega\end{pmatrix},\quad
z_1=i,\ z_2=e,\ z_3=z_4=1.
\end{equation}

Call the resulting phase matrix \(H_0\). In row order \((0,2,4,1,5,3)\) and column order \((0,2,4,1,3,5)\) it equals

\begin{equation}
\begin{pmatrix}F_3&F_3\\F_3D&-F_3D\end{pmatrix},\qquad
D=\operatorname{diag}(1,i,e),\quad (F_3)_{ij}=\omega^{ij}.
\end{equation}

This direct formula avoids a singular Möbius limit.

For the single comparator, the same enclosure rules give

\begin{equation}
\|C_{13}-H_0/\sqrt6\|_F^2\le
\beta_{\rm dist}<49\cdot10^{-10}.
\end{equation}

Here \(\beta_{\rm dist}\) is the exact upper endpoint in the supplementary certificate, with denominator \(2^{126}\). It is computed as the sum of the squared phase-entry differences divided by \(6\), so the normalization is exact. All inequalities use rational endpoints. Taking nonnegative square roots gives the claimed distance bound.

The certificate also proves \(\|C_{13}-H_0/\sqrt6\|_F<7/100000\). Every member of this Fourier family has a zero positional three-minor, since its two corresponding columns agree on three upper coordinates. The transposed family has the corresponding row property. Row and column permutations and unit-modulus diagonal scalings preserve whether a positional minor is zero. Equation \eqref{eq:C13.5} therefore places this example outside both Fourier classes under unitary monomial equivalence, while the displayed comparator shows its proximity to one of them. Neither this family exclusion nor proximity is needed for \hyperref[result:theorem-3.1]{Theorem 3.1}.

The all-minor inequalities and comparator distance follow from the directed rational enclosure method above. Their exact endpoints, input data and replay are identified in \hyperref[sec:computational-material]{Section~10}. The algebraic construction and flatness proof are separate from these certificate-dependent inequalities. The comparator and proximity are additional properties of the example, not hypotheses of \hyperref[result:theorem-3.1]{Theorem~3.1}.

\section{Complete companions cannot escape}\label{appendix-dvr-complete-companions-cannot-escape}

This appendix proves \hyperref[result:theorem-5.1]{Theorem 5.1}: under the type-II and middle-minor-unit hypotheses, the coordinate algebras of complete six-column companions are finite as modules over the coefficient discrete valuation ring (DVR). An individual point of a finite projective family can specialize to the boundary \(t=0\) of its affine chart; this is what escape means here. For a complete reciprocal-orthogonal collection, the inverse identity of its six-column matrix prevents such specialization. We classify the boundary and prove independence of its compatible rank-one matrices. A boundary specialization would give a forbidden dependence among those matrices. The resulting open-and-closed argument proves module finiteness with the entire scheme structure. This argument concerns algebraic specialization; its physical interpretation over the complex numbers appears at the end.

Throughout this appendix \ensuremath{C} is the DVR matrix and \ensuremath{T=C^{-1}}. The labels type A and type B denote the two boundary support orientations. The local three-dimensional span is \(\mathcal L\). In the physical sections, \ensuremath{H} denotes the phase matrix \ensuremath{H=\sqrt n\,C}; no such normalization is imposed on the general DVR matrix here. An overbar in the valuation argument denotes residue-field reduction. In the final complex subsection it denotes complex conjugation.

\subsection{Algebraic setup}\label{appendix-dvr1-algebraic-setup}

Let \ensuremath{R} be a DVR, with uniformizer \ensuremath{\pi}, fraction field \ensuremath{K}, and residue field \ensuremath{k}, and suppose 6 is a unit. Let \ensuremath{C} be an invertible 6 by 6 matrix over \ensuremath{R}, let \(T=C^{-1}\), and suppose

\begin{equation}
C_{ij}T_{ji}=\frac16\qquad(0\le i,j\le5),
\end{equation}

and every positional 3 by 3 minor of \ensuremath{C} is a unit. The generic fibre is the family over the fraction field \(K\), and the special fibre is the family over the residue field \(k\); their geometric fibres are obtained by extending these fields to algebraic closures. The three-minor unit condition permits the support and tangent arguments in both fibres. No condition on 2-minors is added. No assumptions beyond the displayed ring and matrix hypotheses are made; in particular no conjugation, positivity, first moments, reducedness, or companion existence is assumed.

Indices are 0 through 5. In the Laurent algebra with \(x_0=1\) put

\begin{equation}
\begin{aligned}F_j(x)&=\left(\sum_iT_{ji}x_i\right)\left(\sum_iC_{ij}x_i^{-1}\right)-1,\\ A_R&=R[x_1^{\pm1},\ldots,x_5^{\pm1}]/(F_1,\ldots,F_5).\end{aligned}
\end{equation}

The identity \(\sum_jF_j=0\) holds in the entire Laurent algebra: the sum of the products is \(\sum_{i,l}(CT)_{li}x_i/x_l=6\). Thus \ensuremath{F_{0}} belongs to the displayed ideal, scheme-theoretically.

For six independent copies, indexed \(r=0,\ldots,5\), set \(x_{0r}=1\) and

\begin{equation}
q_{rs}=\sum_ix_{ir}^{-1}x_{is}.
\end{equation}

The exact full ideal of the ordered companion algebra is

\begin{equation}
P_R=A_R^{\otimes_R6}/(q_{rs}:0\le r,s\le5,\ r\ne s).
\end{equation}

There are all 30 ordered off-diagonal generators. Neither their radical nor a selected solution set is substituted.

For another complete copy \ensuremath{y_{:s}}, define

\begin{equation}
\begin{aligned}q^{xy}_{rs}&=\sum_ix_{ir}^{-1}y_{is},\\ q^{yx}_{sr}&=\sum_iy_{is}^{-1}x_{ir},\\ Q_R&=(P_R\otimes_RP_R)/(q^{xy}_{rs}q^{yx}_{sr}-6:0\le r,s\le5).\end{aligned}
\end{equation}

All 36 displayed mutual generators are imposed. Our aim is finiteness of \ensuremath{P_{R}} and \ensuremath{Q_{R}} as R-modules, including zero and nonreduced algebras. Finite tensor products and quotients will then give finiteness for every quotient by additional explicitly supplied equations; Nakayama will give the geometric-special-fiber emptiness implication.

The auxiliary projective family \ensuremath{Z_{R}} is the closed subscheme of \ensuremath{\mathbb P^5_{z}\times_{R}\mathbb P^5_{w}} defined by the ten equations

\begin{equation}
\begin{aligned}z_iw_i&=t\qquad(i=1,\ldots,5),\\ v_ju_j&=t\qquad(j=1,\ldots,5),\\ v&=Tz,\qquad u=C^{\mathsf T}w,\qquad t=z_0w_0.\end{aligned}
\end{equation}

The missing \(v_0u_0=t\) follows identically: \(\sum_jv_ju_j=\sum_iz_iw_i=6t\). Hence the ten-equation scheme is exactly the scheme on which all twelve products \ensuremath{z_{i}\,w_{i}} and \ensuremath{v_{j}\,u_{j}} equal t. Support quadruples below mean \ensuremath{(|\operatorname{supp}z|,|\operatorname{supp}w|,|\operatorname{supp}v|,|\operatorname{supp}u|)}.

\subsection{Consequences over algebraically closed fields}\label{appendix-dvr2-preliminary-consequences-over-geometric-fields}

We first work over any algebraically closed field \ensuremath{F} obtained from an R-fiber, or more generally any algebraically closed field of characteristic different from 2 and 3 with the same matrix hypotheses. Every \ensuremath{C} entry and \ensuremath{T} entry is nonzero, since their product is 1/6. Jacobi's complementary-minor identity gives every 3-minor of \ensuremath{T} as a signed complementary 3-minor of \ensuremath{C} divided by \ensuremath{\det C}; hence all are nonzero. Over \ensuremath{R} those minors are units for the same reason.

A useful support bound is this: if a nonzero vector \ensuremath{h} has at most three nonzero coordinates, then \ensuremath{Th} has at least four nonzero coordinates. Otherwise at least three rows of \ensuremath{Th} vanish. Select those rows, extend \ensuremath{\operatorname{supp}h} to three column indices, and apply the invertible selected 3 by 3 minor of \ensuremath{T} to \ensuremath{h} with zeros in the added positions. This would force \ensuremath{h=0}. The analogous statement holds for \ensuremath{C}, \ensuremath{C^T}, and \ensuremath{T^T}. In particular a vector and its image under any of these maps cannot both have support at most three. A vector with support one has an image with support six because all entries are nonzero.

No assertion that every 2-minor is nonzero has entered this argument.

\subsection{Boundary classification, uniqueness, and reducedness}\label{appendix-dvr3-boundary-classification-uniqueness-and-reducedness}

The boundary products vanish in both coordinate systems. Their disjoint supports, together with the preceding support bound, determine all possible support sizes. At a point of \ensuremath{Z_{F}} with \ensuremath{t=0}, write

\begin{equation}
\begin{gathered}A=\operatorname{supp}z,\quad B=\operatorname{supp}w,\quad D=\operatorname{supp}v,\quad E=\operatorname{supp}u,\\ a=|A|,\quad b=|B|,\quad c=|D|,\quad d=|E|.\end{gathered}
\end{equation}

All four supports are nonempty. The equations give \(A\cap B=\varnothing\) and \(D\cap E=\varnothing\). None has size one: for example \ensuremath{a=1} would imply \ensuremath{c=6}, forcing \ensuremath{u=0}; the other cases follow using the inverse/transposed maps. Thus every size is between two and four.

If \ensuremath{a=3}, then \ensuremath{b\le 3}, so the support bound gives \ensuremath{d\ge 4}, hence \ensuremath{c\le 2}, which forces \ensuremath{a\ge 4}, a contradiction. If \ensuremath{a=2}, the same bounds force \ensuremath{c=4,d=2,b=4}. If \ensuremath{a=4}, then \ensuremath{b=2,d=4,c=2}. The exhaustive types are therefore

\begin{equation}
\begin{aligned}\text{type A}:&(2,4,4,2),\qquad B=A^c,\quad D=E^c;\\ \text{type B}:&(4,2,2,4),\qquad A=B^c,\quad E=D^c.\end{aligned}
\end{equation}

Here the words type A/type B are just labels for the two displayed cases.

For a type A point, the 2 by 2 block \ensuremath{T[E,A]} is singular, since it kills \ensuremath{z_{A}}, and has rank one because all its entries are nonzero. Its right kernel determines \ensuremath{z_{A}} up to scale; its left kernel determines \ensuremath{u_{E}} up to scale, since \(u_E^{\mathsf T}T[E,A]=w_A^{\mathsf T}=0\). Then \(v=Tz\) and \(w=T^{\mathsf T}u\) determine the point. Thus the support pair (A,E) determines at most one type A point. For type B, the block \ensuremath{C[B,D]} has rank one and its right and left kernels determine \ensuremath{v_{D}} and \ensuremath{w_{B}}; the pair (B,D) determines at most one such point. There are finitely many support pairs, so the boundary has finite support.

Projection of the set of boundary points to either \ensuremath{\mathbb P^5_{z}} or \ensuremath{\mathbb P^5_{w}} is injective. Indeed \ensuremath{z} determines its supports \ensuremath{A,D}, and \ensuremath{w} is in the relevant kernel of \ensuremath{C[B,D]^{\mathsf T}}. In the type A case that 4 by 4 matrix has rank exactly three: a nonzero kernel exists and every 3-minor is nonzero. In the type B case it is a nonzero singular 2 by 2 matrix, of rank one. Thus \ensuremath{w} is unique up to scale. The assertion for a fixed \ensuremath{w} follows by the same argument with the inverse/transposed variables. It also follows that distinct points with the same small \ensuremath{z} support have distinct \ensuremath{z} lines, and analogously for all four projections.

To prove scheme-theoretic reducedness, take a tangent vector to the ten-equation scheme at a boundary point, using projective tangent vectors represented by \ensuremath{dz,dw} and writing \(dv=T\,dz\), \(du=C^{\mathsf T}dw\). The derivative \ensuremath{dt} is a well-defined scalar after choosing affine normalizations. For every i and j, including zero, the differentiated equations are

\begin{equation}
\begin{aligned}w_i\,dz_i+z_i\,dw_i&=dt,\\ u_j\,dv_j+v_j\,du_j&=dt.\end{aligned}
\end{equation}

In type A, multiplication and summation on the nonzero supports give

\begin{equation}
\begin{aligned}w^{\mathsf T}dz&=b\,dt=4\,dt,\\ u^{\mathsf T}dv&=d\,dt=2\,dt.\end{aligned}
\end{equation}

But \(w^{\mathsf T}dz=u^{\mathsf T}T\,dz=u^{\mathsf T}dv\), so \ensuremath{2dt=0} and \ensuremath{dt=0}. The same calculation with \ensuremath{b=2,d=4} handles type B. Consequently \(dz_B=0\), \(dw_A=0\), \(dv_E=0\), and \(du_D=0\). The kernels just identified imply \ensuremath{dz} is a scalar multiple of \ensuremath{z} and \ensuremath{dw} a scalar multiple of \ensuremath{w}. These represent zero tangent vectors in the projective factors. Thus the projective tangent space is zero at every boundary point.

In an affine chart the ambient space has ten coordinates and the scheme has the displayed ten equations. Their Jacobian therefore has rank ten at a boundary point. Equivalently, its local maximal ideal has \(\mathfrak m/\mathfrak m^2=0\). Nakayama applied to that finitely generated ideal gives \(\mathfrak m=0\). The local ring is the residue field, so it has no nonzero nilpotent elements. Thus the point is reduced and isolated in the full projective scheme, not merely in the \ensuremath{t=0} subscheme.

\subsection{Finiteness of the projective family}\label{appendix-dvr4-finite-projective-geometric-fibers-and-the-projective-dvr-family}

Under the Segre embedding, \ensuremath{t=z_{0}w_{0}} is a hyperplane section. Any positive-dimensional projective irreducible component meets that hyperplane: projective intersection dimension gives a nonempty intersection with a hyperplane unless the component is contained in it, in which case intersection is automatic. Such an intersection would put a positive-dimensional component through one of the boundary points. \hyperref[appendix-dvr3-boundary-classification-uniqueness-and-reducedness]{Appendix B.3} shows each boundary point has a field as local ring, so this is impossible. Hence \ensuremath{Z_{F}} has dimension zero and finite support. Interior points may still be nonreduced.

The structure morphism \(Z_R\longrightarrow\operatorname{Spec}R\) is projective, hence proper, and of finite type over the noetherian ring \ensuremath{R}. Both its geometric fibres have finite support, so the morphism is quasi-finite. The standard proper-quasi-finite theorem proves that this morphism is finite. This conclusion concerns the full ten-equation family, including multiplicities.

The open locus \ensuremath{t\ne 0} is canonically \ensuremath{\operatorname{Spec}A_{R}}: all \ensuremath{z_{i},w_{i}} are nonzero there, choose \(z_0=w_0=1\), then \(w_i=z_i^{-1}\), and the five \(v_ju_j=1\) equations become precisely \ensuremath{F_{j}=0}. This is an isomorphism of schemes with the stated Laurent quotient. An open subscheme of a finite R-scheme need not be finite over \ensuremath{R}, so this does not by itself assert finiteness of \ensuremath{A_{R}}.

\subsection{A rank-one block obstruction}\label{appendix-dvr5-a-rank-one-block-obstruction}

The tensor-independence argument must exclude a configuration with a strip of rank-one blocks. \hyperref[result:block-lemma]{The next lemma} does so. Unlike the support argument, it uses the constant entry products: they give row-ratio identities in both directions and constrain the blocks simultaneously.

\phantomsection\label{result:block-lemma}\textbf{Block lemma.} There cannot be two disjoint pairs of rows and a partition of the six columns into three pairs such that all six corresponding 2 by 2 blocks of \ensuremath{C} have rank one. The transposed configuration is also impossible.

Proof. In this algebraic argument, row and column permutations and diagonal rescalings by nonzero elements of \(F\) preserve the entry-product condition for \ensuremath{C} and its inverse, and preserve nonvanishing of all 3-minors. Normalize the first of the four selected rows to all ones. On column pair j, write the next selected row as \ensuremath{(r_{j},r_{j})}. Invertibility gives, for distinct rows \ensuremath{h,l},

\begin{equation}
\sum_m\frac{C_{hm}}{C_{\ell m}}=0\qquad(h\ne\ell),
\end{equation}

because \(T_{m\ell}=1/(6C_{\ell m})\). Applied in both directions to the first two selected rows, this says \(\sum_jr_j=0\) and \(\sum_jr_j^{-1}=0\). The three nonzero numbers \ensuremath{r_{j}} are consequently the roots of \(X^3-r_0r_1r_2\). Since characteristic is not three, they are distinct. Scale and order so \(r_j=1,\omega,\omega^2\), with \ensuremath{\omega} a primitive cube root of unity.

Write the other two selected rows on pair \(j\) as \((a_j,b_j)\) and \((s_ja_j,s_jb_j)\). Applying the same two row-ratio identities shows that, after scaling the fourth row, \(s_j=r_j\) or \(s_j=r_j^{-1}\). Indeed every permutation of the three cube roots is a scalar multiple of either of those two indexed triples.

Put \(A_j=a_j+b_j\). Orthogonality in the row-ratio sense between the first two rows and the next two gives

\begin{equation}
\sum_jA_j=0,\qquad\sum_j\frac{A_j}{r_j}=0,\qquad\sum_js_jA_j=0,\qquad\sum_j\frac{s_j}{r_j}A_j=0.
\end{equation}

For either choice of \ensuremath{s}, these include all three Fourier equations on three distinct cube roots, whose coefficient matrix is invertible. Therefore \(A_j=0\) and \(b_j=-a_j\).

Take either of the two remaining rows and write its entries on pair j as \ensuremath{(p_{j}^+,p_{j}^-)}. Its row-ratio equations against the first four rows imply

\begin{equation}
\begin{aligned}p_j^+&=hr_j^2+\ell a_js_j^2,\\ p_j^-&=hr_j^2-\ell a_js_j^2.\end{aligned}
\end{equation}

for scalars \(h,\ell\). This follows by applying the same invertible three-point Fourier matrix separately to \(p_j^++p_j^-\) and \((p_j^+-p_j^-)/a_j\). If \(\ell=0\), that row together with the first two rows has proportional columns on each pair, giving a zero 3-minor. If \(h=0\), the same is true for that row together with the third and fourth selected rows. Thus the nonzero-three-minor hypothesis forces \(h\) and \(\ell\) both nonzero.

Set \(D_j=h^2r_j^4-\ell^2a_j^2s_j^4\). Every \ensuremath{D_{j}} is nonzero because it is \ensuremath{p_{j}^+p_{j}^-} and all entries are nonzero. The reversed row-ratio equations, using the first and second rows divided by this remaining row, give

\begin{equation}
\sum_j\frac{hr_j^2}{D_j}=0,\qquad\sum_j\frac{r_jhr_j^2}{D_j}=0.
\end{equation}

Thus \(hr_j^2/D_j=\gamma r_j\) for one nonzero \ensuremath{\gamma}. As \(r_j^4=r_j\) and \(s_j^4=s_j\), it follows that

\begin{equation}
a_j^2s_j=\kappa r_j
\end{equation}

for a nonzero \ensuremath{\kappa}. Hence \((a_js_j^2/r_j^2)^2=\kappa\). Over the algebraically closed field choose \ensuremath{\eta} with \(\eta^2=\kappa\). Swap the two columns within each pair when necessary, which replaces \ensuremath{a_{j}} by \ensuremath{-a_{j}} without changing \ensuremath{r_{j}} or \ensuremath{s_{j}}, to obtain

\begin{equation}
a_j=\eta\frac{r_j^2}{s_j^2}\qquad\text{for every }j.
\end{equation}

If \(s_j=r_j\), then \(a_j=\eta\); the first and third selected rows are proportional on the three first columns of the pairs. If \(s_j=r_j^{-1}\), then \(a_j=\eta r_j\) and \(s_ja_j=\eta\); the first and fourth selected rows are proportional on those three columns. In either case, adding any other row gives a zero positional three-minor. This contradiction proves the lemma. Applying it to \(C^{\mathsf T}\) proves the transposed assertion. All denominators used were nonzero, and only invertibility of two and three and the matrix hypotheses were used.

\subsection{Independence of at most six compatible boundary matrices}\label{appendix-dvr6-independence-of-at-most-six-compatible-boundary-matrices}

The valuation argument will produce a linear relation among the rank-one matrices of the boundary points with the largest valuation. To rule it out, we prove independence for every compatible collection of at most six such matrices. Compatibility first organizes the small supports into a grid of disjoint coordinate pairs. In a minimal dependence, neither the z-vectors nor the w-vectors can have a two-dimensional span, and both support orientations must occur. The only remaining six-point configuration is excluded by the \hyperref[result:block-lemma]{block lemma} or by the type-II trace identity.

The case order is: one support type; mixed types with either span of dimension two; then six terms with both spans of dimension three. In the last case repeated support pairs lead to the \hyperref[result:block-lemma]{block lemma}, while distinct row and column partitions lead to the trace contradiction.

For a coordinate index set \(A\), write \(F^A\subseteq F^6\) for the subspace of vectors supported on \(A\); the same convention applies to other index sets.

\phantomsection\label{result:boundary-tensor-independence}Let distinct boundary points be represented by \((z_r,w_r)\), and let \(N_r=z_rw_r^{\mathsf T}\) be their rank-one matrices, or outer products. Compatibility means that both ordered bilinear pairings \(w_r^{\mathsf T}z_s\) and \(w_s^{\mathsf T}z_r\) vanish for each \(r\ne s\). The self-pairings vanish as well, since \(t=0\). Put \(v_r=Tz_r\) and \(u_r=C^{\mathsf T}w_r\); then \(u_r^{\mathsf T}v_s=w_r^{\mathsf T}z_s=0\) for all \(r,s\).

For a type A point call \ensuremath{(A_{r},E_{r})} its small support pair; for a type B point call \ensuremath{(B_{r},D_{r})} its small support pair. All small row supports occurring in this list are equal or disjoint. For two type A points, an intersection of size one between \ensuremath{A_{r}} and the complement of \ensuremath{A_{s}} would make \ensuremath{w_{s}^{\mathsf T}z_{r}} a single nonzero term. Thus \ensuremath{A_{r}=A_{s}} or \ensuremath{A_{r}} and \ensuremath{A_{s}} are disjoint. The type B/type B argument uses \ensuremath{w_{r}} and \ensuremath{z_{s}}. A mixed pair uses the two small supports of \ensuremath{z_{A}} and \ensuremath{w_{B}}, whose intersection cannot have size one. The same proof with \ensuremath{v,u} gives equal-or-disjoint small column supports.

Distinct points of the same type cannot have the same small support pair, by \hyperref[appendix-dvr3-boundary-classification-uniqueness-and-reducedness]{Appendix B.3}. A mixed pair cannot have the same small support pair either. To see this directly, write \(T[E,A]=ab^{\mathsf T}\) with all a,b entries nonzero. The reciprocal entry identity gives

\begin{equation}
C[A,E]=\frac16b^{[-1]}(a^{[-1]})^{\mathsf T},
\end{equation}

where {[}-1{]} means coordinatewise inverse. The type A \ensuremath{z} line in A is \ensuremath{\ker b^{\mathsf T}}, represented by \ensuremath{(b_{2},-b_{1})}, whereas the type B \ensuremath{w} line in A annihilates \ensuremath{b^{[-1]}}, represented by \ensuremath{(b_{1},-b_{2})}. Their pairing is \ensuremath{2b_{1}b_{2}}, nonzero. This contradicts compatibility. A support-grid cell is specified by one small row-support pair and one small column-support pair. The points therefore occupy distinct cells, with at most three disjoint row pairs and at most three disjoint column pairs.

First, a compatible collection consisting of only one type has independent N's, without any six-point bound. For type A, the matrices \(N_rC=z_ru_r^{\mathsf T}\) have supports \ensuremath{A_{r}\times E_{r}}, and those rectangles are pairwise disjoint: unequal row supports are disjoint, and equal row supports force distinct disjoint column supports. Matrices with disjoint nonempty supports are independent. Type B is identical using \(TN_r=v_rw_r^{\mathsf T}\).

Suppose now there is a dependence using at most six distinct N's, and choose a minimal one with \ensuremath{m} terms and all coefficients nonzero. Let \ensuremath{Z} and \ensuremath{W} have columns \ensuremath{z_{r}} and \ensuremath{w_{r}}, and put \(p=\operatorname{rank}Z\), \(q=\operatorname{rank}W\). With \ensuremath{\Lambda} the invertible diagonal matrix of dependence coefficients,

\begin{equation}
Z\Lambda W^{\mathsf T}=0,\qquad W^{\mathsf T}Z=0.
\end{equation}

Sylvester's rank inequality gives \(p+q\le m\). Injectivity of both boundary projections shows \(p,q\ge2\) when \ensuremath{m\ge 2}. (Two rank-one matrices with proportional nonzero tensors would be the same projective point.)

We exclude \ensuremath{p=2}. The same argument with the inverse/transposed roles excludes \ensuremath{q=2}. If all points have one type, independence is already proved. In a mixed collection with \ensuremath{p=2}, two type A points cannot have the same small row support: their distinct \ensuremath{z} lines would span that entire coordinate 2-plane, leaving no room in \ensuremath{\operatorname{span}Z} for the support-four \ensuremath{z} of a type B point. Hence there are at most two type A points, with disjoint small row supports. Similarly \(\operatorname{rank}[v_r]=2\), and two type B points cannot have the same small \ensuremath{v} support; there are at most two type B points, with disjoint small \ensuremath{v} supports. Thus \ensuremath{m\le 4}. For \ensuremath{m\le 3}, \(p+q\ge4>m\) already contradicts Sylvester. If \ensuremath{m=4}, there are exactly two of each type.

In that last case the two type A \ensuremath{z} lines span the coordinate-pair lines on disjoint sets \ensuremath{A_{1},A_{2}}. Both type B \ensuremath{z} vectors are nontrivial linear combinations and have support \ensuremath{A_{1}\cup A_{2}}; their two distinct \ensuremath{w} lines therefore span the coordinate 2-plane on \ensuremath{B=(A_{1}\cup A_{2})^c}. Dually, the two type B \ensuremath{v} lines have disjoint supports \ensuremath{D_{1},D_{2}}, and the two type A \ensuremath{u} lines span the coordinate 2-plane on \ensuremath{E=(D_{1}\cup D_{2})^c}. Thus the type A \(w\)-span is \(T^{\mathsf T}F^E\). Its intersection with \ensuremath{F^B} is zero: a nonzero vector in the intersection would have both \ensuremath{w} and \ensuremath{C^{\mathsf T}w} supported on at most two coordinates, contrary to the support bound. These two two-dimensional \ensuremath{w} spans have direct sum of dimension four. Hence \ensuremath{q=4}, contradicting \(p+q\le4\).

It follows that \(p,q\ge3\), so \(m=6\) and \(p=q=3\). Write \(\mathcal L=\operatorname{span}\{z_r\}\) and \(\mathcal W=\operatorname{span}\{w_r\}\). Here \(\mathcal L^\perp=\{w:w^{\mathsf T}z=0\text{ for every }z\in\mathcal L\}\) is the annihilator for the standard bilinear pairing, without conjugation. Compatibility gives \(\mathcal W=\mathcal L^\perp\). Likewise \(\operatorname{span}\{v_r\}=T\mathcal L\) and \(\operatorname{span}\{u_r\}= (T\mathcal L)^\perp\).

In this three-dimensional situation each type occurs at most three times. For type A, if a small row support is repeated, the two distinct \ensuremath{z} lines span the whole coordinate 2-plane. Any type A point on a different, disjoint row pair would have \ensuremath{w} nonzero on that plane and could not annihilate \(\mathcal L\). Thus all type A points have that same row pair; their small column pairs must be distinct and disjoint, so there are at most three. If no small row support repeats, there are at most three disjoint row pairs. For type B the same argument starts with repeated small \ensuremath{w} supports, uses \(\mathcal W=\mathcal L^\perp\), and then its small \ensuremath{v} supports. Thus there are exactly three points of each type.

If the three type A row supports are the same pair A, the three type B row supports must also be the same pair B, disjoint from A. Indeed \(\mathcal L\) contains \(F^A\), so every type B \ensuremath{w} vanishes on A; three disjoint two-element supports would not fit in \ensuremath{A^c}. A repetition among the type B supports forces all three equal by the preceding argument. Each type then has three disjoint small column pairs, necessarily a partition of all six coordinates. Across the two types these partitions agree, since every pair from one partition is equal to or disjoint from each pair in the other. All six occupied blocks \ensuremath{C[A,E_{j}]} and \ensuremath{C[B,E_{j}]} have rank one: for type A, singularity of \ensuremath{T[E_{j},A]} is equivalent to singularity of its entrywise reciprocal transpose \ensuremath{C[A,E_{j}]}; for type B it is part of its boundary description. The \hyperref[result:block-lemma]{block lemma} excludes this case.

Otherwise the three type A row pairs are all distinct. The three type B row pairs must also all be distinct: a repeated type B pair would be its entire common pair and would put its full coordinate 2-plane in \(\mathcal W\), contradicting the type A \ensuremath{z} supported on that pair. Hence both types use the same partition of the six rows into three pairs. Apply the analogous argument to the column supports, interchanging the \ensuremath{v} and \ensuremath{u} roles. Either all three small column pairs of each type are constant, in two distinct pairs, which gives the transposed \hyperref[result:block-lemma]{block lemma}, or both types use the same partition into three distinct column pairs.

It remains to exclude this last case. On every row pair \ensuremath{A_{i}}, \(\mathcal L\) contains the type A \ensuremath{z} line and its annihilator \(\mathcal W\) contains the type B \ensuremath{w} line. These are nonzero orthogonal lines in a coordinate 2-plane. Therefore coordinate projection onto \ensuremath{A_{i}} preserves \(\mathcal L\), and \(\mathcal L\) meets that 2-plane in a line. The same argument shows that coordinate projection onto every column pair \ensuremath{E_{j}} preserves \(T\mathcal L\), again with one-dimensional intersection.

Choose bases on each row coordinate 2-plane whose first vectors span \(\mathcal L\)'s lines, and on each column coordinate 2-plane whose first vectors span \(T\mathcal L\)'s lines. In these adapted bases every 2 by 2 block of \ensuremath{C} and \ensuremath{T} is upper triangular, since the global maps preserve the corresponding three-dimensional subspaces. Take a type A point in cell (i,j). Its \ensuremath{z} is the first-vector line on \ensuremath{A_{i}} and its \ensuremath{v} vanishes on \ensuremath{E_{j}}; hence the first diagonal entry of \ensuremath{T[E_{j},A_{i}]} is zero. Its \ensuremath{u} is the covector on \ensuremath{E_{j}} annihilating \(T\mathcal L\), and \ensuremath{w=T^{\mathsf T}u} vanishes on \ensuremath{A_{i}}; hence the second diagonal entry is zero as well. The block is strictly upper triangular. Its product with the upper-triangular block \ensuremath{C[A_{i},E_{j}]} has trace zero. Trace is independent of the chosen bases on these two coordinate 2-planes. But in the original coordinates,

\begin{equation}
\operatorname{tr}\bigl(C[A_i,E_j]T[E_j,A_i]\bigr)=\sum_{a\in A_i,\,e\in E_j}C_{ae}T_{ea}=\frac46=\frac23\ne0.
\end{equation}

This final contradiction excludes the six-term dependence and proves the entire claimed independence assertion.

\subsection{Boundary specialization of a complete basis}\label{appendix-dvr7-boundary-lifts-cannot-collide-and-a-complete-basis-cannot-escape}

To use matrix independence after reduction, we must exclude a collision: two distinct generic-fibre points specializing to the same boundary point in the special fibre. The boundary Jacobian gives uniqueness of such lifts, where a lift is an \(O\)-point reducing to the given point. We then extract a relation from the members with maximal positive valuation, contradicting \hyperref[result:boundary-tensor-independence]{the preceding independence theorem}.

There is a useful uniqueness consequence of \hyperref[appendix-dvr3-boundary-classification-uniqueness-and-reducedness]{Appendix B.3} that avoids any implicit smoothness or henselian assumption. Let \ensuremath{O} be a DVR over \ensuremath{R} and suppose two O-points of \ensuremath{Z_{R}} have the same special-fiber boundary point. Choose unit projective coordinates in both factors at that point and normalize them to one. Both points lie in the same affine chart with ten coordinates. The ten equations have invertible 10 by 10 Jacobian modulo the maximal ideal, by \hyperref[appendix-dvr3-boundary-classification-uniqueness-and-reducedness]{Appendix B.3}. Subtract the two systems of polynomial equations. The difference is \(M(a-b)=0\) for a matrix M over \ensuremath{O} whose reduction is that Jacobian. Its determinant is a unit, so \ensuremath{a=b}. Thus there is at most one O-point specializing to that boundary point. This proves uniqueness directly over \ensuremath{O}, including ramified DVR extensions.

Now take six generic-fiber common-ray pairs over the fraction field \ensuremath{L} of such an \ensuremath{O} which form a complete companion basis, meaning all thirty ordered off-diagonal pairings vanish and every \ensuremath{t_{r}} is nonzero. Their projective pairs extend to O-points by properness of \ensuremath{Z_{R}}. Choose primitive integral representatives \ensuremath{z_{r},w_{r}}. Put \(N_r=z_rw_r^{\mathsf T}\) and \(t_r=z_{0r}w_{0r}\). Then

\begin{equation}
W^{\mathsf T}Z=6\operatorname{diag}(t_0,\ldots,t_5).
\end{equation}

Every \ensuremath{t_{r}} is nonzero in \ensuremath{L}. Determinant/adjugate inversion gives

\begin{equation}
\sum_r\frac{N_r}{t_r}=6I_6.
\end{equation}

Each \ensuremath{N_{r}} is integral and has at least one unit entry, because each of \ensuremath{z_{r},w_{r}} is primitive. The off-diagonal pairings remain zero under specialization.

If some specialization were boundary, let \(e=\max_r\operatorname{val}(t_r)>0\). Multiply the last identity by \(\pi_O^e\) and reduce. This gives

\begin{equation}
\sum_{\operatorname{val}(t_r)=e}\overline{\pi_O^e/t_r}\;\overline{N_r}=0,
\end{equation}

with nonzero coefficients. All participating points are boundary points. They are distinct: if two coincided in the special fiber, the preceding Jacobian argument would make their O-points, and hence their generic projective pairs, equal. Distinct members of a generic complete basis cannot coincide, since their off-diagonal pairing is zero but their self-pairing is \(6t_r\ne0\). The resulting dependence is therefore among at most six distinct compatible boundary matrices. \hyperref[appendix-dvr6-independence-of-at-most-six-compatible-boundary-matrices]{Appendix B.6} forbids it. No member of a complete generic companion basis specializes to the boundary.

The reduction may be checked after extending the residue field to its algebraic closure. Independence there implies independence over the original residue field. No assumption of residue-field rationality or positivity occurs.

\subsection{\texorpdfstring{Module finiteness of \(P_{R}\)}{Module finiteness of P\_R}}\label{appendix-dvr8-finiteness-of-the-exact-full-algebra-p_r}

The no-escape argument has so far concerned points over valuation fields. To obtain module finiteness for the full algebra, including torsion and nilpotents, we apply it to the Laurent chart where every \(t_r\ne0\) inside the finite projective scheme of compatible tuples. Excluding every boundary specialization makes this open locus closed.

The notation \(Z_R^6\) means the sixfold fibre product over \(\operatorname{Spec}R\). Inside this finite R-scheme let \ensuremath{W_{R}} be the closed subscheme defined by all thirty projective ordered-pairing equations \(w_r^{\mathsf T}z_s=0\) for \ensuremath{r\ne s}. These are well-defined vanishing conditions of the appropriate line-bundle sections. \ensuremath{W_{R}} is finite over \ensuremath{R}. On its open subset where every \ensuremath{t_{r}} is nonzero, the chart identification from \hyperref[appendix-dvr4-finite-projective-geometric-fibers-and-the-projective-dvr-family]{Appendix B.4} identifies the equations with exactly the thirty \ensuremath{q_{rs}} generators. Thus this open subscheme \ensuremath{U} is precisely \(\operatorname{Spec}P_R\), with its full scheme structure.

We show \ensuremath{U} is also closed in \ensuremath{W_{R}}. If a point of the topological closure of \ensuremath{U} lay outside \ensuremath{U}, it would be a boundary specialization of a point of \ensuremath{U} over \ensuremath{K}. Here is the finite-scheme justification. A finite R-scheme has finite fibers, with no nontrivial specializations within either field fiber. The only possible nontrivial specialization is from a generic-fiber point to a closed-fiber point. Since \ensuremath{U} is open, a generic-fiber boundary point cannot lie in its closure, and a closed-fibre point in its closure but outside it must therefore be a specialization of a generic-fibre point in \ensuremath{U}.

The residue field \ensuremath{L} of that generic point is a finite extension of \ensuremath{K}. Take the scheme-theoretic closure of this point in \ensuremath{W_{R}} and choose an extension of the valuation of \ensuremath{R} to \ensuremath{L} centered at the indicated special point; equivalently choose a prime above that point in the integral closure of the finite domain defining the closure, and its corresponding valuation ring. Extension of DVR valuations through finite field extensions supplies a DVR \ensuremath{O} with this center. The resulting map \ensuremath{\operatorname{Spec}O} to the closure gives six generic common-ray pairs forming a complete basis and specializing to a boundary tuple. \hyperref[appendix-dvr7-boundary-lifts-cannot-collide-and-a-complete-basis-cannot-escape]{Appendix B.7} rules this out.

For clarity about this standard valuation use: a finite domain over a DVR is integral over it, its fraction field is finite over \ensuremath{K}, and primes over the closed point are dominated by extensions of that discrete valuation. Passing to its integral closure and a prime over the chosen maximal ideal chooses the required center. The proof does not require that integral closure to be finite over the original DVR; the valuation ring in the finite field extension is enough \cite{Stacks}.

Therefore \ensuremath{U} is open and closed. The open-and-closed subscheme of a finite affine scheme is a direct factor of its finite coordinate algebra. Hence \ensuremath{P_{R}} is finite as an R-module. This argument preserves the given ideal and every nilpotent and vertical component on \ensuremath{U}. It does not replace \ensuremath{U} by its reduction or replace the Laurent chart by a different saturation. If \ensuremath{U} is empty its algebra is zero, which is included. It also makes no finiteness assertion about the one-ray algebra \ensuremath{A_{R}}.

\subsection{\texorpdfstring{Finiteness of \(Q_{R}\) and vanishing after reduction}{Finiteness of Q\_R and vanishing after reduction}}\label{appendix-dvr9-q_r-further-exact-quotients-and-geometric-emptiness}

The tensor product of two finite R-modules is finite, and every quotient of a finite R-module is finite. Thus \ensuremath{Q_{R}}, with all thirty-six generators from \hyperref[appendix-dvr1-algebraic-setup]{Appendix B.1}, is finite as an R-module, as is any quotient \(B=Q_R/J\) where J is generated by additional explicitly specified equations. No physical predicate is silently added to J, and no assertion of vanishing on nilpotents is inferred from validity on physical characters.

If the complete geometric special fibre of B is empty, the finite-dimensional \(\overline k\)-algebra \(B\otimes_R\overline k\) is zero. A nonzero finite-dimensional algebra over an algebraically closed field has a maximal ideal with residue that field, so it would have a geometric point. Faithfulness of extension of fields gives \(B/\pi B=0\). Nakayama for the finite R-module B now gives \ensuremath{B=0}, hence \(B\otimes_RK=0\). This applies to \ensuremath{Q_{R}} itself by taking \ensuremath{J=0}. The absence of k-rational points is not substituted for emptiness after algebraic closure.

\subsection{Complex involution and physical companion bases}\label{appendix-dvr10-complex-involution-and-exact-physical-bijections}

We finish by recovering the measurement meaning of these companion equations. The argument over a DVR used inverse coordinates without a conjugation. Over the complex numbers, unitarity supplies the conjugation that selects physical points from the algebraic companion schemes. This subsection identifies those fixed characters and checks that normalization turns reciprocal orthogonality into an orthonormal measurement basis, and the mixed products into uniform cross-basis probabilities.

Now fix a normalized flat unitary complex matrix \ensuremath{C} with every 3-minor nonzero. For this matrix, let \(A_{\mathbb C},P_{\mathbb C},Q_{\mathbb C}\) denote the algebras defined by the equations of \hyperref[appendix-dvr1-algebraic-setup]{Appendix B.1} over \(\mathbb C\), consistently with \hyperref[compatible-algebraic-collections]{Section 5}. Then \(T=C^*\), and \(C_{ij}T_{ji}=|C_{ij}|^2=1/6\).

On the Laurent algebra define the conjugate-linear involution \ensuremath{\tau} by conjugating coefficients and sending every \ensuremath{x_{ir}} to \ensuremath{x_{ir}^{-1}}, and similarly for y. Because \(T_{ji}=\overline{C_{ij}}\), \ensuremath{\tau} interchanges the two factors of each \ensuremath{F_{j}}, so \(\tau(F_j)=F_j\). Also \(\tau(q_{rs})=q_{sr}\), so the full thirty-generator ideal is stable. In \(Q_{\mathbb C}\), \ensuremath{\tau} interchanges \ensuremath{q^{xy}_{rs}} and \ensuremath{q^{yx}_{sr}}, preserving each of the thirty-six products minus six. Thus \ensuremath{\tau} descends to these algebras.

The induced involution on complex characters is

\begin{equation}
\chi\longmapsto\bigl[f\longmapsto\overline{\chi(\tau(f))}\bigr].
\end{equation}

It fixes a character exactly when every ray coordinate has modulus one, since the condition on a generator is \(x=\overline{x^{-1}}\). For such a point x, its normalized column \(X=x/\sqrt6\) is a unit vector flat against \ensuremath{I_{6}}, and

\begin{equation}
F_j(x)=|(C^*x)_j|^2-1.
\end{equation}

Thus \ensuremath{F_{j}=0} is exactly flatness of \ensuremath{C^*X}, i.e.~unbiasedness of that column to \ensuremath{C}. For six columns,

\begin{equation}
q_{rs}=x_{:r}^*x_{:s}=6(X^*X)_{rs}.
\end{equation}

All thirty off-diagonal zero equations are exactly orthogonality of the ordered six unit columns. Hence fixed characters of \(P_{\mathbb C}\) correspond bijectively to ordered column-dephased physical companion bases X for \ensuremath{I_{6},C}. The dephasing convention is \(X_{0r}=1/\sqrt6\), or equivalently \(x_{0r}=1\). Every original physical companion can be put into this convention uniquely by multiplying each column by a unit complex phase. Conversely the displayed formulas recover the unique dephased basis from the character.

For a fixed character of \(Q_{\mathbb C}\), write \(Y=y/\sqrt6\). Its additional equations are

\begin{equation}
q^{xy}_{rs}q^{yx}_{sr}=\left|\sum_i\overline{x_{ir}}y_{is}\right|^2=6,
\end{equation}

or, after normalization,

\begin{equation}
|(X^*Y)_{rs}|^2=\frac16\qquad(0\le r,s\le5).
\end{equation}

The raw overlap-squared threshold is exactly 6; the raw orthogonality threshold is exactly 0. Consequently fixed characters of \(Q_{\mathbb C}\) correspond bijectively to ordered column-dephased physical quartets \ensuremath{(I_{6},C,X,Y)}, retaining all thirty-six cross relations. Nilpotents in the coordinate algebra do not alter this character-level bijection, and this does not assert that the algebra is reduced.

\section{Failure of entrywise phase powers to preserve transition composition}\label{app:prescribed-phase-powers}

The transition matrices of three mutually unbiased bases satisfy a composition
identity. Raising their entry phases separately need not preserve it. This
appendix bounds the error when the same original closing transition is
prescribed at both the cubic and quintic powers. These entrywise phase
powers are algebraic transforms of transition amplitudes; they are not
operator powers or asserted physical operations.

Let \(B_a,B_c,B_b\) be unitary column matrices of three mutually unbiased
bases in dimension six, and put \(U_{xy}=B_x^*B_y\). Then
\begin{equation}\label{eq:phase-power-original-cocycle}
 U_{ac}U_{cb}=U_{ab},\qquad |(U_{xy})_{ij}|=6^{-1/2}\quad(x\ne y).
\end{equation}
We call \(U_{ab}\) the prescribed closing transition, fixed by the same original bases.
For \(m=3,5\), define
\begin{equation}\label{eq:phase-power-defect-definition}
 \Phi_m(U)=6^{(m-1)/2}U^{\circ m},\qquad
 \Delta_m=\Phi_m(U_{ac})\Phi_m(U_{cb})-\Phi_m(U_{ab}),
\end{equation}
where \(U^{\circ m}\) is the entrywise power. The symbol \(\Phi_m\) here
is distinct from the arrangement function \(\Phi_H\), and \(\Delta_m\)
is distinct from the clock operator \(D\). Thus \(\Delta_m\) measures the error in retaining \(\Phi_m(U_{ab})\) as the closing transition.

\phantomsection\label{prop:phase-power-gap}\textbf{Proposition (uniform prescribed-closing defect).}
For every original ordered entry \((i,j)\),
\begin{equation}\label{eq:phase-power-entrywise-gap}
 \max\{|(\Delta_3)_{ij}|,|(\Delta_5)_{ij}|\}
       >\frac{1}{6\,2^{21}}.
\end{equation}
Consequently,
\begin{equation}\label{eq:phase-power-frobenius-gap}
 \|\Delta_3\|_F^2+\|\Delta_5\|_F^2>2^{-42}.
\end{equation}
The exponent attaining the entrywise maximum may depend on \((i,j)\).
The estimate holds for every triple satisfying the displayed hypotheses,
and the powered matrices need not be unitary.

For each original entry, we express the two transition defects through power sums of six unit complex numbers. Newton identities and the unit-circle reciprocal relations show that small cubic and quintic defects force the polynomial with these numbers as roots close to a fixed comparison polynomial. That comparison polynomial has a real root outside the unit circle, where the original polynomial must be bounded away from zero. Comparing this lower bound with the coefficient-error upper bound yields the stated strict quantitative gap.

\subsection{Unit path phases and their power sums}\label{app:phase-power-path-phases}

Fix \((i,j)\). Divide each of the six path amplitudes through the intermediate
basis by the original closing amplitude, with the following normalization:
\begin{equation}\label{eq:phase-power-path-normalization}
 a=\sqrt6,\qquad
 w_k=a\frac{(U_{ac})_{ik}(U_{cb})_{kj}}{(U_{ab})_{ij}},\qquad
 \lambda=a(U_{ab})_{ij},\qquad p_m=\sum_{k=0}^5w_k^m.
\end{equation}
Here \(p_m\) is the unnormalized power sum of the six unit path phases,
not a probability expectation. Flatness gives \(|w_k|=|\lambda|=1\), and
the original composition identity gives \(p_1=a\). Direct substitution into
\eqref{eq:phase-power-defect-definition} yields
\begin{equation}\label{eq:phase-power-moment-defect}
 [\Phi_m(U_{ac})\Phi_m(U_{cb})]_{ij}=\frac{\lambda^mp_m}{6},
 \qquad [\Phi_m(U_{ab})]_{ij}=\frac{\lambda^ma}{6},
 \qquad (\Delta_m)_{ij}=\frac{\lambda^m(p_m-a)}6.
\end{equation}
Thus the estimate reduces to six unit complex numbers with first sum
\(\sqrt6\). The displayed formula is called piercing-phase normalization
in \cite[Proposition~3.1(iv)]{MMVW}. The Newton identities and complementary-subset
identity used below have a zero-sum antecedent \cite[Lemma~4.4]{MMVW}; here
we prove the quantitative estimate for the nonzero power-sum target.
The Fourier-certificate preprint also uses Newton identities, unit-circle
reciprocity and mixed-transition moments \cite{OAIfourier}. Its single-matrix
balanced characters and complete-family mixed certificate have different
hypotheses and targets; neither is a premise of this estimate.

\subsection{Symmetric coefficients and quantitative stability}\label{app:phase-power-stability}

Let \(e_j\) be the elementary symmetric functions of the six phases,
with \(e_0=1\), and set
\begin{equation}\label{eq:phase-power-errors}
 b=e_2,\quad c=e_3,\quad d=e_4,\quad t=e_6,\qquad
 e=p_3-a,\quad f=p_5-a.
\end{equation}
For unit phases, the product over a complementary subset equals the product
of all six phases times the conjugate of the product over the original subset.
Summing gives the unit-circle reciprocity identity
\(e_{6-j}=t\overline{e_j}\). In particular,
\begin{equation}\label{eq:phase-power-reciprocity}
 |t|=1,\qquad e_1=a,\qquad e_5=at,\qquad
 d=t\bar b,\qquad c=t\bar c.
\end{equation}
These identities concern a multiset and retain repeated phases.

The signed Newton identities through degree five are
\begin{equation}\label{eq:phase-power-newton-identities}
\begin{aligned}
 p_2-ap_1+2b&=0,\\
 p_3-ap_2+bp_1-3c&=0,\\
 p_4-ap_3+bp_2-cp_1+4d&=0,\\
 p_5-ap_4+bp_3-cp_2+dp_1-5e_5&=0.
\end{aligned}
\end{equation}
Using \(p_1=a\), \(a^2=6\), and \(p_3=a+e\), successive substitution
gives the complete expansions
\begin{equation}\label{eq:phase-power-newton-expansions}
\begin{aligned}
 p_2&=6-2b,\\
 c&=ab-\frac{5a}{3}+\frac e3,\\
 p_4&=2b^2-4-4d+\frac{4a}{3}e,\\
 p_5&=\frac{25a}{3}b-14a-5ad+5at
             +\frac{5(6-b)}3e.
\end{aligned}
\end{equation}
In the last line the quadratic terms cancel, but the error coefficient
still depends on \(b\). Since \(|p_2|\le6\) and \(b=3-p_2/2\),
\begin{equation}\label{eq:phase-power-second-coefficient-bounds}
 |b|\le6,\qquad |6-b|\le6.
\end{equation}

Apply \(c=t\bar c\) to the second line of
\eqref{eq:phase-power-newton-expansions}. Apply \(p_5=a+f\) and
\(d=t\bar b\) to the last line. The two resulting equations are
\begin{equation}\label{eq:phase-power-two-coefficient-equations}
\begin{aligned}
 b-t\bar b&=\frac53(1-t)+\varepsilon,
 &\varepsilon&=\frac{t\bar e-e}{3a},\\
 \frac53b-t\bar b+t-3&=\delta,
 &\delta&=\frac{f}{5a}-\frac{(6-b)e}{3a}.
\end{aligned}
\end{equation}
The bounds just established give
\begin{equation}\label{eq:phase-power-epsilon-delta-bounds}
 |\varepsilon|\le\frac{2|e|}{3a},\qquad
 |\delta|\le\frac{|f|}{5a}+\frac{2|e|}{a}.
\end{equation}
Subtracting the first coefficient equation from the second shows
\begin{equation}\label{eq:phase-power-g-equation}
 b=t+2+g,\qquad g=\frac32(\delta-\varepsilon),\qquad
 |g|\le\frac{4|e|+(3/10)|f|}{a}.
\end{equation}
Substituting this expression for \(b\) back into the first equation gives
\begin{equation}\label{eq:phase-power-t-stability}
 \frac23(1-t)=g-t\bar g-\varepsilon,\qquad
 |t-1|\le3|g|+\frac32|\varepsilon|
       \le\frac{13|e|+(9/10)|f|}{a}.
\end{equation}
Hence
\begin{equation}\label{eq:phase-power-b-stability}
 |b-3|\le|t-1|+|g|
       \le\frac{17|e|+(6/5)|f|}{a}.
\end{equation}
No step divides by a phase difference, a possibly vanishing error or a
spectral quantity.

Put \(E_0=\max\{|e|,|f|\}\). The preceding estimates imply
\begin{equation}\label{eq:phase-power-basic-coefficient-errors}
 |g|\le\frac{43E_0}{10a},\qquad
 |t-1|\le\frac{139E_0}{10a},\qquad
 |b-3|\le\frac{91E_0}{5a}.
\end{equation}
The remaining coefficients are controlled without further assumptions:
\begin{equation}\label{eq:phase-power-all-coefficient-errors}
\begin{aligned}
 \left|c-\frac{4a}{3}\right|
  &=\left|a(b-3)+\frac e3\right|
    \le\left(\frac{91}{5}+\frac13\right)E_0
    =\frac{278E_0}{15},\\
 |d-3|&=|t(\bar b-3)+3(t-1)|
    \le\left(\frac{91}{5}+3\frac{139}{10}\right)\frac{E_0}{a}
    =\frac{599E_0}{10a},\\
 |e_5-a|&=a|t-1|\le\frac{139E_0}{10}.
\end{aligned}
\end{equation}

\subsection{A comparison root outside the unit circle}\label{app:phase-power-root-gap}

The coefficient values forced by zero cubic and quintic errors would give
the comparison polynomial
\begin{equation}\label{eq:phase-power-comparison-polynomial}
 P_0(z)=z^6-az^5+3z^4-\frac{4a}{3}z^3+3z^2-az+1.
\end{equation}
Its reciprocal form makes a root outside the unit circle visible. With
\(x=z+z^{-1}\),
\begin{equation}\label{eq:phase-power-reciprocal-cubic}
 z^{-3}P_0(z)=f_0(x),\qquad f_0(x)=x^3-ax^2+\frac{2a}{3}.
\end{equation}
To obtain a fixed separation suitable for the quantitative estimate, use
\(22/9<a<5/2\), which follows from \(484/81<6<25/4\). Then
\begin{equation}\label{eq:phase-power-root-sign-brackets}
\begin{aligned}
 f_0(41/20)&=\frac{68921}{8000}-\frac{4243}{1200}a
  <\frac{68921}{8000}-\frac{4243}{1200}\frac{22}{9}
  =-\frac{6053}{216000}<0,\\
 f_0(5/2)&=\frac{125}{8}-\frac{67}{12}a
  >\frac{125}{8}-\frac{67}{12}\frac52=\frac53>0.
\end{aligned}
\end{equation}
The map \(z\mapsto z+z^{-1}\) is continuous and strictly increasing for
\(z>1\), and its values at \(5/4\) and \(2\) are \(41/20\) and
\(5/2\). Therefore \(P_0\) has a positive real root \(r\) with
\(5/4<r<2\).

By contrast, the actual polynomial
\(P(z)=\prod_{k=0}^5(z-w_k)\) has all roots on the unit circle. Thus
\begin{equation}\label{eq:phase-power-actual-polynomial-lower-bound}
 |P(r)|\ge\prod_{k=0}^5(r-|w_k|)=(r-1)^6>2^{-12}.
\end{equation}
The \(z^5\) coefficient of \(P-P_0\) is zero. Using the five other
coefficient errors, \(P_0(r)=0\), and \(r<2\), we obtain
\begin{equation}\label{eq:phase-power-polynomial-upper-bound}
\begin{aligned}
 |P(r)|&=|P(r)-P_0(r)|\\
 &\le16|b-3|+8\left|c-\frac{4a}{3}\right|
      +4|d-3|+2|e_5-a|+|t-1|\\
 &\le C_0E_0,\qquad C_0=\frac{5447}{10a}+\frac{2641}{15}<450.
\end{aligned}
\end{equation}
For clarity, the coefficient collection is exactly
\begin{equation}\label{eq:phase-power-bound-constant}
 16\frac{91}{5}+4\frac{599}{10}+\frac{139}{10}=\frac{5447}{10},
 \qquad
 8\frac{278}{15}+2\frac{139}{10}=\frac{2641}{15}.
\end{equation}
Since \(a>2\),
\(C_0<5447/20+2641/15=26905/60<450\).
If \(E_0\le2^{-21}\), then
\begin{equation}\label{eq:phase-power-uniform-contradiction}
 |P(r)|\le C_0E_0\le C_0\,2^{-21}
           <450\,2^{-21}<512\,2^{-21}=2^{-12},
\end{equation}
contradicting \eqref{eq:phase-power-actual-polynomial-lower-bound}.
This chain includes \(E_0=0\) without a strict inequality proportional
to zero. Consequently \(E_0>2^{-21}\).

The entry \((i,j)\) was arbitrary, so
\eqref{eq:phase-power-moment-defect} proves
\eqref{eq:phase-power-entrywise-gap} for all thirty-six original entries.
Each contributes
\(|(\Delta_3)_{ij}|^2+|(\Delta_5)_{ij}|^2>(6\,2^{21})^{-2}\).
Their sum is strictly greater than
\(36(6\,2^{21})^{-2}=2^{-42}\), proving
\eqref{eq:phase-power-frobenius-gap}. \(\square\)

\subsection{Coherent gauges, reversal and scope}\label{app:phase-power-scope}

For a unitary monomial \(M\), let \(M^{[m]}\) retain its permutation and
raise each nonzero phase to its \(m\)-th power. There is only one summand
in each entry of a monomial product, so
\begin{equation}\label{eq:phase-power-monomial-covariance}
 \Phi_m(M_x^*U_{xy}M_y)
   =(M_x^{[m]})^*\Phi_m(U_{xy})M_y^{[m]}.
\end{equation}
If all three transitions are changed coherently, the same intermediate
\(M_c^{[m]}\) cancels and
\begin{equation}\label{eq:phase-power-defect-covariance}
 \Delta_m'=(M_a^{[m]})^*\Delta_mM_b^{[m]}.
\end{equation}
Frobenius norms are preserved and entry moduli are permuted. For the
reversed triangle \((b,c,a)\), the transitions are the adjoints in reverse
order. Since \(\Phi_m(U^*)=\Phi_m(U)^*\), its defect is \(\Delta_m^*\).
Neither a coherent ray gauge nor reversal removes the estimate.

The conclusion excludes simultaneous cubic and quintic reuse of the
prescribed closing transition. The original \(m=1\) identity remains
exact. The proposition does not assert cubic failure alone, and it does
not exclude a different construction in which a new closing transition is
defined by the product \(\Phi_m(U_{ac})\Phi_m(U_{cb})\). Whether that new
closing is flat is a separate question. No global bound on the number of
MU bases, completion theorem or extension obstruction is inferred from
this estimate; its explicit constant is not claimed to be optimal.

\section{Computer-assisted proof near a fixed Hadamard matrix}\label{app:source-neighborhood}

This appendix gives the reference Hadamard matrix \(H_{\star}\), the interval
proof and the completed subdivision used in \hyperref[result:theorem-9.3]{Theorem 9.3}.
Fix an arbitrary exact Hadamard matrix \(H\) in the closed entrywise neighborhood
\(\mathcal U\) \hyperref[eq:source-neighborhood]{defined in Section 9.3}. The second
measurement basis is \(H/\sqrt6\); all bounds below hold uniformly for these
fixed matrices \(H\). The supplement \texttt{anc/source\_neighborhood/} contains
the inputs, reference-matrix checker, interval implementation and successful
full-run root, pair and coverage outputs. The proof uses physical conjugation
and requires no middle-minor, dephasing, companion-family or
companion-orthogonality assumption.

\subsection{The reference matrix and supplementary character sums}\label{sec:source-marks}

With \(F_2\) in unnormalized Hadamard form, use positive radicals and set
\begin{equation}
F_2=\begin{pmatrix}1&1\\1&-1\end{pmatrix},\quad
Y=\begin{pmatrix}0&-i\\i&0\end{pmatrix},\quad
A_0=\tfrac12F_2(-I+i\sqrt3Y),\quad B_0=-F_2-A_0,
\end{equation}
and define
\begin{equation}
z=\frac{3+4i}{5},\quad q=\frac{8+3i\sqrt3}{\sqrt{91}},\quad
Z_1=\begin{pmatrix}1&1\\z&-z\end{pmatrix},\quad Z_2=\overline{Z_1},\quad
Z_3=\begin{pmatrix}1&q\\1&-q\end{pmatrix},\quad Z_4=\overline{Z_3}.
\end{equation}
The reference Hadamard matrix is
\begin{equation}
H_{\star}=\begin{pmatrix}
F_2&Z_1&Z_2\\
Z_3&Z_3A_0Z_1/2&Z_3B_0Z_2/2\\
Z_4&Z_4B_0Z_1/2&Z_4A_0Z_2/2
\end{pmatrix}.
\label{eq:source-neighborhood-center}
\end{equation}
This is a point of Karlsson's known block construction \cite{K}. Exact rational arithmetic in \(\mathbb Q(\sqrt3,\sqrt{91},i)\) checks all 36 unit-entry identities and both full Gram tables, giving \(H_{\star}H_{\star}^*=H_{\star}^*H_{\star}=6I\). No classification is needed for those finite identities.

For a three-element row set \(S\subset\{0,\ldots,5\}\), define the character sums
\begin{equation}
f_S(H)=\sum_{c=0}^5
\left(\prod_{r\in S}H_{rc}\prod_{r\notin S}\overline{H_{rc}}\right)^3,
\qquad g_S(H)=f_S(H^*).
\end{equation}
For each column, \(f_S\) cubes the balanced product of entries in the selected
rows and conjugates in the complementary rows, then sums over all six columns;
\(g_S\) applies the same sum to \(H^*\). A subscript such as \(014\) abbreviates
the set \(\{0,1,4\}\), and \(012\) abbreviates \(\{0,1,2\}\). The reference-matrix
checker gives
\begin{equation}
\begin{aligned}
f_{014}(H_{\star})&=\frac{4175226}{3767855}
-\frac{1911888}{18839275}i\sqrt3,&
|f_{014}(H_{\star})|^2&=\frac{592882092}{470981875}>1,\\
g_{012}(H_{\star})&=-\frac{824736}{1421875}
+\frac{168696}{109375}i,&
|g_{012}(H_{\star})|^2&=\frac{351337536}{129390625}>2.
\end{aligned}
\label{eq:source-neighborhood-marks}
\end{equation}
Each summand is a product of eighteen unit-modulus factors. Telescoping a
product when every entry changes by at most \(\varepsilon\) costs at most
\(18\varepsilon\); summing the six terms costs \(108\varepsilon\). Taking the
adjoint preserves the entrywise distance. Thus, for every \(H\in\mathcal U\),
\begin{equation}
|f_{014}(H)|>1-108\cdot10^{-6}>0,\qquad
|g_{012}(H)|>\sqrt2-108\cdot10^{-6}>0.
\end{equation}
These supplementary character sums are not measured probabilities or
assumptions in the ray-count argument. Their complex values can acquire a phase
under row rephasing; the conclusion is their nonvanishing on \(\mathcal U\).
Consequently the sums \(\sum_{|S|=3}|f_S(H)|^2\) and
\(\sum_{|S|=3}|g_S(H)|^2\) are strictly positive, and so is their product.

\subsection{Five phase variables cover every physical candidate}

The closed five-angle cube parametrizes the physical unit torus, with
opposite endpoints representing the same phase. Every coordinate-unbiased
projective unit ray has a representative
\begin{equation}
v(\theta)=6^{-1/2}(1,e^{i\theta_1},\ldots,e^{i\theta_5}),\qquad
\theta_0=0,\quad\theta_r\in[-\pi,\pi],
\end{equation}
with the usual endpoint identifications. Writing \(x=\sqrt6v\), unbiasedness to \(H/\sqrt6\) is \(|(H^*x)_j|^2=6\). Parseval gives \(\sum_j|(H^*x)_j|^2=36\), so the first five equations imply the sixth. After subtracting the diagonal contribution, use the real residuals
\begin{equation}
F_j(\theta)=2\sum_{0\le r<s\le5}
\operatorname{Re}\bigl(\overline{H_{rj}}H_{sj}e^{i(\theta_r-\theta_s)}\bigr),
\quad j=0,\ldots,4.
\label{eq:source-neighborhood-residual}
\end{equation}
Only the ray's global phase was removed. Each column of any \(K\) in \eqref{eq:source-neighborhood-companion} belongs to this full five-angle problem after division by \(\sqrt6\), independently of the other columns.

For fixed \(H\), the residuals contain 75 coefficients
\(c_{jrs}=\overline{H_{rj}}H_{sj}\). Each differs from its value at
\(H_{\star}\) by at most \(2\varepsilon\). For two unit complex numbers at
chord distance \(d\), the principal angular distance is
\(2\arcsin(d/2)\le(\pi/2)d\le2d\), since
\(\arcsin u\le\pi u/2\) on \([0,1]\). This also holds at \(d=0\).
Every coefficient therefore has a real phase lift within the closed radius
\(4\varepsilon\) of its reference phase. The code encloses all 75 phases
independently in a larger box that contains every correlated physical input.
The enclosures are uniform over \(H\in\mathcal U\), but \(H\) stays fixed
during each contraction argument.

\subsection{Exact arithmetic and trigonometric enclosures}

All endpoints are dyadic numbers: integers divided by the power of two
\(Q=2^{44}\). Products take the minimum and maximum of all four integer
endpoint products, then round outward by mathematical floor and ceiling
division by \(Q\). Outward rounding enlarges the enclosure so that it contains
the exact result; negative truncation is corrected before accepting an endpoint.
Intermediate calculations use arbitrary-precision integers or an
overflow-checked 128-bit backend. Conversion to signed 64-bit endpoint storage
checks the storage range and rejects overflow. This conversion is distinct
from contracting an interval box in the search.

The input checker represents radical coefficients as sparse rational linear
combinations of powers of \(i,\sqrt3,\sqrt{91}\). It brackets the positive
\(\sqrt3,\sqrt{91}\) by integer square roots at denominator \(10^{60}\), evaluates
the coefficients outward, and checks every consumed complex rectangle, whose
real and imaginary intervals bound the two components separately. It also
regenerates the fixed-input header and compares its bytes with
\texttt{fixed\_inputs.h}. The proposed coefficient angles and 48 root centers are inputs to be validated,
not assumed exact phases or solutions. Their preserved provenance and the scope
of implementation independence are explained in Appendix~D.5.

The bound for \(\pi\) uses Machin's identity
\begin{equation}
\pi=16\arctan(1/5)-4\arctan(1/239).
\end{equation}
Rational tangent doubling and subtraction verify tangent one for
\(4\arctan(1/5)-\arctan(1/239)\). Alternating bounds place that angle between zero and \(4/5<\pi/2\), selecting \(\pi/4\). One hundred terms of the alternating series enclose \(\pi\) strictly between \(\mathrm{PI}_{\rm LO}/Q\) and \(\mathrm{PI}_{\rm HI}/Q\), where
\begin{equation}
\mathrm{PI}_{\rm LO}=55267482437722,\qquad
\mathrm{PI}_{\rm HI}=55267482437723.
\end{equation}

For dyadic arguments in \([-32,32]\), the evaluator reduces by the nearest quarter turn to a remainder of modulus below \(0.786\), then evaluates degree-19 sine and degree-18 cosine Taylor polynomials by fixed-point Horner arithmetic. The factorial coefficients, Horner products and squared remainder are rounded on the same grid. Since the squared remainder is below \(5/8\), coefficient error costs less than \(3/Q\), propagated multiplication error less than \(3/Q\), and squared-argument error less than \(8/Q\), using
\begin{equation}
\sum_{j\ge0}(5/8)^j<3,\qquad
\sum_{j\ge1}j(5/8)^{j-1}<8.
\end{equation}
The last sine multiplication costs less than \(1/Q\), and either Taylor tail at remainder modulus at most \(4/5\) is below \(1/Q\). At most 21 quarter turns occur; the rational \(\pi\) bracket and integer halving cost less than \(12/Q\) in the argument. Sine and cosine are 1-Lipschitz. The total error is below \(28/Q\), whereas the code pads each endpoint by \(1024/Q\). These inequalities are checked rationally by the input checker.

Interval trigonometric ranges include the padded endpoint evaluations and every possible extremum \(k\pi/2\). For an interval with integer endpoints \(l,h\), the inspected index bounds are
\begin{equation}
\begin{aligned}
k_{\min}&=\min\{\lfloor2l/\mathrm{PI}_{\rm LO}\rfloor,
\lfloor2l/\mathrm{PI}_{\rm HI}\rfloor\}-1,\\
k_{\max}&=\max\{\lceil2h/\mathrm{PI}_{\rm LO}\rceil,
\lceil2h/\mathrm{PI}_{\rm HI}\rceil\}+1.
\end{aligned}
\end{equation}
Both signs are covered. Each critical point is itself enclosed using the
two signed \(\pi\) endpoints. Width at least \(2\mathrm{PI}_{\rm LO}/Q\) returns
the conservative full range \([-1,1]\). Thus no untested interior extremum or
floating trigonometric value supplies an exclusion. For fixed \(H\), differentiating
a term in \eqref{eq:source-neighborhood-residual} with respect to \(\theta_r\)
gives \(-2\sin(\arg(c_{jrs})+\theta_r-\theta_s)\), while differentiation with
respect to \(\theta_s\) gives the positive version. The coordinate
\(\theta_0=0\) is fixed and is not differentiated.

\subsection{Uniform existence, uniqueness and complete coverage}

Let \(X\) be a closed box of phase angles and \(m\) its dyadic midpoint.
The interval Jacobian \(J(X)\) has entries enclosing the corresponding
derivatives throughout \(X\), uniformly over the allowed fixed matrices \(H\).
Choose an exact dyadic \(5\times5\) preconditioning matrix \(M\), typically
proposed as an approximation to the inverse Jacobian. Define the Krawczyk
enclosure, a box obtained by enclosing each component separately,
\begin{equation}
N(X)=m-MF(m)+(I-MJ(X))(X-m).
\label{eq:source-neighborhood-krawczyk}
\end{equation}
If \(F(x)=0\) at \(x\in X\), the segment integral identity for that fixed
\(H\) gives \(x=m-MF(m)+(I-M\overline J)(x-m)\), with
\(\overline J\in J(X)\). Hence every zero in \(X\) lies in \(N(X)\).
Disjointness excludes all zeros in \(X\), while intersection with \(N(X)\)
preserves them. Approximate inversion merely selects \(M\); its accuracy is
not an acceptance premise.

If \(N(X)\subset\operatorname{int}X\) and the infinity-norm bound for
\(I-MJ(X)\) is below one, \(x\mapsto x-MF(x)\) is a contraction of \(X\)
into itself. This norm bound is the maximum, over rows, of the sum of absolute
entry bounds. Iteration gives a unique fixed point. The same bound makes
\(MJ(m)\) invertible by its Neumann series, hence makes the square \(M\)
invertible; the fixed point therefore satisfies \(F=0\). This proves existence
and uniqueness for every fixed \(H\in\mathcal U\).

For each label \(0,\ldots,47\), a proposed center locates two candidate
boxes: a small box of radius \(\lceil Q/1000\rceil/Q\) and a large box of
radius \(\lceil Q/250\rceil/Q\). All 96 tests satisfy strict inclusion and
contraction. Writing \(\beta_{\rm ctr}\) for the largest recorded large-box
row-sum bound, every such bound is at most \(\beta_{\rm ctr}<3/8<1\).
Its dyadic endpoint is preserved in the supplement and checked in
\hyperref[sec:computational-material]{Section~10}; strict image inclusion is
a separate acceptance condition. Each small box is inside its large box;
the large boxes are pairwise disjoint and lie strictly in \((-\pi,\pi)^5\).
The validated small box supplies existence and the validated large box supplies
uniqueness of the same root. Thus every \(H\in\mathcal U\) has at least 48
distinct projective rays, with no endpoint duplication. Excluding all other
roots requires the complete subdivision below.

For exhaustion, start afresh from the full closed cube
\([-\mathrm{PI}_{\rm HI}/Q,\mathrm{PI}_{\rm HI}/Q]^5\). A leaf is a terminal
subdivision box accepted by a proved rule: an interval for one residual
\(F_j\) omits zero, \(N(X)\) is disjoint from \(X\), or \(X\) or \(N(X)\)
lies inside one validated large uniqueness domain. Otherwise intersect
outward with the image when useful and bisect into two closed children,
keeping their shared boundary. At each step, every possible initial root
remains in a pending box or is the unique root of a validated domain. This
induction proves coverage when no pending box remains. An unresolved small
box or uncatalogued contraction root causes failure.

The saved full execution of the verification implementation described in
Appendix~D.5 evaluated 2,902,327 nodes, with 868,092 residual-exclusion leaves, 582,905 disjointness
leaves, 167 certified-domain leaves, 1,451,163 binary splits and 238,023
root-preserving contractions; its maximum depth was 77. It read no checkpoint
and completed the full cube. The identities
\(\text{nodes}=2\,\text{splits}+1\) and \(\text{leaves}=\text{splits}+1\)
are consistency checks. The sound predicates, root-preserving steps and
completed execution together prove the upper bound of 48 rays; the counters
alone do not.

\subsection{Overlap predicates, evidence and reproduction}

The 1128 distinct label pairs enclose the overlap of unit-modulus phase vectors:
\begin{equation}
L_{ab}=1+\sum_{r=1}^5e^{i(\theta_{b,r}-\theta_{a,r})}.
\end{equation}
An edge is retained precisely when both real and imaginary overlap intervals
can contain zero. This is necessary for orthogonality and does not certify it.
For each of the 1092 nonedges, one real or imaginary coordinate of \(L_{ab}\)
has distance from zero at least \(\beta_{\rm sep}>1/20\), the minimum recorded
coordinate separation. This is a bound on an overlap coordinate, not a
distance between states. The dyadic endpoint and strict comparison are
reproduced by the supplementary displayed-bound check. Unit-ray overlaps
are \(L_{ab}/6\), giving \(>1/120\). The possible-orthogonality graph has
36 edges and no triangle, with structure \(K_{4,4}\sqcup20K_2\) as explained
in \hyperref[sec:source-neighborhood]{Section 9.3}. The six-position argument
there handles repeated labels and gives the strict Gram gap.

The supplement preserves the input data and successful reference-matrix,
root, pair and coverage outputs. It includes the sparse-radical checker and
integer interval code used for this proof, with the fixed-input header.
The verifier was implemented separately from the original producer search,
while reusing the proposed coefficient phases and 48 root centers in
\path{anc/source_neighborhood/inputs/data/input_certificate.json} and the
paired fixed-input headers, as well as the elementary Taylor/Krawczyk framework.
The exact input checks, root inclusion and contraction predicates, and pair
bounds validate the required assertions rather than presuming proposal accuracy;
the completed fresh full cover excludes additional roots. Independence here
concerns the implementation of these checks, not independent generation of the
inputs or reconstruction of the analytic proof without access to its existing
derivation. The public guide \path{anc/source_neighborhood/README.md}
explains this scope.

The arbitrary-precision and checked 128-bit backends have identical local
root/pair outputs; complete coverage was performed with the checked 128-bit
backend. The program and its completed execution form part of the
computer-assisted proof.

\hyperref[sec:computational-material]{Section~10} identifies the public source
code, inputs, stored outputs and replay commands. A roots-only replay checks
local existence, uniqueness and pair predicates but does not establish full
coverage. The saved full run and its input bindings support the theorem;
hashes and successful packaging alone do not prove it.

The physical conclusions transport under unitary monomial equivalence.
For unitary monomials \(L,R\), put \(\widetilde H=LHR\) and
\(\widetilde K=LK\). Then \(\widetilde H^*\widetilde K=R^*H^*K\) has the
same entry moduli, while \(\widetilde K^*\widetilde K=K^*K\). The common
rays are transported by \(L\), preserving their count and overlaps. Thus
the count, separation, no-third and Gram-gap conclusions hold on the unitary
monomial orbit of \(\mathcal U\). The indexed character sums above refer
to the fixed entry labels of \(\mathcal U\); coherent row and column
relabeling also relabels their indices. This transport does not classify
matrices outside that orbit.

\pagebreak[3]
\section{An exact example of companion rays without inherited anchor symmetry}\label{app:nonsetwise-witness}

A unitary can permute the rays of two MU measurements without permuting
the rays of a selected third measurement. The following exact example
separates this question from the given-companion hypothesis in
\hyperref[result:theorem-8.2]{Theorem~8.2}. We display the
matrices, normalization and support certificate so that the operator and
its representations in different bases can be checked together.
The example illustrates a limitation of symmetry inference; no priority
or classification claim is made for the construction.

\subsection{Radical data and the physical bases}\label{app:nonsetwise-radicals}

All square roots in this appendix are positive real roots, and
\(\sqrt{21}=\sqrt3\sqrt7\). Put
\begin{equation}\label{eq:nonsetwise-cycle-convention}
 \zeta=\frac{1+i\sqrt3}{2},\quad \omega=\zeta^2,\qquad
 P_3=\begin{pmatrix}0&0&1\\1&0&0\\0&1&0\end{pmatrix},\qquad
 R=\operatorname{diag}(P_3,P_3).
\end{equation}
Thus \(P_3\) is the three-cycle \(P_3e_0=e_1\), \(P_3e_1=e_2\),
\(P_3e_2=e_0\), with this orientation fixed.
The phase matrix of the first anchor is
\begin{equation}\label{eq:nonsetwise-source-matrix}
 S=\begin{pmatrix}
 1&1&\omega&1&1&\bar\omega\\
 \omega&1&1&\bar\omega&1&1\\
 1&\omega&1&1&\bar\omega&1\\
 1&\omega&1&-1&-\bar\omega&-1\\
 1&1&\omega&-1&-1&-\bar\omega\\
 \omega&1&1&-\bar\omega&-1&-1
 \end{pmatrix}.
\end{equation}
It is a two-circulant representative of the order-six construction
in \cite[Section~4]{S}. The following data specify the selected companion:
\begin{equation}\label{eq:nonsetwise-six-phases}
\begin{aligned}
 p_1&=-\frac{7+9\sqrt{21}}{56}
            -\frac{3i(7\sqrt3-\sqrt7)}{56},\\
 p_2&=\frac{-7+9\sqrt{21}}{56}
            -\frac{3i(7\sqrt3+\sqrt7)}{56},\\
 p_3&=\frac{35-3\sqrt{21}}{56}
            +\frac{i(7\sqrt3+15\sqrt7)}{56},\\
 p_4&=\frac{35+3\sqrt{21}}{56}
            +\frac{i(7\sqrt3-15\sqrt7)}{56},\\
 p_5&=\frac{-14+3\sqrt{21}}{28}
            +\frac{i(7\sqrt3+6\sqrt7)}{28},\\
 p_6&=\frac{-14-3\sqrt{21}}{28}
            +\frac{i(7\sqrt3-6\sqrt7)}{28},
\end{aligned}
\end{equation}
and
\begin{equation}\label{eq:nonsetwise-extra-phases}
 h=\frac{\sqrt3-5i}{2\sqrt7}
       =\frac{\sqrt{21}-5i\sqrt7}{14},\qquad
 q_+=\frac{3+\sqrt{21}}8-\frac{i(3\sqrt3-\sqrt7)}8,\qquad
 q_-=\frac{3-\sqrt{21}}8-\frac{i(3\sqrt3+\sqrt7)}8.
\end{equation}
The companion phase matrix is
\begin{equation}\label{eq:nonsetwise-companion-matrix}
 \widehat C=\begin{pmatrix}
 1&1&1&1&1&1\\
 p_1&p_2&p_3&p_4&p_5&p_6\\
 1&1&\bar\omega&\bar\omega&\omega&\omega\\
 h&-h&h\zeta&-h\zeta&h\zeta^2&-h\zeta^2\\
 h&-h&-h&h&h&-h\\
 q_+&q_-&\bar q_+&\bar q_-&(-3-i\sqrt7)/4&(-3+i\sqrt7)/4
 \end{pmatrix}.
\end{equation}
Define
\begin{equation}\label{eq:nonsetwise-normalization}
 U=\frac S{\sqrt6},\qquad C=\frac{\widehat C}{\sqrt6},\qquad
 E=S^*\widehat C,\qquad D_{\rm aux}=\frac E6=U^*C,
 \qquad V=C^*,\quad W=D_{\rm aux}^*.
\end{equation}
The notation \(D_{\rm aux}\) distinguishes this transition from the clock
operator elsewhere in the paper. The matrix \(E/\sqrt6\) is a phase
Hadamard with unit-modulus entries; the physical basis matrix is \(E/6\).
Together, \eqref{eq:nonsetwise-source-matrix}--\eqref{eq:nonsetwise-normalization}
specify every radical entry of these matrices. Expanded entries
of \(E\) and of the action below are also supplied in the ancillary exact
coefficient data.

For completeness, the denominator-cleared identities establishing the
physical MU conditions are
\begin{equation}\label{eq:nonsetwise-flatness-identities}
\begin{gathered}
 |S_{ij}|^2=|\widehat C_{ij}|^2=1,\qquad
 SS^*=S^*S=6I,\qquad
 \widehat C\widehat C^*=\widehat C^*\widehat C=6I,\\
 |E_{ij}|^2=6,\qquad EE^*=E^*E=36I,\qquad
 \widehat C E^*=6S,\qquad SE=6\widehat C.
\end{gathered}
\end{equation}
These are exact equalities for every entry, obtained by multiplication in
\(\mathbb Q(i,\sqrt3,\sqrt7)\). Hence \(U,C,D_{\rm aux},V,W\) are flat
unitaries. The original triple \((I,U,C)\) is MU because
\(U^*C=D_{\rm aux}\) is flat. Also
\begin{equation}\label{eq:nonsetwise-second-triple}
 V^*W=U,
\end{equation}
so \((I,V,W)\) is an actual MU triple as well.

\subsection{The action in the original and rebased descriptions}\label{app:nonsetwise-action}

The permutation \(R\) is unitary of exact order three, and direct
multiplication gives \(SR=RS\), hence \(UR=RU\). Thus it permutes the
six column rays of both original anchors \(I\) and \(U\), in two cycles
of length three. Its matrix in the \(C\) basis is
\begin{equation}\label{eq:nonsetwise-shared-conjugate}
 A=C^*RC=\frac{\widehat C^*R\widehat C}{6}
       =D_{\rm aux}^*RD_{\rm aux}=\frac{E^*RE}{36}.
\end{equation}
The second equality of conjugates also follows algebraically from
\(D_{\rm aux}=U^*C\) and \(UR=RU\). The corresponding covariance
identities are
\begin{equation}\label{eq:nonsetwise-original-covariance}
 RC=CA,\qquad RD_{\rm aux}=D_{\rm aux}A.
\end{equation}
Since \(A\) is conjugate to \(R\), it is unitary and has exact order
three, with
\begin{equation}\label{eq:nonsetwise-action-spectrum}
 \chi_A(t)=(t^3-1)^2,\qquad
 \operatorname{tr}A=\operatorname{tr}A^2=0.
\end{equation}
Here \(\chi_A(t)=\det(tI-A)\) is the characteristic polynomial. Its eigenvalues \(1,\omega,\bar\omega\) each have multiplicity two.

This representation must be distinguished from its action after a common
coordinate change. Left multiplication by \(C^*\) sends the whole
original triple \((I,U,C)\) to \((V,W,I)\) and sends the operator \(R\)
to \(A\). In that description,
\begin{equation}\label{eq:nonsetwise-rebased-covariance}
 A=VRV^*=WRW^*,\qquad AV=VR,\qquad AW=WR.
\end{equation}
Thus \(A\) \emph{does} preserve the ray sets of the rebased \(V\) and
\(W\). The failure established below concerns \(R\) on the original
\(C\) and \(D_{\rm aux}\) rays, or equivalently \(A\) on the coordinate
rays after this common rebasing.

\subsection{Exact support and squared moduli}\label{app:nonsetwise-support}

The support can be read from the full squared-modulus matrix. Put
\begin{equation}\label{eq:nonsetwise-modulus-values}
 s=\frac1{16},\qquad v=\frac9{28},\qquad
 u_- =\frac{33-3\sqrt{21}}{224},\qquad
 u_+ =\frac{33+3\sqrt{21}}{224}.
\end{equation}
Exact evaluation of \eqref{eq:nonsetwise-shared-conjugate} gives
\begin{equation}\label{eq:nonsetwise-full-modulus-matrix}
 (|A_{ij}|^2)_{i,j=0}^5=
 \begin{pmatrix}
 s&0&v&u_+&u_-&v\\
 0&s&u_-&v&v&u_+\\
 v&u_+&s&0&v&u_-\\
 u_-&v&0&s&u_+&v\\
 u_+&v&v&u_-&s&0\\
 v&u_-&u_+&v&0&s
 \end{pmatrix}.
\end{equation}
All four displayed nonzero values are positive: for the only potentially
questionable one, \(33>3\sqrt{21}\) follows from \(11^2>21\).
There are exactly six zero entries, at the zero-based positions
\begin{equation}\label{eq:nonsetwise-zero-positions}
 (0,1),\ (1,0),\ (2,3),\ (3,2),\ (4,5),\ (5,4).
\end{equation}
Each row and column has five nonzero entries, so \(A\) has thirty nonzero entries in total.
The complete multiplicities are
\begin{equation}\label{eq:nonsetwise-modulus-counts}
\begin{array}{c|ccccc}
 \text{squared modulus}&0&1/16&9/28&u_-&u_+\\\hline
 \text{number of entries}&6&6&12&6&6.
\end{array}
\end{equation}

To interpret this calculation, let \(T\) be a unitary and \(B\) an
orthonormal column basis. The six rays of \(B\) are preserved setwise
precisely when \(Tb_j=\lambda_jb_{\pi(j)}\) for a permutation \(\pi\)
and unit phases \(\lambda_j\). This is equivalent to \(B^*TB\) being
a unitary monomial matrix. A six-dimensional unitary monomial matrix has one unit-modulus nonzero entry in each row and each column, hence six in total. Therefore \(A\) is not monomial, and
\eqref{eq:nonsetwise-original-covariance} shows that \(R\) preserves
neither the original \(C\) rays nor the original \(D_{\rm aux}\) rays.
Column phases and permutations conjugate this representation by a
unitary monomial matrix and preserve its support count. The conclusion therefore
does not depend on the labeling or phasing of those same ray sets.

\subsection{Distinct pair tests and the limit of their conclusions}\label{app:nonsetwise-pair-controls}

The ancillary pair tests use two different predicates. For two distinct rows of a phase matrix, take the six entrywise ratios in matching columns; these are unit phases. The strong common-phase predicate means a multiset
\(\lambda\{1,1,1,-1,-1,-1\}\); the general antipodal predicate means
invariance of its multiplicity function under \(z\mapsto-z\), allowing
three arbitrary antipodal pairs. The latter is weaker. Both existence
tests are invariant under row and column unitary monomial equivalence, and there
are analogous tests on column quotients. Exhausting the fifteen unordered
pairs in each orientation gives
\begin{equation}\label{eq:nonsetwise-pair-profiles}
\begin{array}{c|cc}
 \text{phase matrix}&\text{strong (rows, columns)}
                    &\text{general (rows, columns)}\\\hline
 \widehat C&(3,0)&(5,9)\\
 E/\sqrt6&(0,0)&(9,9).
\end{array}
\end{equation}
For \(E\), the equivalent products of one entry and the conjugate of the other are divided by its squared entry modulus six. In particular, absence of a strong common-phase pair must
not be read as absence of arbitrary antipodal pairs.

For reference, use the transposed-Fourier phase convention
\begin{equation}\label{eq:nonsetwise-fourier-convention}
 F^T(x,y)=\begin{pmatrix}
 1&1&1&1&1&1\\
 1&\bar\omega&\omega&x&\bar\omega x&\omega x\\
 1&\omega&\bar\omega&y&\omega y&\bar\omega y\\
 1&1&1&-1&-1&-1\\
 1&\bar\omega&\omega&-x&-\bar\omega x&-\omega x\\
 1&\omega&\bar\omega&-y&-\omega y&-\bar\omega y
 \end{pmatrix}.
\end{equation}
Every member has three strong row pairs; its transpose has three strong
column pairs. Thus the strong \((0,0)\) profile excludes those two
Fourier orientations and their unitary monomial equivalence classes for
\(D_{\rm aux}\). This uses only a necessary condition. On the other hand,
reordering \(\widehat C\)'s rows as \((0,2,1,4,3,5)\) and columns as
\((0,3,4,1,2,5)\), then dephasing, gives
\begin{equation}\label{eq:nonsetwise-companion-fourier-witness}
 F^T\left(1,\frac{-1+3i\sqrt7}{8}\right).
\end{equation}
The second parameter has modulus one. The ancillary data also give the
explicit Fourier equivalence of \(S\):
\begin{equation}\label{eq:nonsetwise-source-fourier-witness}
 S_{ij}=\zeta^{r_i+c_j}F^T(\omega,\omega)_{\rho_i,\kappa_j},\qquad
 \begin{aligned}
 \rho&=(0,1,2,5,3,4),&\kappa&=(3,4,5,1,0,2),\\
 r&=(0,0,4,1,3,3),&c&=(0,0,2,0,0,4).
 \end{aligned}
\end{equation}
Individual equivalences of these matrices are not one common change of
the anchored triple and its operator.

Finally, all 225 positional two-by-two cross ratios of \(E\) are retained
in the exact coefficient data. Exactly 33 equal \(-1\), including
\begin{equation}\label{eq:nonsetwise-pinned-cross-ratio}
 \frac{E_{00}E_{31}}{E_{01}E_{30}}=-1.
\end{equation}
All cross ratios in a Tao-equivalent matrix are cube roots of unity, so
this excludes the Tao class for \(D_{\rm aux}\). Neither this test nor
the general profiles imply membership or nonmembership in the
two-circulant family. The symmetry conclusion above needs none of these
classification controls.

\subsection{Exact reproduction and relation to the clock hypothesis}\label{app:nonsetwise-reproduction}

The exact checker and expanded data are identified in \hyperref[sec:computational-material]{Section~10}. The checker uses only the Python standard library. Each field element has eight rational
coefficients in the ordered basis
\begin{equation}\label{eq:nonsetwise-field-basis}
 1,\ i,\ \sqrt3,\ i\sqrt3,\ \sqrt7,\ i\sqrt7,
 \sqrt{21},\ i\sqrt{21}.
\end{equation}
Products use \(i^2=-1\), \((\sqrt3)^2=3\) and \((\sqrt7)^2=7\),
and complex conjugation changes the signs of the four coefficients
containing \(i\). Rational coefficient equality is exact; no tolerance
or random choice is used. The checker reconstructs the displayed radical matrices, checks every entry of the Gram and covariance identities, and enumerates all support decisions, pair profiles and positional cross ratios. The data file records all entries of \(S,\widehat C,E,R,A\), the thirty support positions, their squared moduli and the complete finite pair and cross-ratio results.

This example supplies no simultaneous change of coordinates carrying
its anchors and operator to the data \((I,K_\eta,D)\) of
\hyperref[result:theorem-8.2]{Theorem~8.2}. Such a change must transport
the operator and all bases together. Unitary similarity of \(R\) to an
order-three clock, or an individual Fourier equivalence of an anchor,
does not establish that simultaneous correspondence. The theorem applies to a given MU triple whose third basis has \(D\)-invariant original rays and concludes that no extra common-unbiased ray exists. An extra ray is only a temporary contradiction hypothesis in its proof, where it enables the conditional companion-identification step. The conclusion of this example is that symmetry of an anchor pair does not force setwise invariance of a selected companion's original rays. It therefore does not contradict the theorem or settle a global extension problem.

\phantomsection
\end{document}